\providecommand\PaperColumns{twocolumn}
\edef\PaperOptions{aps,prd,10pt,\PaperColumns,superscriptaddress,nofootinbib,floatfix}
\expandafter\documentclass\expandafter[\PaperOptions]{revtex4-2}

\usepackage{amsmath,amssymb}
\allowdisplaybreaks                
\usepackage{graphicx}
\usepackage{booktabs}
\usepackage[dvipsnames]{xcolor}
\makeatletter
\newcommand{\raggedcolumns}{\let\rebar@parboxrestore\@arrayparboxrestore
  \def\@arrayparboxrestore{\rebar@parboxrestore\let\rebar@cr\\\raggedright\let\\\rebar@cr}}
\makeatother
\makeatletter\AtBeginDocument{\AtEndDocument{\par\close@column}}\makeatother
\usepackage{pgfplots}               
\pgfplotsset{compat=1.15}
\usepackage[colorlinks=true,linkcolor=MidnightBlue,citecolor=MidnightBlue,urlcolor=MidnightBlue,
            breaklinks=true]{hyperref}
\hypersetup{pdftitle={Ten-loop fixed-dimension renormalization-group functions of the three-dimensional O(N) model
  from distributed GPU-accelerated parametric integration},pdfauthor={Paolo Ribeca}}

\newcommand{\TBD}[1]{\textcolor{red}{\bfseries[TBD\if\relax\detokenize{#1}\relax\else: #1\fi]}}

\renewcommand{\acknowledgmentsname}{Acknowledgements}

\newcommand{\ubar}{\ensuremath{\bar u}}
\newcommand{\Rbar}{\ensuremath{\bar R}}

\makeatletter
\protected\def\check@link#1{\hyperlink{check.#1}{#1}}
\DeclareRobustCommand{\check}[1]{{\small[check~\check@link{#1}]}}
\DeclareRobustCommand{\checks}[1]{{\edef\check@ids{\zap@space#1 \@empty}\def\check@sep{}\def\check@list{}%
  \@for\check@id:=\check@ids\do{\edef\check@list{\check@list\check@sep\noexpand\check@link{\check@id}}%
    \def\check@sep{,\space}}%
  \small[checks~\check@list]}}
\newcommand*{\checkid}[1]{\hypertarget{check.#1}{#1}}
\makeatother

\newcommand{\relBseven}{$2.8\times10^{-6}$}
\newcommand{\nclBseven}{3150}

\newcommand{\nclBnine}{166\,047}

\newcommand{\nclBten}{1\,377\,678}

\newcommand{\relHseven}{$4.9\times10^{-6}$}
\newcommand{\nclHseven}{209}

\newcommand{\relHeight}{$4.4\times10^{-6}$}

\newcommand{\nclHten}{46\,102}

\newcommand{\relCseven}{$4.9\times10^{-6}$}
\newcommand{\nclCseven}{265}

\newcommand{\nclCten}{53\,397}

\newcommand{\etaEight}{0.002910571(13)}

\newcommand{\etaEightNzeroK}{3.475878(13)}

\newcommand{\etaEightNoneK}{2.910571(13)}

\newcommand{\etaEightNtwoK}{2.264839(12)}

\newcommand{\etaEightNthreeK}{1.721335(12)}

\newcommand{\etaEightNfourK}{1.302520(11)}
\newcommand{\etaNineRel}{$1\times10^{-5}$}

\newcommand{\etaNineNzeroK}{$-6.61920(5)$}

\newcommand{\etaNineNoneK}{$-5.10339(5)$}

\newcommand{\etaNineNtwoK}{$-3.68159(5)$}

\newcommand{\etaNineNthreeK}{$-2.60888(4)$}

\newcommand{\etaNineNfourK}{$-1.84985(4)$}
\newcommand{\betaNineRel}{$2\times10^{-6}$}
\newcommand{\betaNine}{$-2.366420(6)$}
\newcommand{\betaNineNzero}{$-3.834306(8)$}
\newcommand{\betaNineNone}{$-2.366420(6)$}
\newcommand{\betaNineNtwo}{$-1.510105(5)$}
\newcommand{\betaNineNthree}{$-0.995275(4)$}
\newcommand{\betaNineNfour}{$-0.675610(3)$}
\newcommand{\betaTenRel}{$1\times10^{-5}$}
\newcommand{\betaTen}{4.78115(5)}
\newcommand{\betaTenNzero}{8.41939(6)}
\newcommand{\betaTenNone}{4.78115(5)}
\newcommand{\betaTenNtwo}{2.83206(4)}
\newcommand{\betaTenNthree}{1.74500(3)}
\newcommand{\betaTenNfour}{1.11399(2)}

\newcommand{\nuEight}{$-0.0317869(4)$}

\newcommand{\nuEightNzeroC}{$-3.89149(4)$}

\newcommand{\nuEightNoneC}{$-3.17869(4)$}

\newcommand{\nuEightNtwoC}{$-2.41430(4)$}

\newcommand{\nuEightNthreeC}{$-1.79059(4)$}

\newcommand{\nuEightNfourC}{$-1.32058(4)$}
\newcommand{\nuNineRel}{$1\times10^{-5}$}

\newcommand{\nuNineNzeroC}{8.08899(8)}

\newcommand{\nuNineNoneC}{6.13316(8)}

\newcommand{\nuNineNtwoC}{4.35887(7)}

\newcommand{\nuNineNthreeC}{3.04851(7)}

\newcommand{\nuNineNfourC}{2.13758(7)}

\newcommand{\etaTenRel}{$1\times10^{-5}$}
\newcommand{\etaTen}{0.00995358(14)}

\newcommand{\etaTenNzeroK}{13.96951(15)}

\newcommand{\etaTenNoneK}{9.95358(14)}

\newcommand{\etaTenNtwoK}{6.69457(12)}

\newcommand{\etaTenNthreeK}{4.45525(11)}

\newcommand{\etaTenNfourK}{2.98472(10)}
\newcommand{\nuTenRel}{$4\times10^{-5}$}
\newcommand{\nuTen}{$-0.127571(5)$}

\newcommand{\nuTenNzeroC}{$-18.1875(5)$}

\newcommand{\nuTenNoneC}{$-12.7571(5)$}

\newcommand{\nuTenNtwoC}{$-8.4522(4)$}

\newcommand{\nuTenNthreeC}{$-5.5437(4)$}

\newcommand{\nuTenNfourC}{$-3.6611(4)$}

\newcommand{\betaElevenRel}{$1\times10^{-5}$}
\newcommand{\betaEleven}{$-10.38718(12)$}
\newcommand{\betaElevenNzero}{$-19.91716(17)$}
\newcommand{\betaElevenNone}{$-10.38718(12)$}
\newcommand{\betaElevenNtwo}{$-5.70235(9)$}
\newcommand{\betaElevenNthree}{$-3.28063(7)$}
\newcommand{\betaElevenNfour}{$-1.96751(5)$}
\newcommand{\betaratios}{$1.32$, $1.51$, $1.70$, $1.86$, $2.02$ and $2.17$}
\newcommand{\betaratiokmin}{5}
\newcommand{\betaratiokmax}{10}
\newcommand{\etaEightPull}{$+0.8$}
\newcommand{\etaEightGain}{120}
\newcommand{\nuEightPull}{$-1.0$}
\newcommand{\nuEightGain}{4}
\newcommand{\precBsix}{0.82}

\newcommand{\precHsix}{2.4}

\newcommand{\precCsix}{0.88}

\newcommand{\precBseven}{1.5}

\newcommand{\precHseven}{11}

\newcommand{\precCseven}{2.9}

\newcommand{\fourrelmax}{$7.6\times10^{-7}$}
\newcommand{\fourjointdev}{$(-0.2\pm4.2)\times10^{-7}$}
\newcommand{\fourbiasbound}{$8.5\times10^{-7}$}
\newcommand{\hfourbiasbound}{$1.5\times10^{-6}$}
\newcommand{\newrelmin}{$2\times10^{-6}$}
\newcommand{\newrelmax}{$4\times10^{-5}$}
\newcommand{\kurtmaxseven}{$1.3\times10^{6}$}
\newcommand{\kurtmaxsevenorder}{$B_{7}$}
\newcommand{\wmaxmaxseven}{$2.2\times10^{6}$}
\newcommand{\wmaxmaxsevenorder}{$H_{7}$}
\newcommand{\kurtmaxhigh}{$1.6\times10^{7}$}
\newcommand{\kurtmaxhighorder}{$H_{10}$}
\newcommand{\wmaxmaxhigh}{$6.8\times10^{9}$}
\newcommand{\wmaxmaxhighorder}{$C_{10}$}
\newcommand{\rmaxcountBseven}{144}
\newcommand{\rmaxmaxBseven}{0.051}
\newcommand{\rmaxcountHseven}{54}

\newcommand{\rmaxcountCseven}{76}

\newcommand{\rmsmaxlow}{$1.8\times10^{-5}$}
\newcommand{\rmscounthigh}{668}
\newcommand{\rmsmaxhigh}{2.8}
\newcommand{\rmsmaxhighorder}{$B_{10}$}
\newcommand{\rcohmax}{0.17}
\newcommand{\rcohmaxorder}{$C_{8}$}
\newcommand{\valncmp}{15}
\newcommand{\valzmax}{1.35}
\newcommand{\valchisq}{6.5}
\newcommand{\valpval}{0.97}
\newcommand{\valNchisqmin}{5.8}
\newcommand{\valNchisqmax}{7.4}
\newcommand{\valNzmax}{1.71}
\newcommand{\mergeruns}{386}
\newcommand{\mergeorders}{17}
\newcommand{\mergeLmin}{3}
\newcommand{\mergeLmax}{8}
\newcommand{\mergechisq}{123\,786}
\newcommand{\mergedof}{124\,342}
\newcommand{\mergeratio}{0.996}
\newcommand{\mergezmean}{$-0.07$}
\newcommand{\mergezsd}{0.99}
\newcommand{\mergesumn}{386}
\newcommand{\mergesumchisq}{370.3}
\newcommand{\mergesump}{0.71}
\newcommand{\refinputs}{$B_{3}$--$B_{6}$, $H_{3}$, $H_{4}$ and $C_{3}$--$C_{6}$}
\newcommand{\ledgeronlyzmax}{0.0075}

\newcommand*{\lofitkmin}{6}

\newcommand*{\lofitresbeta}{4\times10^{-4}}\newcommand*{\lofitresphi}{3\times10^{-3}}
\newcommand*{\lofitdonemin}{-4.9}\newcommand*{\lofitdonemax}{-3.0}
\newcommand*{\lofitdtwomin}{13}\newcommand*{\lofitdtwomax}{17}
\newcommand*{\lofitcomin}{1.2}\newcommand*{\lofitcomax}{1.8}
\newcommand*{\lofitpredbeta}{1\times10^{-3}}\newcommand*{\lofitpredbetaone}{4\times10^{-4}}\newcommand*{\lofitpredphi}{5\times10^{-3}}
\newcommand*{\newpolybetaeta}{%
\begin{ruledtabular}
\begin{tabular}{rllllll}
$j$ & $p_{\beta,9}$ & $p_{\beta,10}$ & $p_{\beta,11}$ & $p_{\eta,8}$ & $p_{\eta,9}$ & $p_{\eta,10}$ \\
\hline
0 & $-6.432899(13)\times10^{7}$ & $1.130031(9)\times10^{9}$ & $-2.138589(19)\times10^{10}$ & 58315.6(2) & $-8.88414(7)\times10^{5}$ & $1.499965(16)\times10^{7}$ \\
1 & $-3.225258(10)\times10^{7}$ & $6.08032(7)\times10^{8}$ & $-1.224609(16)\times10^{10}$ & 52164.0(2) & $-8.28881(8)\times10^{5}$ & $1.46114(2)\times10^{7}$ \\
2 & $-5.02956(3)\times10^{6}$ & $1.07233(2)\times10^{8}$ & $-2.39487(6)\times10^{9}$ & 13685.34(10) & $-2.36573(4)\times10^{5}$ & $4.54534(10)\times10^{6}$ \\
3 & $-2.53755(5)\times10^{5}$ & $6.9100(5)\times10^{6}$ & $-1.86296(14)\times10^{8}$ & 1113.67(2) & $-22922.3(9)$ & $5.3144(3)\times10^{5}$ \\
4 & $-1755.7(7)$ & $1.1017(7)\times10^{5}$ & $-4.715(2)\times10^{6}$ & 11.847(3) & $-379.02(15)$ & 18224(5) \\
5 & 0.56(6) & $-392(8)$ & $-4.1(3)\times10^{3}$ & 0.1561(3) & 11.036(18) & $-41.7(6)$ \\
6 & 0.331(5) & 2.5(7) & $-1.1(3)\times10^{2}$ & $-0.07039(3)$ & $-0.3953(18)$ & 4.15(8) \\
7 & 0.0076(2) & 0.24(5) & 2(2) & $-0.0034718(13)$ & $-0.07014(15)$ & $-0.687(7)$ \\
8 &  & 0.0070(19) & 0.18(16) &  & $-0.002790(7)$ & $-0.0673(6)$ \\
9 &  &  & 0.009(6) &  &  & $-0.00228(2)$ \\
\end{tabular}
\end{ruledtabular}
}
\newcommand*{\newpolyphinu}{%
\begin{ruledtabular}
\begin{tabular}{rllllll}
$j$ & $p_{\eta_{\phi^2},8}$ & $p_{\eta_{\phi^2},9}$ & $p_{\eta_{\phi^2},10}$ & $p_{\nu,8}$ & $p_{\nu,9}$ & $p_{\nu,10}$ \\
\hline
0 & $3.10364(3)\times10^{6}$ & $-5.07683(4)\times10^{7}$ & $9.0136(2)\times10^{8}$ & $-6.52885(6)\times10^{5}$ & $1.085686(10)\times10^{7}$ & $-1.95287(5)\times10^{8}$ \\
1 & $2.90212(3)\times10^{6}$ & $-4.95301(5)\times10^{7}$ & $9.1367(3)\times10^{8}$ & $-5.67424(7)\times10^{5}$ & $9.93775(12)\times10^{6}$ & $-1.87088(6)\times10^{8}$ \\
2 & $8.33826(13)\times10^{5}$ & $-1.54661(2)\times10^{7}$ & $3.06457(14)\times10^{8}$ & $-1.39066(3)\times10^{5}$ & $2.72429(6)\times10^{6}$ & $-5.6224(3)\times10^{7}$ \\
3 & 82917(3) & $-1.82787(6)\times10^{6}$ & $4.1482(4)\times10^{7}$ & $-9076.3(7)$ & $2.39743(16)\times10^{5}$ & $-6.0652(10)\times10^{6}$ \\
4 & 1734.1(4) & $-6.5507(10)\times10^{4}$ & $2.0285(8)\times10^{6}$ & 115.87(10) & 2450(3) & $-1.5181(19)\times10^{5}$ \\
5 & $-30.08(4)$ & 28.1(13) & $1.349(11)\times10^{4}$ & 9.238(11) & 20.6(3) & $2.67(3)\times10^{3}$ \\
6 & $-1.154(4)$ & $-22.46(13)$ & $-2.05(13)\times10^{2}$ & 2.2339(9) & 16.90(4) & 49(3) \\
7 & $-0.02393(19)$ & $-0.835(10)$ & $-17.5(12)$ & 0.11383(5) & 1.941(3) & 20.9(3) \\
8 &  & $-0.0169(5)$ & $-0.64(9)$ & $2^{-9}$ & 0.07480(13) & 1.58(2) \\
9 &  &  & $-0.012(4)$ &  & $2^{-10}$ & 0.0489(10) \\
10 &  &  &  &  &  & $2^{-11}$ \\
\end{tabular}
\end{ruledtabular}
}
\newcommand*{\largeNorders}{27}
\newcommand*{\largeNLmin}{2}
\newcommand*{\largeNLmax}{10}
\newcommand*{\largeNchisq}{23.7}
\newcommand*{\largeNpvalue}{0.65}
\newcommand*{\largeNzmax}{2.3}
\newcommand*{\largeNphi}{$3\times10^{-10}$}
\newcommand*{\routesumrows}{%
$\Gamma^{(4)}$ & 7 & 13509.529(19) & 13509.5443(3) & $-0.79$ \\
$\Gamma^{(2)}$ & 8 & 6165.902(10) & 6165.9013(5) & $+0.12$ \\
$\partial_{p^2}\Gamma^{(2)}$ & 8 & $-543.6422(15)$ & $-543.64217(15)$ & $-0.01$ \\
$\Gamma^{(4)}$ & 8 & 84849.82(13) & --- & --- \\
$\Gamma^{(2)}$ & 9 & 32627.58(5) & --- & --- \\
$\partial_{p^2}\Gamma^{(2)}$ & 9 & $-3201.296(8)$ & --- & --- \\
$\Gamma^{(4)}$ & 9 & 559371.9(2) & --- & --- \\
$\Gamma^{(2)}$ & 10 & 185488.29(12) & --- & --- \\
$\partial_{p^2}\Gamma^{(2)}$ & 10 & $-19767.96(2)$ & --- & --- \\
}
\newcommand*{\routecoefrows}{%
$\beta_9$ & & $-2.366420(6)$ & $-2.3671(16)$ & $+0.40$ \\
$10^3\eta_9$ & & $-5.10339(5)$ & $-5.121(15)$ & $+1.16$ \\
$\eta_{\phi^2,9}$ & & $-0.303695(3)$ & $-0.3037(3)$ & $+0.15$ \\
$10^2\nu_9$ & & 6.13316(8) & 6.134(8) & $-0.08$ \\
$\beta_{10}$ & & 4.78115(5) & 4.781(8) & $-0.01$ \\
$10^3\eta_{10}$ & & 9.95358(14) & 9.84(7) & $+1.53$ \\
$\eta_{\phi^2,10}$ & & 0.620920(18) & 0.6206(15) & $+0.22$ \\
$10^2\nu_{10}$ & & $-12.7571(5)$ & $-12.75(4)$ & $-0.14$ \\
}
\newcommand*{\routeBhours}{26.8}
\newcommand*{\routeBzmax}{1.5}
\newcommand*{\routeBratiomin}{90}
\newcommand*{\routeBratiomax}{500}
\newcommand*{\routeBrelmin}{$7\times10^{-4}$}
\newcommand*{\routeBrelmax}{$7\times10^{-3}$}
\newcommand*{\routeBzphibeta}{$-1.2$}
\newcommand*{\routeBzphieta}{$+4.0$}
\newcommand*{\routeBzphietaphi}{$+1.4$}
\newcommand*{\routeBzphinu}{$-1.2$}
\newcommand*{\routeBzphietaeight}{$+2.9$}
\newcommand*{\routeBquadratio}{3.3}
\newcommand*{\exprows}{%
0 & 7th order, Ref.~\cite{GuidaZinnJustin1998} & 1.413(6) & 0.5882(11) & 0.0284(25) & 0.812(16) \\%
 & 8th order, Ref.~\cite{Phi4tools2024} and this work & 1.4175(64) & 0.58825(66) & 0.0278(26) & 0.798(24) \\%
 & 9th order, this work & 1.4137(54) & 0.58806(57) & 0.0281(20) & 0.807(25) \\%
 & 10th order, this work & 1.4098(41) & 0.58788(45) & 0.0283(18) & 0.818(26) \\%
 & self-avoiding walks, Refs.~\cite{ClisbyDunweg2016,Clisby2017} & --- & 0.58759700(40) & 0.0310434(21) & 0.899(14) \\%
\hline%
1 & 7th order, Ref.~\cite{GuidaZinnJustin1998} & 1.411(4) & 0.6304(13) & 0.0335(25) & 0.799(11) \\%
 & 8th order, Ref.~\cite{Phi4tools2024} and this work & 1.4141(48) & 0.63024(94) & 0.0326(26) & 0.791(17) \\%
 & 9th order, this work & 1.4116(33) & 0.63006(66) & 0.0329(20) & 0.796(15) \\%
 & 10th order, this work & 1.4093(22) & 0.62997(51) & 0.0333(18) & 0.803(14) \\%
 & conformal bootstrap, Refs.~\cite{ChangEtAl2025,SimmonsDuffin2017,Reehorst2022} & --- & 0.62997097(12) & 0.036297612(48) & 0.82968(23) \\%
\hline%
2 & 7th order, Ref.~\cite{GuidaZinnJustin1998} & 1.403(3) & 0.6703(15) & 0.0354(25) & 0.789(11) \\%
 & 8th order, Ref.~\cite{Phi4tools2024} and this work & 1.4049(31) & 0.6698(14) & 0.0344(23) & 0.785(11) \\%
 & 9th order, this work & 1.4038(12) & 0.66998(94) & 0.0347(17) & 0.7869(54) \\%
 & 10th order, this work & 1.40297(50) & 0.67019(74) & 0.0352(16) & 0.7895(31) \\%
 & Monte Carlo, Ref.~\cite{Hasenbusch2019} & --- & 0.67169(7) & 0.03810(8) & 0.789(4) \\%
 & conformal bootstrap, Ref.~\cite{ChesterEtAl2020} & --- & 0.67175(10) & 0.038176(44) & 0.794(8) \\%
\hline%
3 & 7th order, Ref.~\cite{GuidaZinnJustin1998} & 1.390(4) & 0.7073(35) & 0.0355(25) & 0.782(13) \\%
 & 8th order, Ref.~\cite{Phi4tools2024} and this work & 1.3913(16) & 0.7062(19) & 0.0343(18) & 0.7816(58) \\%
 & 9th order, this work & 1.39117(58) & 0.7070(17) & 0.0347(14) & 0.7818(24) \\%
 & 10th order, this work & 1.39138(16) & 0.7075(13) & 0.0352(13) & 0.78148(82) \\%
 & Monte Carlo, Ref.~\cite{Hasenbusch2020} & --- & 0.71164(10) & 0.03784(5) & 0.759(2) \\%
 & conformal bootstrap, Ref.~\cite{ChesterEtAl2021} & --- & 0.71168(41) & 0.03787(13) & --- \\%
}%
\newcommand*{\rsPullNuNzero}{+0.6}\newcommand*{\rsPullEtaNzero}{-1.5}\newcommand*{\rsPullOmegaNzero}{-2.7}%
\newcommand*{\rsUstarBNone}{1.4141(48)}\newcommand*{\rsNuBNone}{0.63024(94)}\newcommand*{\rsEtaBNone}{0.0326(26)}\newcommand*{\rsOmegaBNone}{0.791(17)}%
\newcommand*{\rsUstarDNone}{1.4093(22)}\newcommand*{\rsNuDNone}{0.62997(51)}\newcommand*{\rsEtaDNone}{0.0333(18)}\newcommand*{\rsOmegaDNone}{0.803(14)}%
\newcommand*{\rsUstarDtenNone}{1.4106(21)}%
\newcommand*{\rsPullNuNone}{0.0}\newcommand*{\rsPullEtaNone}{-1.6}\newcommand*{\rsPullOmegaNone}{-1.9}%
\newcommand*{\rsPullNuNtwo}{-2.0}\newcommand*{\rsPullEtaNtwo}{-1.8}\newcommand*{\rsPullOmegaNtwo}{+0.1}%
\newcommand*{\rsOmegaDNthree}{0.78148(82)}%
\newcommand*{\rsPullNuNthree}{-3.1}\newcommand*{\rsPullEtaNthree}{-2.0}\newcommand*{\rsPullOmegaNthree}{+10.4}%
\newcommand*{\rsBenchMCEtaNtwo}{0.03810(8)}\newcommand*{\rsBenchMCNuNtwo}{0.67169(7)}%
\newcommand*{\rsBenchMCEtaNthree}{0.03784(5)}\newcommand*{\rsBenchMCNuNthree}{0.71164(10)}\newcommand*{\rsBenchMCOmegaNthree}{0.759(2)}%
\newcommand*{\rsBenchCBEtaNone}{0.036297612(48)}\newcommand*{\rsBenchCBNuNone}{0.62997097(12)}\newcommand*{\rsBenchCBOmegaNone}{0.82968(23)}%
\newcommand*{\rsBenchCBEtaNtwo}{0.038176(44)}\newcommand*{\rsBenchCBNuNtwo}{0.67175(10)}%
\newcommand*{\rsBenchCBEtaNthree}{0.03787(13)}\newcommand*{\rsBenchCBNuNthree}{0.71168(41)}%
\newcommand*{\rsStatMaxUstar}{5\times10^{-5}}%
\newcommand*{\rsStatMaxNu}{7\times10^{-6}}%
\newcommand*{\rsStatMaxEta}{3\times10^{-6}}%
\newcommand*{\rsStatMaxOmega}{1\times10^{-4}}%
\newcommand*{\rsStatRatio}{0.1}%
\newcommand*{\rsStatRatioA}{8\times10^{-4}}%
\newcommand*{\rsLinDev}{4\%}%
\newcommand*{\rsClassDev}{12\%}%
\newcommand*{\rsABdiff}{1\times10^{-6}}%
\newcommand*{\rsABmax}{1\times10^{-3}}%
\newcommand*{\rsGainUstar}{1.6--10.5}%
\newcommand*{\rsGainNu}{1.4--1.8}%
\newcommand*{\rsGainEta}{1.4}%
\newcommand*{\rsGainOmega}{0.9--7.0}%
\newcommand*{\rsEtaWindow}{2\times10^{-4}}%
\newcommand*{\rsNuDirMax}{1\times10^{-3}}%
\newcommand*{\rsDraws}{200}%
\newcommand*{\rsBmax}{12}\newcommand*{\rsAlphaMin}{-2}\newcommand*{\rsAlphaMax}{10}%
\newcommand*{\rsAone}{0.1477742}\newcommand*{\rsInstanton}{113.3835}%

\newcommand*{\costhistB}{%
  \addplot[const plot, thick, mark=none, color=colL6] coordinates {(-8.00,0.3984) (-7.75,1.1952) (-7.50,2.1912) (-7.25,3.1873) (-7.00,7.5697) (-6.75,12.7490) (-6.50,20.1195) (-6.25,22.9084) (-6.00,17.5299) (-5.75,8.5657) (-5.50,3.3865) (-5.25,0.1992) (-5.00,0)};
  \addlegendentry{$L=6$ (502 classes)}
  \addplot[const plot, thick, mark=none, color=colL7] coordinates {(-9.75,0.0317) (-9.50,0.1905) (-9.25,0.0635) (-9.00,0.0635) (-8.75,0.1587) (-8.50,0.6984) (-8.25,1.0159) (-8.00,3.1111) (-7.75,4.7937) (-7.50,8.1905) (-7.25,14.1905) (-7.00,20.2857) (-6.75,23.4603) (-6.50,14.2540) (-6.25,7.4603) (-6.00,1.7778) (-5.75,0.2540) (-5.50,0)};
  \addlegendentry{$L=7$ (3\,150 classes)}
  \addplot[const plot, thick, mark=none, color=colL8] coordinates {(-11.50,0.0046) (-11.25,0.0046) (-11.00,0.0046) (-10.75,0.0000) (-10.50,0.0092) (-10.25,0.0092) (-10.00,0.0184) (-9.75,0.0830) (-9.50,0.1245) (-9.25,0.4518) (-9.00,0.6732) (-8.75,1.8350) (-8.50,3.5686) (-8.25,6.1967) (-8.00,9.9175) (-7.75,15.9390) (-7.50,25.2248) (-7.25,20.3467) (-7.00,10.5491) (-6.75,4.0159) (-6.50,1.0143) (-6.25,0.0092) (-6.00,0)};
  \addlegendentry{$L=8$ (21\,689 classes)}
  \addplot[const plot, thick, mark=none, color=colL9] coordinates {(-15.25,0.0006) (-15.00,0.0000) (-14.75,0.0000) (-14.50,0.0006) (-14.25,0.0018) (-14.00,0.0006) (-13.75,0.0030) (-13.50,0.0024) (-13.25,0.0030) (-13.00,0.0012) (-12.75,0.0048) (-12.50,0.0084) (-12.25,0.0066) (-12.00,0.0175) (-11.75,0.0181) (-11.50,0.0175) (-11.25,0.0331) (-11.00,0.0512) (-10.75,0.1006) (-10.50,0.1734) (-10.25,0.2108) (-10.00,0.3595) (-9.75,0.6920) (-9.50,1.4833) (-9.25,3.0582) (-9.00,5.6598) (-8.75,8.8939) (-8.50,13.9051) (-8.25,24.1685) (-8.00,21.6613) (-7.75,12.5808) (-7.50,5.4467) (-7.25,1.3255) (-7.00,0.1096) (-6.75,0.0006) (-6.50,0)};
  \addlegendentry{$L=9$ (166\,047 classes)}
  \addplot[const plot, thick, mark=none, color=colL10] coordinates {(-15.50,0.0001) (-15.25,0.0005) (-15.00,0.0006) (-14.75,0.0012) (-14.50,0.0026) (-14.25,0.0023) (-14.00,0.0038) (-13.75,0.0049) (-13.50,0.0086) (-13.25,0.0084) (-13.00,0.0126) (-12.75,0.0143) (-12.50,0.0207) (-12.25,0.0290) (-12.00,0.0459) (-11.75,0.0722) (-11.50,0.1232) (-11.25,0.2061) (-11.00,0.3378) (-10.75,0.5428) (-10.50,0.9252) (-10.25,1.5897) (-10.00,2.8883) (-9.75,5.4316) (-9.50,8.0674) (-9.25,13.4840) (-9.00,22.0255) (-8.75,21.2039) (-8.50,13.8772) (-8.25,6.8916) (-8.00,1.8954) (-7.75,0.2770) (-7.50,0.0054) (-7.25,0.0001) (-7.00,0)};
  \addlegendentry{$L=10$ (1\,377\,678 classes)}
}
\newcommand*{\costhistH}{%
  \addplot[const plot, thick, mark=none, color=colL6] coordinates {(-7.00,4.0000) (-6.75,2.0000) (-6.50,0.0000) (-6.25,2.0000) (-6.00,2.0000) (-5.75,12.0000) (-5.50,12.0000) (-5.25,18.0000) (-5.00,16.0000) (-4.75,20.0000) (-4.50,10.0000) (-4.25,2.0000) (-4.00,0)};
  \addlegendentry{$L=6$ (50 classes)}
  \addplot[const plot, thick, mark=none, color=colL7] coordinates {(-7.25,0.9569) (-7.00,1.4354) (-6.75,2.8708) (-6.50,1.9139) (-6.25,5.2632) (-6.00,9.5694) (-5.75,14.8325) (-5.50,14.8325) (-5.25,22.0096) (-5.00,20.0957) (-4.75,4.3062) (-4.50,1.9139) (-4.25,0)};
  \addlegendentry{$L=7$ (209 classes)}
  \addplot[const plot, thick, mark=none, color=colL8] coordinates {(-9.25,0.0864) (-9.00,0.0000) (-8.75,0.2593) (-8.50,0.3457) (-8.25,0.0000) (-8.00,0.2593) (-7.75,0.5186) (-7.50,2.0743) (-7.25,2.3336) (-7.00,3.1115) (-6.75,7.6059) (-6.50,9.8531) (-6.25,12.9646) (-6.00,17.0268) (-5.75,23.0769) (-5.50,14.1746) (-5.25,4.9265) (-5.00,1.2100) (-4.75,0.1729) (-4.50,0)};
  \addlegendentry{$L=8$ (1\,157 classes)}
  \addplot[const plot, thick, mark=none, color=colL9] coordinates {(-10.75,0.0148) (-10.50,0.0148) (-10.25,0.0000) (-10.00,0.0148) (-9.75,0.0295) (-9.50,0.0590) (-9.25,0.0738) (-9.00,0.1919) (-8.75,0.2804) (-8.50,0.3247) (-8.25,0.6789) (-8.00,1.6086) (-7.75,2.6712) (-7.50,5.0915) (-7.25,8.1021) (-7.00,10.5519) (-6.75,14.0496) (-6.50,23.6275) (-6.25,18.9935) (-6.00,8.6924) (-5.75,4.1913) (-5.50,0.6789) (-5.25,0.0590) (-5.00,0)};
  \addlegendentry{$L=9$ (6\,776 classes)}
  \addplot[const plot, thick, mark=none, color=colL10] coordinates {(-13.50,0.0022) (-13.25,0.0000) (-13.00,0.0043) (-12.75,0.0043) (-12.50,0.0065) (-12.25,0.0130) (-12.00,0.0174) (-11.75,0.0130) (-11.50,0.0304) (-11.25,0.0347) (-11.00,0.0477) (-10.75,0.0499) (-10.50,0.0672) (-10.25,0.1171) (-10.00,0.0911) (-9.75,0.1367) (-9.50,0.2061) (-9.25,0.2798) (-9.00,0.5900) (-8.75,1.1778) (-8.50,1.9305) (-8.25,3.4988) (-8.00,6.8132) (-7.75,9.3553) (-7.50,11.6134) (-7.25,21.8125) (-7.00,21.9643) (-6.75,12.5960) (-6.50,5.6917) (-6.25,1.4446) (-6.00,0.3904) (-5.75,0)};
  \addlegendentry{$L=10$ (46\,102 classes)}
}
\newcommand*{\costhistC}{%
  \addplot[const plot, thick, mark=none, color=colL6] coordinates {(-6.50,1.6129) (-6.25,1.6129) (-6.00,6.4516) (-5.75,8.0645) (-5.50,11.2903) (-5.25,17.7419) (-5.00,22.5806) (-4.75,12.9032) (-4.50,14.5161) (-4.25,1.6129) (-4.00,1.6129) (-3.75,0)};
  \addlegendentry{$L=6$ (62 classes)}
  \addplot[const plot, thick, mark=none, color=colL7] coordinates {(-7.50,0.3774) (-7.25,0.7547) (-7.00,0.7547) (-6.75,1.8868) (-6.50,3.7736) (-6.25,4.9057) (-6.00,9.4340) (-5.75,16.9811) (-5.50,16.9811) (-5.25,20.7547) (-5.00,14.3396) (-4.75,5.2830) (-4.50,3.0189) (-4.25,0.7547) (-4.00,0)};
  \addlegendentry{$L=7$ (265 classes)}
  \addplot[const plot, thick, mark=none, color=colL8] coordinates {(-8.50,0.0721) (-8.25,0.0000) (-8.00,0.1442) (-7.75,0.5768) (-7.50,1.5141) (-7.25,2.4513) (-7.00,3.1723) (-6.75,5.8399) (-6.50,10.6705) (-6.25,13.9149) (-6.00,16.3663) (-5.75,21.1247) (-5.50,13.6265) (-5.25,5.7678) (-5.00,3.7491) (-4.75,0.8652) (-4.50,0.1442) (-4.25,0)};
  \addlegendentry{$L=8$ (1\,387 classes)}
  \addplot[const plot, thick, mark=none, color=colL9] coordinates {(-11.00,0.0124) (-10.75,0.0000) (-10.50,0.0000) (-10.25,0.0000) (-10.00,0.0124) (-9.75,0.0249) (-9.50,0.0249) (-9.25,0.0249) (-9.00,0.0249) (-8.75,0.0249) (-8.50,0.2364) (-8.25,0.4105) (-8.00,1.2565) (-7.75,2.0030) (-7.50,3.3590) (-7.25,7.4272) (-7.00,10.7489) (-6.75,13.8592) (-6.50,22.2568) (-6.25,18.3255) (-6.00,11.3834) (-5.75,5.2501) (-5.50,2.6001) (-5.25,0.6842) (-5.00,0.0498) (-4.75,0)};
  \addlegendentry{$L=9$ (8\,038 classes)}
  \addplot[const plot, thick, mark=none, color=colL10] coordinates {(-13.25,0.0037) (-13.00,0.0019) (-12.75,0.0019) (-12.50,0.0019) (-12.25,0.0019) (-12.00,0.0000) (-11.75,0.0019) (-11.50,0.0037) (-11.25,0.0037) (-11.00,0.0150) (-10.75,0.0225) (-10.50,0.0243) (-10.25,0.0562) (-10.00,0.0449) (-9.75,0.1011) (-9.50,0.1461) (-9.25,0.2622) (-9.00,0.3502) (-8.75,0.6836) (-8.50,1.2473) (-8.25,2.2323) (-8.00,4.6688) (-7.75,8.6203) (-7.50,10.5100) (-7.25,17.6752) (-7.00,23.1492) (-6.75,16.2912) (-6.50,7.6914) (-6.25,4.7943) (-6.00,0.9345) (-5.75,0.4438) (-5.50,0.0150) (-5.25,0)};
  \addlegendentry{$L=10$ (53\,397 classes)}
}
\newcommand*{\costcurves}{%
  \addplot[color=blue, thick, mark=*] coordinates {(4,0.03669) (5,0.1516) (6,0.5248) (7,1.612) (8,4.603) (9,10.92) (10,31.56)};
  \addlegendentry{$B_L$, GPU}
  \addplot[color=blue, thick, dashed, mark=o] coordinates {(4,4.301e-06) (5,3.9e-05) (6,0.0008205) (7,0.02598) (8,0.9889) (9,59.33) (10,2968)};
  \addlegendentry{$B_L$, CPU}
  \addplot[color=red, thick, mark=square*] coordinates {(4,0.03804) (5,1.32) (6,1.497) (7,6.625) (8,16.9) (9,39.65) (10,87.95)};
  \addlegendentry{$H_L$, GPU}
  \addplot[color=red, thick, dashed, mark=square] coordinates {(4,9.276e-07) (5,6.168e-06) (6,0.0001107) (7,0.00193) (8,0.06192) (9,2.259) (10,137.8)};
  \addlegendentry{$H_L$, CPU}
  \addplot[color=brown!60!black, thick, mark=triangle*] coordinates {(4,0.1171) (5,0.6977) (6,3.22) (7,12.21) (8,37.72) (9,97.74) (10,287.6)};
  \addlegendentry{$C_L$, GPU}
  \addplot[color=brown!60!black, thick, dashed, mark=triangle] coordinates {(4,1.461e-06) (5,1.477e-05) (6,0.0002546) (7,0.00653) (8,0.2355) (9,9.797) (10,975.1)};
  \addlegendentry{$C_L$, CPU}
}
\newcommand*{\costcpu}[1]{\csname costcpu@#1\endcsname}
\expandafter\def\csname costcpu@b10\endcsname{2.97e+03}
\expandafter\def\csname costcpu@b4\endcsname{4.3e-06}
\expandafter\def\csname costcpu@b5\endcsname{3.9e-05}
\expandafter\def\csname costcpu@b6\endcsname{0.00082}
\expandafter\def\csname costcpu@b7\endcsname{0.026}
\expandafter\def\csname costcpu@b8\endcsname{0.989}
\expandafter\def\csname costcpu@b9\endcsname{59.3}
\expandafter\def\csname costcpu@c10\endcsname{975}
\expandafter\def\csname costcpu@c4\endcsname{1.46e-06}
\expandafter\def\csname costcpu@c5\endcsname{1.48e-05}
\expandafter\def\csname costcpu@c6\endcsname{0.000255}
\expandafter\def\csname costcpu@c7\endcsname{0.00653}
\expandafter\def\csname costcpu@c8\endcsname{0.235}
\expandafter\def\csname costcpu@c9\endcsname{9.8}
\expandafter\def\csname costcpu@h10\endcsname{138}
\expandafter\def\csname costcpu@h4\endcsname{9.28e-07}
\expandafter\def\csname costcpu@h5\endcsname{6.17e-06}
\expandafter\def\csname costcpu@h6\endcsname{0.000111}
\expandafter\def\csname costcpu@h7\endcsname{0.00193}
\expandafter\def\csname costcpu@h8\endcsname{0.0619}
\expandafter\def\csname costcpu@h9\endcsname{2.26}
\newcommand*{\costcpus}[1]{\csname costcpus@#1\endcsname}
\newcommand*{\buildpc}[1]{\csname buildpc@#1\endcsname}
\expandafter\def\csname costcpus@b10\endcsname{11\,000\,000}
\expandafter\def\csname buildpc@b10\endcsname{7.8}
\expandafter\def\csname costcpus@b4\endcsname{0}
\expandafter\def\csname buildpc@b4\endcsname{0.00081}
\expandafter\def\csname costcpus@b5\endcsname{0}
\expandafter\def\csname buildpc@b5\endcsname{0.0015}
\expandafter\def\csname costcpus@b6\endcsname{3}
\expandafter\def\csname buildpc@b6\endcsname{0.0059}
\expandafter\def\csname costcpus@b7\endcsname{94}
\expandafter\def\csname buildpc@b7\endcsname{0.03}
\expandafter\def\csname costcpus@b8\endcsname{3\,600}
\expandafter\def\csname buildpc@b8\endcsname{0.16}
\expandafter\def\csname costcpus@b9\endcsname{210\,000}
\expandafter\def\csname buildpc@b9\endcsname{1.3}
\expandafter\def\csname costcpus@c10\endcsname{3\,500\,000}
\expandafter\def\csname buildpc@c10\endcsname{66}
\expandafter\def\csname costcpus@c4\endcsname{0}
\expandafter\def\csname buildpc@c4\endcsname{0.0011}
\expandafter\def\csname costcpus@c5\endcsname{0}
\expandafter\def\csname buildpc@c5\endcsname{0.0033}
\expandafter\def\csname costcpus@c6\endcsname{0}
\expandafter\def\csname buildpc@c6\endcsname{0.015}
\expandafter\def\csname costcpus@c7\endcsname{24}
\expandafter\def\csname buildpc@c7\endcsname{0.089}
\expandafter\def\csname costcpus@c8\endcsname{850}
\expandafter\def\csname buildpc@c8\endcsname{0.61}
\expandafter\def\csname costcpus@c9\endcsname{35\,000}
\expandafter\def\csname buildpc@c9\endcsname{4.4}
\expandafter\def\csname costcpus@h10\endcsname{500\,000}
\expandafter\def\csname buildpc@h10\endcsname{11}
\expandafter\def\csname costcpus@h4\endcsname{0}
\expandafter\def\csname buildpc@h4\endcsname{0.00083}
\expandafter\def\csname costcpus@h5\endcsname{0}
\expandafter\def\csname buildpc@h5\endcsname{0.002}
\expandafter\def\csname costcpus@h6\endcsname{0}
\expandafter\def\csname buildpc@h6\endcsname{0.008}
\expandafter\def\csname costcpus@h7\endcsname{6}
\expandafter\def\csname buildpc@h7\endcsname{0.033}
\expandafter\def\csname costcpus@h8\endcsname{220}
\expandafter\def\csname buildpc@h8\endcsname{0.19}
\expandafter\def\csname costcpus@h9\endcsname{8\,100}
\expandafter\def\csname buildpc@h9\endcsname{1.2}
\newcommand*{\costtablerows}{%
  $B_L$ & 4 & 0.07 & 1.15 & 0.037 & 3.47 & 0.12 \\
        & 5 & 0.06 & 2.34 & 0.15 & 5.44 & 0.30 \\
        & 6 & 0.14 & 4.35 & 0.52 & 7.89 & 0.62 \\
        & 7 & 0.28 & 7.62 & 1.6 & 11.5 & 1.3 \\
        & 8 & 0.25 & 12.9 & 4.6 & 15.6 & 2.4 \\
        & 9 & 0.98 & 19.8 & 11 & 20.8 & 4.3 \\
        & 10 & 1.15 & 33.7 & 32 & 28.6 & 8.2 \\
  $H_L$ & 4 & 0.07 & 1.17 & 0.038 & 2.80 & 0.079 \\
        & 5 & 0.42 & 6.89 & 1.3 & 11.8 & 1.4 \\
        & 6 & 0.45 & 7.34 & 1.5 & 9.36 & 0.88 \\
        & 7 & 0.49 & 15.4 & 6.6 & 15.2 & 2.3 \\
        & 8 & 0.44 & 24.7 & 17 & 19.7 & 3.9 \\
        & 9 & 0.99 & 37.8 & 40 & 27.5 & 7.5 \\
        & 10 & 1.42 & 56.3 & 88 & 34.2 & 12 \\
  $C_L$ & 4 & 0.08 & 2.05 & 0.12 & 3.72 & 0.14 \\
        & 5 & 0.12 & 5.01 & 0.70 & 5.92 & 0.35 \\
        & 6 & 0.28 & 10.8 & 3.2 & 8.83 & 0.78 \\
        & 7 & 0.49 & 21.0 & 12 & 12.5 & 1.6 \\
        & 8 & 0.99 & 36.9 & 38 & 17.4 & 3.0 \\
        & 9 & 1.00 & 59.3 & 98 & 23.0 & 5.3 \\
        & 10 & 2.89 & 102 & 288 & 32.1 & 10 \\
}
\newcommand*{\effortrows}{%
  This work, six to ten loops & \begin{tabular}[t]{@{}l@{}}$B_{6}$, $H_{6}$, $C_{6}$: 614 classes \\ $B_{7}$, $H_{7}$, $C_{7}$: 3\,624 classes \\ $B_{8}$, $H_{8}$, $C_{8}$: 24\,233 classes \\ $B_{9}$, $H_{9}$, $C_{9}$: 180\,861 classes \\ $B_{10}$, $H_{10}$, $C_{10}$: 1\,477\,177 classes\end{tabular} & NVIDIA H200 GPUs, Intel Xeon Platinum 8568Y+ CPUs & \begin{tabular}[t]{@{}l@{}}0.52, 1.5, 3.2 GPU h \\ 1.6, 6.6, 12 GPU h \\ 4.6, 17, 38 GPU h; 1.3 CPU h \\ 11, 40, 98 GPU h; 71 CPU h \\ 32, 88, 288 GPU h; 4\,100 CPU h\end{tabular} & $10^{-5}$ relative, per order sum \\
}
\newcommand*{\costTm}[1]{\csname costTm@#1\endcsname}
\expandafter\def\csname costTm@b10\endcsname{32}
\expandafter\def\csname costTm@b2\endcsname{0.00065}
\expandafter\def\csname costTm@b3\endcsname{0.0066}
\expandafter\def\csname costTm@b4\endcsname{0.037}
\expandafter\def\csname costTm@b5\endcsname{0.15}
\expandafter\def\csname costTm@b6\endcsname{0.52}
\expandafter\def\csname costTm@b7\endcsname{1.6}
\expandafter\def\csname costTm@b8\endcsname{4.6}
\expandafter\def\csname costTm@b9\endcsname{11}
\expandafter\def\csname costTm@c10\endcsname{288}
\expandafter\def\csname costTm@c2\endcsname{0.00068}
\expandafter\def\csname costTm@c3\endcsname{0.0070}
\expandafter\def\csname costTm@c4\endcsname{0.12}
\expandafter\def\csname costTm@c5\endcsname{0.70}
\expandafter\def\csname costTm@c6\endcsname{3.2}
\expandafter\def\csname costTm@c7\endcsname{12}
\expandafter\def\csname costTm@c8\endcsname{38}
\expandafter\def\csname costTm@c9\endcsname{98}
\expandafter\def\csname costTm@h10\endcsname{88}
\expandafter\def\csname costTm@h2\endcsname{0.000059}
\expandafter\def\csname costTm@h3\endcsname{0.041}
\expandafter\def\csname costTm@h4\endcsname{0.038}
\expandafter\def\csname costTm@h5\endcsname{1.3}
\expandafter\def\csname costTm@h6\endcsname{1.5}
\expandafter\def\csname costTm@h7\endcsname{6.6}
\expandafter\def\csname costTm@h8\endcsname{17}
\expandafter\def\csname costTm@h9\endcsname{40}
\expandafter\def\csname costgrowth@b\endcsname{2.7}
\expandafter\def\csname costratios@b\endcsname{4.1, 3.5, 3.1, 2.9, 2.4 and 2.9}
\expandafter\def\csname costratiorange@b\endcsname{four to ten}
\expandafter\def\csname costgrowth@h\endcsname{2.4}
\expandafter\def\csname costratios@h\endcsname{35, 1.1, 4.4, 2.6, 2.3 and 2.2}
\expandafter\def\csname costratiorange@h\endcsname{four to ten}
\expandafter\def\csname costgrowth@c\endcsname{2.9}
\expandafter\def\csname costratios@c\endcsname{6, 4.6, 3.8, 3.1, 2.6 and 2.9}
\expandafter\def\csname costratiorange@c\endcsname{four to ten}
\newcommand*{\costgrowth}[1]{\csname costgrowth@#1\endcsname}
\newcommand*{\costratios}[1]{\csname costratios@#1\endcsname}
\newcommand*{\costratiorange}[1]{\csname costratiorange@#1\endcsname}
\newcommand*{\costgrowthrange}{2.4--2.9}
\newcommand*{\evalgrowthrange}{1.7--1.9}

\newcommand*{\costrefbsix}{1.1}\newcommand*{\costmatchbsix}{41}\newcommand*{\costmatchbsixfour}{10}
\newcommand*{\costkappadev}{11}
\newcommand*{\cpugrowthrange}{32 to 100}
\newcommand*{\kappagrowthrange}{1.5 to 2.1}
\newcommand*{\graphgrowthrange}{4.2 to 8.3}
\newcommand*{\kappashift}{0.6}
\newcommand*{\costtenthgpu}{89}\newcommand*{\costtenthcpu}{1\,200}
\newcommand*{\costtenthnodecpu}{37}\newcommand*{\costtenthnodegpu}{11}
\newcommand*{\costN}[1]{\csname costN@#1\endcsname}
\newcommand*{\costmedian}[1]{\csname costmedian@#1\endcsname}
\newcommand*{\costsum}[1]{\csname costsum@#1\endcsname}
\newcommand*{\costT}[1]{\csname costT@#1\endcsname}
\newcommand*{\costninety}[1]{\csname costninety@#1\endcsname}
\newcommand*{\costninetyrange}{54--62}
\expandafter\def\csname costN@b10\endcsname{1377678}
\expandafter\def\csname costmedian@b10\endcsname{1.54 \times 10^{-9}}
\expandafter\def\csname costsum@b10\endcsname{3.22 \times 10^{-3}}
\expandafter\def\csname costT@b10\endcsname{28.9}
\expandafter\def\csname costninety@b10\endcsname{59}
\expandafter\def\csname costN@b6\endcsname{502}
\expandafter\def\csname costmedian@b6\endcsname{5.93 \times 10^{-7}}
\expandafter\def\csname costsum@b6\endcsname{4.37 \times 10^{-4}}
\expandafter\def\csname costT@b6\endcsname{0.53}
\expandafter\def\csname costninety@b6\endcsname{62}
\expandafter\def\csname costN@b7\endcsname{3150}
\expandafter\def\csname costmedian@b7\endcsname{1.67 \times 10^{-7}}
\expandafter\def\csname costsum@b7\endcsname{7.56 \times 10^{-4}}
\expandafter\def\csname costT@b7\endcsname{1.59}
\expandafter\def\csname costninety@b7\endcsname{61}
\expandafter\def\csname costN@b8\endcsname{21689}
\expandafter\def\csname costmedian@b8\endcsname{4.05 \times 10^{-8}}
\expandafter\def\csname costsum@b8\endcsname{1.28 \times 10^{-3}}
\expandafter\def\csname costT@b8\endcsname{4.54}
\expandafter\def\csname costninety@b8\endcsname{62}
\expandafter\def\csname costN@b9\endcsname{166047}
\expandafter\def\csname costmedian@b9\endcsname{8.12 \times 10^{-9}}
\expandafter\def\csname costsum@b9\endcsname{1.97 \times 10^{-3}}
\expandafter\def\csname costT@b9\endcsname{10.8}
\expandafter\def\csname costninety@b9\endcsname{61}
\expandafter\def\csname costN@c10\endcsname{53397}
\expandafter\def\csname costmedian@c10\endcsname{1.09 \times 10^{-7}}
\expandafter\def\csname costsum@c10\endcsname{9.58 \times 10^{-3}}
\expandafter\def\csname costT@c10\endcsname{255}
\expandafter\def\csname costninety@c10\endcsname{57}
\expandafter\def\csname costN@c6\endcsname{62}
\expandafter\def\csname costmedian@c6\endcsname{1.03 \times 10^{-5}}
\expandafter\def\csname costsum@c6\endcsname{1.08 \times 10^{-3}}
\expandafter\def\csname costT@c6\endcsname{3.27}
\expandafter\def\csname costninety@c6\endcsname{58}
\expandafter\def\csname costN@c7\endcsname{265}
\expandafter\def\csname costmedian@c7\endcsname{4.59 \times 10^{-6}}
\expandafter\def\csname costsum@c7\endcsname{2.09 \times 10^{-3}}
\expandafter\def\csname costT@c7\endcsname{12.1}
\expandafter\def\csname costninety@c7\endcsname{56}
\expandafter\def\csname costN@c8\endcsname{1387}
\expandafter\def\csname costmedian@c8\endcsname{1.54 \times 10^{-6}}
\expandafter\def\csname costsum@c8\endcsname{3.64 \times 10^{-3}}
\expandafter\def\csname costT@c8\endcsname{36.8}
\expandafter\def\csname costninety@c8\endcsname{54}
\expandafter\def\csname costN@c9\endcsname{8038}
\expandafter\def\csname costmedian@c9\endcsname{4.24 \times 10^{-7}}
\expandafter\def\csname costsum@c9\endcsname{5.84 \times 10^{-3}}
\expandafter\def\csname costT@c9\endcsname{94.6}
\expandafter\def\csname costninety@c9\endcsname{56}
\expandafter\def\csname costN@h10\endcsname{46102}
\expandafter\def\csname costmedian@h10\endcsname{8.46 \times 10^{-8}}
\expandafter\def\csname costsum@h10\endcsname{5.60 \times 10^{-3}}
\expandafter\def\csname costT@h10\endcsname{87.3}
\expandafter\def\csname costninety@h10\endcsname{58}
\expandafter\def\csname costN@h6\endcsname{50}
\expandafter\def\csname costmedian@h6\endcsname{9.73 \times 10^{-6}}
\expandafter\def\csname costsum@h6\endcsname{7.34 \times 10^{-4}}
\expandafter\def\csname costT@h6\endcsname{1.5}
\expandafter\def\csname costninety@h6\endcsname{58}
\expandafter\def\csname costN@h7\endcsname{209}
\expandafter\def\csname costmedian@h7\endcsname{5.05 \times 10^{-6}}
\expandafter\def\csname costsum@h7\endcsname{1.54 \times 10^{-3}}
\expandafter\def\csname costT@h7\endcsname{6.55}
\expandafter\def\csname costninety@h7\endcsname{58}
\expandafter\def\csname costN@h8\endcsname{1157}
\expandafter\def\csname costmedian@h8\endcsname{1.45 \times 10^{-6}}
\expandafter\def\csname costsum@h8\endcsname{2.44 \times 10^{-3}}
\expandafter\def\csname costT@h8\endcsname{16.5}
\expandafter\def\csname costninety@h8\endcsname{57}
\expandafter\def\csname costN@h9\endcsname{6776}
\expandafter\def\csname costmedian@h9\endcsname{3.75 \times 10^{-7}}
\expandafter\def\csname costsum@h9\endcsname{3.66 \times 10^{-3}}
\expandafter\def\csname costT@h9\endcsname{37.1}
\expandafter\def\csname costninety@h9\endcsname{57}

\begin{document}

\title{Ten-loop fixed-dimension renormalization-group functions\\ of the three-dimensional
\texorpdfstring{$O(N)$}{O(N)} model\\ from distributed GPU-accelerated parametric integration}

\author{Paolo Ribeca}
\affiliation{Biomathematics and Statistics Scotland (BioSS), Edinburgh, United Kingdom}
\affiliation{UK Health Security Agency (UKHSA), London, United Kingdom}

\date{\today}

\begin{abstract}
The renormalization-group functions $\beta$, $\eta$ and $\nu$ of the three-dimensional $O(N)$ model in fixed
dimension, in the massive zero-momentum scheme, were known to eighth order in the coupling. Here we extend them to
tenth order, and $\beta$ to eleventh, for general $N$---which requires the four-point and two-point functions to ten
loops, 1\,593\,254 graphs at the last order---and we re-evaluate all lower orders independently. Through the
Callan--Symanzik equation every coefficient follows algebraically from per-order sums that are linear in finite,
zero-momentum $\Rbar$-subtracted parametric integrals, one per graph; we numerically sample them on GPUs by tropical Monte Carlo, with densities adapted to the
forest terms. The computation reproduces the known per-order sums from two to seven loops within three standard
deviations, and agrees at ninth and tenth order, at lower precision, with an independent route through the vertex
functions of an intermediate scheme. The new
coefficients have relative uncertainties of a few $10^{-5}$ or better, and the eighth-order coefficients of $\eta$ and
$\nu$ improve by factors of \etaEightGain\ and \nuEightGain. An encouraging result concerns the cost: with this strategy,
the time needed to integrate an order to a fixed relative precision grows from seven loops on by a factor of only
\costgrowthrange\ per loop on average. However, the construction of the integrands still grows factorially, and at tenth order it costs more than the
integration: \costtenthcpu\ CPU core-hours against \costtenthgpu\ GPU hours. Resummed, the tenth-order series give
$\nu=\rsNuDNone$, $\eta=\rsEtaDNone$ and $\omega=\rsOmegaDNone$ for $N=1$, against $\rsBenchCBNuNone$,
$\rsBenchCBEtaNone$ and $\rsBenchCBOmegaNone$ from the conformal bootstrap; for $N=2$ and 3 the deviation of $\nu$ grows, to
\rsPullNuNthree\ standard deviations at $N=3$, where $\omega$ lies \rsPullOmegaNthree\ standard deviations from the
Monte Carlo value, so that the uncertainty is now dominated by the resummation
rather than by the coefficients. Our computational strategy is cooperative
and distributed, and thus open to future improvement: every result presented here is generated automatically from, or
checked against, a public ledger of independent, exactly pooled runs, hosted on GitHub, to which anyone is welcome to contribute
computing time.
\end{abstract}

\maketitle

\section{Introduction and state of the art}
\label{sec:intro}

\subsection{Universality and field-theoretic approaches}
\label{sec:intro-universality}

Continuous phase transitions in systems with short-range interactions and an $N$-component order parameter with
$O(N)$ symmetry fall into the universality classes of the three-dimensional $O(N)$ model: $N=0$ describes polymers
and self-avoiding walks, $N=1$ uniaxial magnets and the liquid--vapour critical point, $N=2$ planar magnets and the
superfluid transition of $^4$He, $N=3$ isotropic magnets~\cite{PelissettoVicari2002,ZinnJustinBook}. Their critical
exponents are a standard benchmark both for field theory and for non-perturbative methods.

Two perturbative field-theoretic approaches have been pushed to high order. In the $\varepsilon$ expansion the
renormalization-group (RG) functions of $\phi^4$ theory are computed by minimal subtraction in $d=4-\varepsilon$ and
the exponents are expanded in powers of $\varepsilon$~\cite{ZinnJustinBook,KleinertBook}. In fixed-dimension
perturbation theory, proposed by Parisi~\cite{Parisi1980}, one works directly in $d=3$, where the theory is
super-renormalizable, renormalizes it by normalization conditions at zero momentum, and obtains the exponents from
the RG functions at the non-trivial zero of the $\beta$ function. In both approaches the series are asymptotic, with
coefficients growing like $k!$ at large order $k$~\cite{Lipatov1977,BrezinLeGuillouZinnJustin1977}, and must be
resummed, typically by a Borel--Leroy transform followed by a conformal mapping built on the known large-order
behaviour~\cite{LeGuillouZinnJustin1977,LeGuillouZinnJustin1980,GuidaZinnJustin1998}. Fixed dimension has the
advantage that Borel summability is established for the correlation functions of the $d=3$ theory~\cite{MagnenSeneor1977}, although not directly for
the RG functions, while for the $\varepsilon$ series it has to be assumed. The precision of the resummed exponents is limited by the number of known terms, by the
numerical uncertainty of the highest ones, and by systematic effects of the resummation; in fixed dimension these
include the expected non-analyticity of $\beta$ at the fixed point, caused by confluent corrections to
scaling~\cite{Nickel1991,PelissettoVicari2002}.

\subsection{The fixed-dimension series: history and status}
\label{sec:intro-fixed}

We consider the Euclidean $O(N)$ $\phi^4$ theory in $d=3$ in the massive zero-momentum scheme, in which the
renormalized mass $m$, the field normalization and the dimensionless coupling $g$ are fixed by conditions on
$\Gamma^{(2)}_R(0)$, $\partial\Gamma^{(2)}_R/\partial p^2(0)$ and $\Gamma^{(4)}_R(0)=m\,g$. We write the series in
$\ubar=(N+8)u/6$ with $u=g/8\pi$, i.e., in the normalized coupling of
Refs.~\cite{BNGM1976,BNM1978,Phi4tools2024} (the $\tilde u$ of Ref.~\cite{Phi4tools2024}). The computation of the
coefficient of $\ubar^{L+1}$ in $\beta$ requires the four-point function $\Gamma^{(4)}$ at $L$ loops; the coefficients of $\ubar^{L}$
in $\eta$ and $\nu$ require the two-point function $\Gamma^{(2)}$, with its $p^2$ derivative or with $\phi^2$
insertions, at $L$ loops.

\paragraph{Nickel and collaborators.}
A numerical evaluation programme was first carried out by Baker, Nickel, Green and Meiron~\cite{BNGM1976}, who took the series for
$N=1$ from three to six loops~\cite{Ribeca2003thesis}, i.e., $\beta$ to $\ubar^{7}$ and $\eta$, $\nu$ to $\ubar^{6}$;
Baker, Nickel and Meiron~\cite{BNM1978} analysed the corresponding series for $N=0$--$3$. The key technical ingredient was that one-loop bubbles, triangles and boxes in $d=3$
have closed forms~\cite{Nickel1978}, so that each graph could be integrated in momentum space with these one-loop
subgraphs replaced by effective vertices. The individual graph values were compiled, together with their weights, in
an unpublished Guelph report~\cite{NMB1977}. The 1042 diagrams needed through six loops~\cite{Ribeca2003thesis} are
given to ten decimals when the integral has dimension at most three and to eight decimals when it has dimension five
or six, with errors ``generally less than $\pm5$ in the last digit quoted'' and up to $\pm20$ in exceptional
cases. The whole evaluation took about
20~hours on an IBM 370/155 and 3~hours on a CDC 6600~\cite{NMB1977}. The integral of each graph was set up by hand, and the
paper on the numerical methods announced in Refs.~\cite{BNGM1976,NMB1977,BNM1978} never appeared. Because the compilation
separates integrals from symmetry factors, it was reused for decades to build the series of other models without
recomputing any integral, for instance for cubic anisotropy~\cite{CarmonaPelissettoVicari2000}; see
Ref.~\cite{PelissettoVicari2002} for a review. Murray and Nickel later extended $\eta$ and $\eta_{\phi^2}$, hence
$\nu$, to seven loops for $N=0$--$3$~\cite{MurrayNickel1991}; this report is also unpublished, its method is
undocumented, and its coefficients are quoted, to about seven significant digits, in the appendix of
Ref.~\cite{GuidaZinnJustin1998}. The resummation of these series by Le~Guillou and Zinn-Justin and by Guida and
Zinn-Justin~\cite{LeGuillouZinnJustin1977,LeGuillouZinnJustin1980,GuidaZinnJustin1998} produced the standard
fixed-dimension estimates of the $O(N)$ exponents.

\paragraph{Automation.}
By 2003 the six-loop integrals had not been reproduced by an independent group for a quarter of a century, and the
next order involves several thousand graphs. References~\cite{Ribeca2003thesis,GuidaRibeca2006} therefore set out to automate
the whole computation: graph generation, weights and $O(N)$ factors, and an automatic search for the cheapest
effective-vertex parametrization of every amplitude, followed by deterministic adaptive cubature. All seven-loop
amplitudes were parametrized with at most nine integration variables~\cite{Ribeca2003thesis}, but the seven-loop
numerical evaluation was not completed at the time.

\paragraph{Phi4tools.}
Sberveglieri and Spada~\cite{Phi4tools2024} (the Phi4tools compilation) recomputed all zero-momentum graphs with up to eight vertices, that is,
$\Gamma^{(4)}$ to seven loops and $\Gamma^{(2)}$ to eight. They essentially followed the strategy proposed in
Ref.~\cite{GuidaRibeca2006}, integrating every graph in momentum space after automatically assigning effective
vertices,
with only slight differences in their own words: separate tools for each step, Mathematica for the integrals of
dimension up to three and the VEGAS algorithm~\cite{Lepage1978} for the rest in place of deterministic adaptive
cubature, some new analytic and numerical effective vertices, and a simpler search for the parametrization of the
amplitudes. Some of the hardest graphs were cross-checked with feyntrop~\cite{feyntrop2024}, an implementation of
tropical Monte Carlo. Their largest relative uncertainties are a few parts in $10^5$ for two- and
four-point graphs and a few parts in $10^4$ for the $p^2$ derivatives of two-point graphs. This provided the first
independent check of Ref.~\cite{NMB1977}: of the 585 six-loop four-point graphs renormalized in the same way, most
values agree to $\pm1$ in the last digit quoted by Nickel et al., with few exceptions, of which only one differs by
about 20~units. From these values
they obtained $\beta$, $\eta$ and $\nu$ to $\ubar^{8}$ for general $N$, and the corresponding series of the cubic
model. For $N=1$ the quoted uncertainties of the $\ubar^8$ coefficients, added linearly over the coefficients of the
$N$ polynomials, are about $4\times10^{-6}$ ($\beta$), $5\times10^{-4}$ ($\eta$) and $5\times10^{-5}$ ($\nu$) in
relative terms. The authors postponed the resummation to future work~\cite{Phi4tools2024}; to our knowledge the
eighth-order fixed-dimension series have not been resummed. Table~\ref{tab:sota} summarizes this history.

\begin{table*}
\caption{\label{tab:sota}%
Perturbative RG functions of the $O(N)$ model: highest known orders. In fixed dimension the term $\ubar^{k}$ of
$\beta$ needs $\Gamma^{(4)}$ at $L=k-1$ loops, and that of $\eta$ and $\nu$ needs $\Gamma^{(2)}$ at $L=k$ loops; the
graph counts are one-particle-irreducible (1PI) graphs without self-loops, one-vertex-reducible ones (with a cut vertex)
included. The precision is the relative
uncertainty of the highest coefficient at $N=1$ or, where no uncertainty is quoted, the number of significant digits
printed (sig.) or the absolute accuracy of the individual integrals (int.). The $\varepsilon$-expansion coefficients are known exactly, as
rational combinations of multiple zeta values and, at seven loops, of one further period that is conjecturally not of
this form~\cite{Schnetz2018}.}
\begin{ruledtabular}
\begin{tabular}{lllll}
Function & Term & Graphs & Reference & Precision \\
\hline
\multicolumn{5}{l}{\emph{Fixed dimension $d=3$, massive zero-momentum scheme}}\\
$\beta$ & $\ubar^{7}$ & $\Gamma^{(4)}$, $L=6$ (660) & \cite{BNGM1976} ($N=1$); \cite{BNM1978,NMB1977} ($N=0$--$3$) & 6 sig.; int.\ $10^{-8}$--$10^{-10}$ \\
        & $\ubar^{8}$ & $\Gamma^{(4)}$, $L=7$ (3986) & \cite{Phi4tools2024}, all $N$ & $4\times10^{-6}$ \\
        & $\ubar^{9}$ & $\Gamma^{(4)}$, $L=8$ (26\,540) & this work, all $N$ & \betaNineRel \\
        & $\ubar^{10}$ & $\Gamma^{(4)}$, $L=9$ (197\,022) & this work, all $N$ & \betaTenRel \\
        & $\ubar^{11}$ & $\Gamma^{(4)}$, $L=10$ (1\,593\,254) & this work, all $N$ & \betaElevenRel \\
$\eta$, $\nu$ & $\ubar^{6}$ & $\Gamma^{(2)}$, $L=6$ (75) & \cite{BNGM1976} ($N=1$); \cite{BNM1978,NMB1977} ($N=0$--$3$) & int.\ $10^{-8}$--$10^{-10}$ \\
        & $\ubar^{7}$ & $\Gamma^{(2)}$, $L=7$ (317) & \cite{MurrayNickel1991,GuidaZinnJustin1998}, $N=0$--$3$ & $\sim7$ sig. \\
        & $\ubar^{8}$ & $\Gamma^{(2)}$, $L=8$ (1622) & \cite{Phi4tools2024}, all $N$ & $5\times10^{-4}$ ($\eta$), $5\times10^{-5}$ ($\nu$) \\
        & $\ubar^{9}$ & $\Gamma^{(2)}$, $L=9$ (9184) & this work, all $N$ & \etaNineRel\ ($\eta$), \nuNineRel\ ($\nu$) \\
        & $\ubar^{10}$ & $\Gamma^{(2)}$, $L=10$ (59\,614) & this work, all $N$ & \etaTenRel\ ($\eta$), \nuTenRel\ ($\nu$) \\
\hline
\multicolumn{5}{l}{\emph{$\varepsilon$ expansion, minimal subtraction: $\beta$, $\gamma_{m^2}$, $\gamma_\phi$}}\\
        & 5 loops & & \cite{ChetyrkinKataevTkachov1981,ChetyrkinGorishnyLarinTkachov1983,GorishnyLarinTkachov1984,Kleinert1991,KleinertBook}, all $N$ & exact \\
        & 6 loops & & \cite{BatkovichChetyrkinKompaniets2016,KompanietsPanzer2017}, all $N$ & exact \\
        & 7 loops & & \cite{Schnetz2018,Schnetz2023}, all $N$ & exact \\
        & 8 loops & $\gamma_\phi$ only & \cite{Schnetz2023}, $N=1$ & exact \\
\end{tabular}
\end{ruledtabular}
\end{table*}

\subsection{The \texorpdfstring{$\varepsilon$}{epsilon} expansion and non-perturbative benchmarks}
\label{sec:intro-eps}

In the $\varepsilon$ expansion the minimal-subtraction RG functions of the $O(N)$ model were extended to five
loops~\cite{ChetyrkinKataevTkachov1981,ChetyrkinGorishnyLarinTkachov1983,GorishnyLarinTkachov1984,Kleinert1991} and
to six~\cite{BatkovichChetyrkinKompaniets2016,KompanietsPanzer2017}, the latter computation also estimating the graphs
without subdivergences to eleven loops and comparing them with the large-order behaviour of $\beta$. Position-space
methods based on graphical functions have since given the $\beta$ function and the anomalous dimensions to seven
loops~\cite{Schnetz2018}, for the $O(N)$ model as well (published for $N=1$, the $O(N)$ results distributed with the
author's software), and for $N=1$ the field anomalous dimension to eight~\cite{Schnetz2023}. Numerical approaches in minimal subtraction remove subdivergences by
subtraction operators acting on the integrands, so that the RG functions follow algebraically from linear sums of
finite integrals~\cite{AdzhemyanKompaniets2011,AdzhemyanKompaniets2014}; the zero-momentum $\Rbar$ operation used here
(Sec.~\ref{sec:methods-rbar}) is the fixed-dimension counterpart of this idea. The six-loop $\varepsilon$ series have been resummed for general
$N$~\cite{KompanietsPanzer2017,KompanietsWiese2020}, and for $N=1$ the seven-loop series and, for $\eta$, the eight-loop
one~\cite{Shapoval2026}. A recent study for $N=1$~\cite{Shapoval2026} finds that the uncertainty of $\eta$ decreases
rapidly with the loop order while those of $\nu$ and of the correction-to-scaling exponent $\omega$ do not, and
reports a slight but systematic tension with the conformal bootstrap.

The most precise determinations of the exponents are now non-perturbative. The numerical conformal bootstrap
determines the Ising exponents with uncertainties far below those of any resummed perturbative series, and
constrains the $O(N)$ exponents as well~\cite{KosPolandSimmonsDuffinVichi2016,PolandRychkovVichi2019}; Monte Carlo
simulations, of the Ising model~\cite{FerrenbergXuLandau2018} and of improved lattice
Hamiltonians~\cite{Hasenbusch2010,Hasenbusch2019,Hasenbusch2020}, also reach high precision. The conformal bootstrap gives $\nu=\rsBenchCBNuNone$ and
$\eta=\rsBenchCBEtaNone$ for $N=1$~\cite{ChangEtAl2025}, $\nu=\rsBenchCBNuNtwo$ and
$\eta=\rsBenchCBEtaNtwo$ for $N=2$~\cite{ChesterEtAl2020}, and $\nu=\rsBenchCBNuNthree$ and $\eta=\rsBenchCBEtaNthree$ for
$N=3$~\cite{ChesterEtAl2021}, in agreement with the Monte Carlo values $\nu=\rsBenchMCNuNtwo$, $\eta=\rsBenchMCEtaNtwo$
for $N=2$~\cite{Hasenbusch2019} and $\nu=\rsBenchMCNuNthree$, $\eta=\rsBenchMCEtaNthree$ for $N=3$~\cite{Hasenbusch2020}.
For $N=1$ in particular, these results make the perturbative series a stringent test of resummation methods, a
test that is only meaningful if the series coefficients and their uncertainties are reliable.

\subsection{This work}
\label{sec:intro-this-work}

Since 1978 each fixed-dimension series has been extended by one order at a
time~\cite{MurrayNickel1991,Phi4tools2024}. Both complete computations~\cite{NMB1977,Phi4tools2024} used one
momentum-space representation per graph, chosen by hand or automatically and integrated to high accuracy, and
obtained the RG functions from the vertex functions of an intermediate scheme through the $Z$ factors (Nickel et
al.\ subtracted, according to Ref.~\cite{Phi4tools2024}, all two-point subgraphs at zero momentum, Phi4tools only the divergent tadpole and sunset
subgraphs). The ninth and tenth orders need 26\,540 and 197\,022 four-point graphs at eight and nine loops, and
9184 and 59\,614 two-point graphs at nine and ten loops, against 3986 and 1622 at the eighth order, each of higher
dimension; the eleventh order of $\beta$ needs 1\,593\,254 four-point graphs at ten loops. Moreover, along this route the order-$L$
contributions largely cancel: at four, five and six loops the intermediate-scheme four-point input to $\beta$ exceeds
the net order-$L$ four-point coefficient $B_L$ (Sec.~\ref{sec:methods}) by factors of about 29, 50 and 74, so that its relative
error must be smaller than the target by the same factor.

In this paper we compute $\beta$, $\eta$ and $\nu$ to $\ubar^{10}$ ($\beta$ to $\ubar^{11}$) for general $N$, and
re-evaluate all lower orders, with a method designed for this scale and for massively parallel hardware: its
precision depends as little as possible on the order, and its cost should grow slowly with the order
(Sec.~\ref{sec:methods-choice}); measured, from seven loops on, the cost grows by a factor of \costgrowthrange\ per loop
(Sec.~\ref{sec:results-performance}).
\begin{enumerate}
\item Through the Callan--Symanzik equation, the coefficients of $\beta$, $\eta$ and $\nu$ follow algebraically from
three functions, $\Gamma^{(4,1)}_R(0)$, $\partial\Gamma^{(2,1)}_R/\partial p^2(0)$ and $\Gamma^{(2,2)}_R(0)$, whose $L$-loop
terms we call $B_L$, $H_L$ and $C_L$ (Sec.~\ref{sec:methods-cs}). Each is a \emph{linear} sum over graphs of finite
integrals, subtracted at zero momentum by Zimmermann's forest
formula~\cite{BogoliubovParasiuk1957,Hepp1966,Zimmermann1969,Zavialov1990}, the
$\Rbar$ operation of Sec.~\ref{sec:methods-rbar}. No $Z$ factors and no series reversion are needed, and the
cancellations that the $Z$-factor route performs between graphs take place inside each integrand. We call this
route A.
\item At zero external momenta and unit masses every graph, with all its forest terms, is a single parametric
integral over a simplex, built from the Kirchhoff polynomials of the graph and of its subgraphs and quotient
graphs~\cite{BergereZuber1974,BergereLam1976,BrownKreimer2013}. Every integrand is sampled by tropical, Hepp-sector Monte
Carlo~\cite{Volkov2017,Borinsky2023tropical,Panzer2022Hepp,feyntrop2024,Borinsky2025global}, with a density adapted to the
forest terms, on any number of CPUs or GPUs from the same source code.
\item The computation is automatic, from graph generation to series coefficients: there is no per-graph choice of
momentum routing, effective vertices or integration variables. Since every graph carries its $O(N)$ polynomial, one
run gives every $N$.
\item The results are checked against exact low-order values, against the published series, against the large-$N$
limit and the large-order behaviour, and, at the ninth and tenth orders, against a second, independent route (route B)
through the vertex functions of an intermediate scheme, its $Z$ factors and the reversion of the coupling series. The
precision of route B, \routeBrelmin--\routeBrelmax\ relative, is far coarser than that of the new coefficients, and the
coefficient of $\ubar^{11}$ in $\beta$ is checked only by the internal statistics, the large-$N$ limit and the
large-order behaviour.
\end{enumerate}
The results are therefore an independent determination of all known orders: the renormalization route, the
integral representation and the numerics share nothing with Refs.~\cite{NMB1977,Phi4tools2024} except the graph
lists, which our own generator reproduces wherever both exist (Sec.~\ref{sec:methods-checks}). The published coefficients combine our new sums with, at each lower order, the more
precise of our sum and the one derived from Ref.~\cite{Phi4tools2024}; with our own lower orders only, they change
by at most \ledgeronlyzmax\ standard deviations (Sec.~\ref{sec:results-new}).

Runs on NVIDIA H200 GPUs reproduce the known per-order sums from
two to seven loops for all three functions within three standard deviations, with error estimates that the
within-run statistics confirm to a few per cent (Sec.~\ref{sec:results-validation}).
At six loops, with the computing time we allocated to it, the method reproduces the published values with similar or
better precision. From seven loops on, where the graphs are many and their integrals of high dimension, it overtakes
the per-graph treatment: all three seven-loop sums are now known more precisely than
from Phi4tools, by factors of about \precBseven, \precHseven\ and \precCseven\ for $B_7$, $H_7$ and $C_7$, and the
complete sum $H_7$, which fixes the $\ubar^7$ term of $\eta$ (209 integrands), reaches a relative uncertainty of
$10^{-5}$ in about \costTm{h7} GPU hours, against $5.5\times10^{-5}$ for the same quantity derived from the Phi4tools
series (Sec.~\ref{sec:results}). The new coefficients, from ninth to eleventh order, have relative uncertainties between
\newrelmin\ and \newrelmax\ at $N=1$ (Tables~\ref{tab:newcoeff}--\ref{tab:newcoeffnu}), and the coefficient of
$\ubar^8$ in $\eta$ is now \etaEightGain\ times more precise than before (Sec.~\ref{sec:results-new}). The
uncertainties thus hardly grow with the order. Along the per-graph route the $p^2$ derivatives are the least precise:
at the eighth order of Ref.~\cite{Phi4tools2024}, $5\times10^{-4}$ for $\eta$ against $4\times10^{-6}$ for $\beta$
(Table~\ref{tab:sota}). A uniform precision ensures that the number of known orders, not their statistical errors,
limits the resummation: at the present orders the statistical errors are already negligible there
(Sec.~\ref{sec:resummation-errors}), while the precision of the highest coefficients matters for the large-order and
consistency tests and for the orders still to come.

The paper is organized as follows. Section~\ref{sec:methods} describes the series, the graphs, the numerical
method and the resummation; Sec.~\ref{sec:results} gives the validation, the performance, the new coefficients and
the exponents, compared with other determinations; Sec.~\ref{sec:summary} summarizes. Five appendices give the conventions, the graphs, the zero-dimensional sum rules, the reproducibility checks, and
the availability of the data and the code.

\section{Methods}
\label{sec:methods}

This section explains how the series coefficients are reduced to integrals (Sec.~\ref{sec:methods-series}), how
the graphs and their integrands are built (Sec.~\ref{sec:methods-graphs}), how the integrals are evaluated
(Sec.~\ref{sec:methods-numerics}), how the series are resummed (Sec.~\ref{sec:methods-resummation}), and how
the results are checked (Sec.~\ref{sec:methods-validation}). A claim that a script of the repository verifies is marked with the identifier of
its check, as in \check{G2}; Appendix~\ref{app:reproducibility} lists the checks.

\subsection{Computing the series}
\label{sec:methods-series}

\subsubsection{Scheme and Callan--Symanzik relations}
\label{sec:methods-cs}

We consider the Euclidean action
\begin{equation}
S[\phi]=\int d^3x\,\Bigl[\tfrac12(\partial_\mu\phi)^2+\tfrac12 m_0^2\,\phi^2+\tfrac{\lambda_0}{4!}\,(\phi^2)^2\Bigr]
\label{eq:action}
\end{equation}
for an $N$-component field $\phi$, and denote by $\Gamma^{(a,b)}$ the one-particle-irreducible (1PI) vertex functions
with $a$ legs and $b$ insertions of $\tfrac12\phi^2$ at zero momentum, with the $O(N)$ tensors factored
out~\cite{ZinnJustinBook,Ribeca2003thesis}. In the massive zero-momentum scheme~\cite{Parisi1980,BNGM1976,BNM1978}
the renormalized mass $m$, the normalizations of $\phi$ and $\phi^2$, and the dimensionless coupling $g$ are fixed by
\begin{gather}
\Gamma^{(2)}_R(p)=m^2+p^2+O(p^4),\nonumber\\
\Gamma^{(4)}_R(0)=m\,g,\qquad \Gamma^{(2,1)}_R(0;0)=1 .
\label{eq:normcond}
\end{gather}
The renormalized functions obey the Callan--Symanzik equation~\cite{ZinnJustinBook,Ribeca2003thesis}
\begin{equation}
\Bigl[m\,\partial_m+\beta\,\partial_g-\tfrac a2\,\eta-b\,(\eta_{\phi^2}-\eta)\Bigr]\Gamma^{(a,b)}_R
=(2-\eta)\,m^2\,\Gamma^{(a,b+1)}_R
\label{eq:cs}
\end{equation}
for $b\le1$, where the right-hand side carries one more insertion at zero momentum, and $\beta=m\,\partial g/\partial m$,
$\eta=m\,\partial\ln Z/\partial m$ and $\eta_{\phi^2}=m\,\partial\ln Z_{\phi^2}/\partial m$ are taken at fixed bare
parameters ($Z$ and $Z_{\phi^2}$ normalize $\phi$ and $\phi^2$). Applied at zero momenta to the conditions~\eqref{eq:normcond}, the first through its
$p^2$ derivative, Eq.~\eqref{eq:cs} gives, with $u=g/8\pi$ and $\beta_u=m\,\partial u/\partial m$,
\begin{gather}
\eta=\frac{2h}{1+h},\qquad \beta_u=-u+2\eta\,u+(2-\eta)\,\tilde b,\nonumber\\
\eta_{\phi^2}=-(2-\eta)\,c,\qquad \frac1\nu=(2-\eta)(1-c),
\label{eq:csrel}
\end{gather}
where
\begin{gather}
\tilde b=\frac{m\,\Gamma^{(4,1)}_R(0)}{8\pi},\qquad c=m^2\,\Gamma^{(2,2)}_R(0),\nonumber\\
h=-m^2\,\frac{\partial\Gamma^{(2,1)}_R}{\partial p^2}\bigg|_{p=0}.
\label{eq:bhc}
\end{gather}
The exponents follow at the non-trivial zero of $\beta_u$ (Sec.~\ref{sec:methods-resummation}).

The three functions~\eqref{eq:bhc} are renormalized vertex functions with insertions, superficially convergent, and
their expansions in $u$,
\begin{gather}
\tilde b=\sum_{L\ge1}(-1)^{L+1}B_L\,u^{L+1},\nonumber\\
h=\sum_{L\ge2}(-1)^L H_L\,u^L,\qquad c=\sum_{L\ge1}(-1)^{L+1}C_L\,u^L,
\label{eq:series}
\end{gather}
are \emph{linear} in the graphs: $B_L$ is a weighted sum over the $L$-loop graphs of $\Gamma^{(4)}$ with one insertion,
$H_L$ and $C_L$ over the $L$-loop graphs of $\Gamma^{(2)}$ with one and two insertions, of integrals in which every
renormalization part is subtracted at zero momentum (Secs.~\ref{sec:methods-param} and~\ref{sec:methods-rbar}). No $Z$ factor
appears, because the subtractions implement the conditions~\eqref{eq:normcond} graph by graph. Together with lower orders, the coefficient of
$\ubar^{L+1}$ in $\beta$ is fixed by $B_L$ and, through the term $2\eta u$, by $H_L$; that of $\ubar^{L}$ in $\eta$
by $H_L$; and that of $\ubar^{L}$ in $\nu$ by $C_L$ and $H_L$ [Eq.~\eqref{eq:linear} in
Appendix~\ref{app:conventions}]. Only products and quotients of series enter Eq.~\eqref{eq:csrel}; no series has to
be reversed. At lowest order, for instance, $B_1=(N+8)/12$ gives $\beta(\ubar)=-\ubar+\ubar^2+O(\ubar^3)$ for every
$N$. We write the RG functions in the normalization of Ref.~\cite{Phi4tools2024},
\begin{align}
\beta(\ubar)&=-\ubar+\ubar^2+\sum_{k\ge3}\frac{p_{\beta,k}(N)}{(N+8)^{k-1}}\,\ubar^{k},\nonumber\\
\eta(\ubar)&=\sum_{k\ge2}\frac{p_{\eta,k}(N)}{(N+8)^{k}}\,\ubar^{k},\label{eq:series-form}\\
\eta_{\phi^2}(\ubar)&=\sum_{k\ge1}\frac{p_{\eta_{\phi^2},k}(N)}{(N+8)^{k}}\,\ubar^{k},\nonumber\\
\nu(\ubar)&=\frac12+\sum_{k\ge1}\frac{p_{\nu,k}(N)}{(N+8)^{k}}\,\ubar^{k},\nonumber
\end{align}
with polynomials $p_{X,k}(N)$, which follow from the polynomials $B_L(N)$, $H_L(N)$, $C_L(N)$ through
Eq.~\eqref{eq:csrel} applied to the polynomials themselves, in exact rational arithmetic, rather than to their values
at integer $N$, whose interpolation is ill-conditioned.
Equation~\eqref{eq:csrel} is the fixed-dimension counterpart of the representation of the minimal-subtraction RG
functions by finite subtracted integrals~\cite{AdzhemyanKompaniets2011,AdzhemyanKompaniets2014}. We call this computation, through
Eqs.~\eqref{eq:csrel}--\eqref{eq:series}, route A.

The previous complete computations~\cite{NMB1977,BNM1978,Phi4tools2024} and the automated approach of
Ref.~\cite{Ribeca2003thesis} instead evaluated the zero-momentum vertex functions of a scheme in which only two-point
subgraphs are subtracted (in Refs.~\cite{Ribeca2003thesis,Phi4tools2024} only the tadpole, set to zero, and the sunset,
subtracted at $p^2=0$; in Ref.~\cite{NMB1977}, according to Ref.~\cite{Phi4tools2024}, all two-point subgraphs),
and obtained the RG functions through the $Z$ factors, the relation between the coupling of that scheme and the
renormalized one, and its inverse~\cite{Ribeca2003thesis}; we use this $Z$-factor route as an independent check
(route B, Sec.~\ref{sec:methods-validation}). Along that route, in the intermediate scheme of
Refs.~\cite{Ribeca2003thesis,Phi4tools2024}, the contribution of the $L$-loop graphs is a small difference of large numbers: the $L$-loop four-point graphs contribute
to the coefficient of $\beta$ an amount larger than the net $B_L$ by factors of 5.5, 14.4, 29.2, 49.5 and 74 at
$L=2,\dots,6$, and the $L$-loop graphs of $\partial\Gamma^{(2)}/\partial p^2$ contribute to $\eta$ an amount larger than
$H_L$ by about 20, 114 and 106 at $L=4$, 5 and 6; the rest is cancelled by products of lower orders. The relative
error of the coefficient is the relative error of the graph sum multiplied by this factor, so that the number of samples
grows with its square. In Eq.~\eqref{eq:series} the same cancellations take place within each integrand, between a
graph and its counterterms evaluated at the same point, so that they cost no additional samples;
Sec.~\ref{sec:results-performance} quantifies the gain.

\subsubsection{Parametric integrands}
\label{sec:methods-param}

At zero external momenta and unit mass, the integral of a graph $G$ with $\ell$ lines and $L$ loops, with one
$d^3k/\pi^2$ per loop (Nickel's normalization~\cite{NMB1977}), is
\begin{equation}
A_G=\frac{\Gamma(\ell-3L/2)}{\pi^{L/2}}\int_\Delta d\mu\;U_G^{-3/2},\quad U_G=\sum_{T}\prod_{e\notin T}x_e ,
\label{eq:param}
\end{equation}
where $T$ runs over the spanning trees of $G$, $\Delta$ is the simplex $x_e\ge0$, $\sum_e x_e=1$, and
$d\mu=\delta(1-\sum_e x_e)\prod_e dx_e$~\cite{BergereZuber1974,Ribeca2003thesis,BrownKreimer2013}; the one-loop bubble, for instance,
equals one. The Kirchhoff polynomial $U_G$ factorizes over the 2-connected components (blocks) of $G$. An insertion of
$\tfrac12\phi^2$ at zero momentum on line $e$ squares its propagator, that is, it multiplies the Schwinger integrand by
the parameter of $e$: in Eq.~\eqref{eq:param} it adds a factor $x_e$ and raises the argument of the Gamma function by
one. For a two-point graph with external vertices $a$ and $b$, the external momentum enters as
$\exp(-p^2\,W_G/U_G)$, where $W_G$ is the Kirchhoff polynomial of the graph with $a$ and $b$ identified (equivalently,
the polynomial of the spanning 2-forests that separate $a$ from $b$; $W_G=0$ if $a=b$). The $p^2$ derivative at
$p=0$ is therefore $-\Gamma(\ell-3L/2+1)\,\pi^{-L/2}\int_\Delta d\mu\,W_G\,U_G^{-5/2}$, a single one-signed integrand,
which replaces the expansion into $n(n-1)/2+2n$ signed terms per chain of $n$ propagators used in momentum
space~\cite{Ribeca2003thesis}.

Since $\ell=2L$ for $\Gamma^{(4)}$ and $\ell=2L-1$ for $\Gamma^{(2)}$, all integrals entering Eq.~\eqref{eq:series} carry the
same prefactor $\Pi_L=\Gamma(L/2+1)\,\pi^{-L/2}$, and the per-order sums are
\begin{align}
B_L&=\Pi_L\sum_{G\in\mathcal G_4(L)}S_G\sum_{e}\int_\Delta d\mu\;x_e\,
      \Rbar_{G,e}\bigl[U_G^{-3/2}\bigr],\nonumber\\
H_L&=\Pi_L\sum_{G\in\mathcal G_2(L)}S_G\sum_{e}\int_\Delta d\mu\;x_e\,
      \Rbar_{G,e}\bigl[W_G\,U_G^{-5/2}\bigr],\label{eq:BHC}\\
C_L&=\Pi_L\sum_{G\in\mathcal G_2(L)}S_G\sum_{(e,e')}\int_\Delta d\mu\;x_e\,x_{e'}\,
      \Rbar_{G,ee'}\bigl[U_G^{-3/2}\bigr],\nonumber
\end{align}
where $\mathcal G_E(L)$ is the list of $L$-loop graphs with $E$ legs, $S_G$ the weight of $G$ including its $O(N)$
factor (Sec.~\ref{sec:methods-graphs}), $e$ runs over the lines and $(e,e')$ over the ordered pairs of lines,
$e=e'$ included, and $\Rbar_{G,e}$ is the subtraction operator of Sec.~\ref{sec:methods-rbar}, which depends on where the
insertions are. The propagators have the renormalized mass, $m=1$. (Refs.~\cite{NMB1977,Ribeca2003thesis,Phi4tools2024}
use propagators of unit mass in the scheme of their integrals instead; individual integrals differ, the RG functions
do not.) The sums are known exactly at
one and two loops [Eq.~\eqref{eq:exactsums}]. On the simplex the integrands can be
multiplied by $(\sum_e x_e)^{-(L/2+1)}$, which equals one there and makes them homogeneous of degree $-\ell$: the
integrals are then projective, and can be evaluated in any coordinates of the projective simplex
(Sec.~\ref{sec:methods-tropical}). The integration dimension is $\ell-1$, that is 15 for the eight-loop four-point and 16
for the nine-loop two-point graphs.

\subsubsection{Zero-momentum subtractions}
\label{sec:methods-rbar}

The renormalization parts of a graph are its 1PI subgraphs $\gamma$ of three types, classified by their legs (the
lines and external legs of $G$ attached to $\gamma$) and by the insertions on their lines: four-point parts without
insertions, subtracted by $t^0_\gamma$, the term of order zero of the Taylor expansion in the external momenta of
$\gamma$; two-point parts without insertions, subtracted by $t^2_\gamma$, the terms up to order two (mass and field
normalization); and two-point parts with one insertion, subtracted by $t^0_\gamma$. These subtractions enforce the
conditions~\eqref{eq:normcond} graph by graph~\cite{BogoliubovParasiuk1957,Hepp1966,Zimmermann1969,BergereLam1976,Zavialov1990}. In $d=3$ only the tadpole and
the sunset subgraphs need renormalization~\cite{Ribeca2003thesis,Phi4tools2024}; all the other subtractions are finite, but the scheme requires them,
and through them the cancellations described above take place inside each integrand. The graphs of Eq.~\eqref{eq:BHC}
are themselves not renormalization parts, except those of $H_L$, whose own subtraction $t^0$ does not depend on $p^2$
and drops out of the derivative, so that the subtraction of the subgraphs alone, $\Rbar$, is the complete
renormalization:
\begin{equation}
\Rbar_G=\sum_{F}\prod_{\gamma\in F}\bigl(-t_\gamma\bigr),
\label{eq:forest}
\end{equation}
where $F$ runs over the forests, sets of renormalization parts that are pairwise nested or vertex-disjoint (the
compatibility condition of the counterterms generated by a Lagrangian), the empty forest included, and inner operators
act first.

In parametric space the subtractions have closed forms. The term $t^0_\gamma$ replaces $U_\theta$, in the parent $\theta$ of
$\gamma$, by $U_\gamma\,U_{\theta/\gamma}$, where $\theta/\gamma$ is $\theta$ with $\gamma$ contracted to a point: this is the
leading term of $U_\theta$ when the parameters of $\gamma$ go to zero. For a two-point part $\gamma$ joined to the rest at
the vertices $a$ and $b$, the classical factorization of the Kirchhoff polynomial across a two-vertex
cut~\cite{Nakanishi1971,BognerWeinzierl2010},
\begin{align}
U_\theta&=U_\gamma\,U_{\theta/\gamma}+W_\gamma\,U_{\theta\setminus\gamma}\nonumber\\
&=U_\gamma\,\bigl(U_{\theta/\gamma}+\varepsilon_\gamma\,U_{\theta\setminus\gamma}\bigr),\qquad
\varepsilon_\gamma=\frac{W_\gamma}{U_\gamma},
\label{eq:effline}
\end{align}
where $\theta\setminus\gamma$ is $\theta$ without the lines of $\gamma$ and $W_\gamma$ is the Kirchhoff polynomial of $\gamma$
with $a$ and $b$ identified (a spanning tree of $\theta$ contains either a spanning tree of $\gamma$, and then a
spanning tree of $\theta/\gamma$, or a spanning two-forest of $\gamma$ that separates $a$ from $b$, and then its other
lines connect $a$ and $b$ through $\theta\setminus\gamma$), shows that $\gamma$ acts on its parent as a massless line
with Schwinger parameter
$\varepsilon_\gamma$; $t^2_\gamma$ keeps the terms of order zero and one in $\varepsilon_\gamma$, and reduces to
$t^0_\gamma$ when $a=b$. The term of a forest $F$ is thus
\begin{equation}
T_F=(-1)^{|F|}\,x_{\mathrm{ins}}\prod_{\theta\in F\cup\{G\}}U_{\theta/F}^{-3/2},
\label{eq:forestterm}
\end{equation}
where $x_{\mathrm{ins}}$ is the monomial of the insertions and $\theta/F$ is the minor of $\theta$ in which every maximal
element of $F$ below $\theta$ is contracted ($t^0$) or replaced by its effective line ($t^2$), with the result expanded to
first order in every $\varepsilon$; for $H_L$ the factor of the root is $W_{G/F}\,U_{G/F}^{-5/2}$. Both terms of
Eq.~\eqref{eq:effline} have the degree of $U_\theta$, so all forest terms are homogeneous of the same degree: they share
the simplex and the prefactor of Eq.~\eqref{eq:BHC}, and their sum, with all positions of the insertions, is integrated
as one function~\cite{BrownKreimer2013}.

\subsection{Graphs and integrands}
\label{sec:methods-graphs}

The pipeline from the Lagrangian to the sampled integrands has three stages, each with its own checks: for each order, the generation
of the graphs, the construction of the subtracted integrand of every graph, and the consistency checks that
tie the two to exact results. All three are automatic and were run here to ten loops.

\subsubsection{Generating the graphs}
\label{sec:methods-generation}

The lists of 1PI graphs of $\Gamma^{(2)}$ and $\Gamma^{(4)}$ are produced by a generator that derives from the one
written for the seven-loop computation of Ref.~\cite{Ribeca2003thesis}, with three improvements: it grows the graphs
by vertex insertion instead of enumerating labelled graphs and screening them by a canonical form, it canonicalizes
them by colour refinement with backtracking, and it takes any list of vertex degrees, the external legs being implicit
in the valences. The class it generates is every connected graph with the given valences, self-loops and bridges
allowed, and the construction is complete: removing a leaf of a spanning tree and re-pairing the half-lines it leaves
is the inverse of an insertion, so that every graph with $V$ vertices descends from one with $V-1$. The class
is grown order by order, each candidate canonicalized and kept in a hash set, and the 1PI graphs are written out at
the end of every order, together with their Nickel indices and weights. The cost is a few hundred insertions per
graph grown, so that it follows the size of the class, growing by a factor of ten to twelve per loop on one core.
Table~\ref{tab:generation} gives the sizes and the times, measured on one core of a cluster node; the
counts through eight loops are those of Ref.~\cite{Ribeca2003thesis} (Table~3.1), the others are new~\check{G1}. Every
complete list satisfies its zero-dimensional sum rule (Sec.~\ref{sec:methods-checks}) through twelve loops for
$\Gamma^{(2)}$ and eleven for $\Gamma^{(4)}$.

Every graph is converted to its Nickel index~\cite{NMB1977} in canonical form---the
lexicographic minimum over vertex labellings---which identifies it across lists and keys its random numbers
(Sec.~\ref{sec:methods-rng}); one-vertex-reducible graphs and graphs with both legs on one vertex are flagged, since
their integrands vanish (Sec.~\ref{sec:methods-construction}), and are not built for $B_L$ and $H_L$.

\begin{table}
\caption{\label{tab:generation}%
Generation of the graph lists: numbers of 1PI graphs without self-loops (one-vertex-reducible ones included), and the
CPU time of the generator on one core of a cluster node (an Intel Xeon Platinum 9242), one list at a time on an
otherwise idle node, the Nickel indices, weights and sum rules included. The eleven- and twelve-loop lists were made
on 96 threads, in 23~s and 275~s of wall time for $\Gamma^{(2)}$ and 1005~s for $\Gamma^{(4)}$ at eleven loops, the
last with 38~GB of memory; their times in the table are core times summed over the threads. Every regenerated list
equals the stored one. The counts of graphs whose integrands survive the pruning (Sec.~\ref{sec:methods-construction})
are in Table~\ref{tab:graphcounts}.}
\begin{ruledtabular}
\setlength{\tabcolsep}{3pt}\footnotesize
\begin{tabular}{lrrrr}
$L$ & $\Gamma^{(4)}$ & time & $\Gamma^{(2)}$ & time \\
\hline
7  & 3986       & 2.1 s & 317    & 0.1 s \\
8  & 26\,540    & 22 s & 1622   & 1.1 s \\
9  & 197\,022   & 4.3 min & 9184   & 9.8 s \\
10 & 1\,593\,254 & 53 min & 59\,614 & 98 s \\
11 & 13\,975\,511 & 14.1 h & 428\,164 & 22 min \\
12 & --- & --- & 3\,393\,754 & 4.2 h \\
\end{tabular}
\end{ruledtabular}
\end{table}

\paragraph{Weights.}
Each graph carries the weight $S_G=T_G(N)\,E!/|\mathrm{Aut}\,G|$, where $E$ is its number of legs, $\mathrm{Aut}\,G$ its
automorphism group at the level of half-edges with unlabelled legs, and $T_G(N)$ its $O(N)$ factor, obtained by
contracting the vertex tensors $(\delta_{ij}\delta_{kl}+\delta_{ik}\delta_{jl}+\delta_{il}\delta_{jk})/3$ over the $3^V$
pairings at the $V$ vertices and projecting on the external tensor, so that $T_G(1)=1$. The factor $E!/|\mathrm{Aut}\,G|$
includes the sum over the labellings of the legs, which is legitimate at zero external
momenta~\cite{Ribeca2003thesis}. The integrals do not depend on $N$: since every graph carries its polynomial
$T_G(N)$, one run gives the coefficients for all $N$.

\subsubsection{Constructing the integrands}
\label{sec:methods-construction}

Every polynomial in Eq.~\eqref{eq:forestterm} is the Kirchhoff polynomial of a minor of $G$, possibly with two vertices
identified, and factorizes into the Kirchhoff polynomials of its blocks, which we call atoms. Expanding all forest terms
into products of half-integer powers of atoms, with binomial coefficients that are dyadic rationals and hence exact in
binary floating point, and adding the coefficients of identical products, removes every combination of forest terms
that cancels identically; we call this step the pruning. Each graph becomes
\begin{equation}
f_G(x)=\sum_k c_k\,x_{\mathrm{ins},k}\prod_a P_a(x)^{p_{ka}/2},
\label{eq:integrand}
\end{equation}
with atoms $P_a$, integer $p_{ka}$ and exact $c_k$; terms with the same atoms, which differ only by the insertions, are
grouped. The result of the expansion is a small program: the list of atoms (each a minor of $G$, evaluated at run time
by a spanning-tree pass and a factorization of its loop matrix), the list of terms (coefficient, insertion monomial,
atom exponents), and the tables of the sampling density of Sec.~\ref{sec:methods-tropical}. One generic kernel
evaluates any such program (Sec.~\ref{sec:methods-kernel}).

The pruning removes up to two thirds of the raw terms. At six loops a four-point graph has on average 250 forests,
which leave 116 terms in 22 groups with 19 distinct atoms (at most 55); at seven loops, in a sample of 300 four-point
graphs, 583 forests, 219 terms, 39 groups and 26 atoms (at most 6128, 766, 126 and 85). The largest programs are
those of two-point graphs with many nested two-point parts: at nine loops the chain of eight bubbles of $C_9$ has
$3.4\times10^5$ forests and 2328 atoms, against at most 269 atoms for $H_8$ and 203 for $B_8$. The expansion is done
once per graph on the host, in parallel over the graphs; its cost grows faster than the number of graphs
(Table~\ref{tab:graphcounts}: about \buildpc{b8} core-seconds per graph for $B_8$, \buildpc{h9} for $H_9$,
\buildpc{c9} for $C_9$, and from \buildpc{b10} for $B_{10}$ to \buildpc{c10} for $C_{10}$ at ten loops), so that at ten
loops it dominates a job. The built integrands of a list are therefore
written once to a cache file, by a CPU-only job, and loaded by every subsequent run; a run with and without the cache
gives bitwise the same results~\check{I4}.

Graphs that vanish identically come out of the pruning with no terms at all, without any structural rule: graphs with a
cut vertex (one-vertex-reducible graphs), which factorize at zero momentum~\cite{Ribeca2003thesis} into pieces of which
one is a renormalization part at exactly zero momentum, cancelled by its own subtraction; graphs with self-loops,
for $L\ge2$; and, for $H_L$, generalized tadpoles, graphs with both external legs on one vertex, for which $W_G=0$. At
three to seven loops these are 21--30\% of the four-point and 15--17\% of the two-point lists (more for $H_L$); at
eight loops 18\% and 14\%, and at nine loops 12\% of the two-point list. In production runs the one-vertex-reducible
graphs of $B_L$ and $H_L$ are dropped before the expansion; their vanishing, and that of the graphs with self-loops, was
verified explicitly up to five loops~\check{I2}. Those of $C_L$ are expanded in every run, and the pruning leaves no
term in any of them, at every order (independently verified up to seven loops~\check{I2}).

\paragraph{Twins.}
At zero momenta and equal masses, Eq.~\eqref{eq:param} depends on $G$ only through its cycle matroid, so that
2-isomorphic graphs in Whitney's sense~\cite{Whitney1933}, which we call twins, have equal integrals in any
representation~\cite{Ribeca2003thesis}: the 3986 seven-loop four-point graphs have 3018 distinct unsubtracted integrals,
and the 3150 non-vanishing ones fall into 2391 classes. We checked that the subtracted integrands of twins, summed over
the positions of the insertion, agree pointwise under the matroid map of the lines (all 3 pairs at four loops and all 25
non-vanishing pairs at five loops for $B_L$)~\check{I3}. Twins have different $O(N)$ factors, so that a merged class carries the
sum of their weights. Twins were not merged in the runs of Sec.~\ref{sec:results}: every class is one graph.

\subsubsection{Consistency checks}
\label{sec:methods-checks}

The graphs, their weights and their integrands are checked, before any sampling, against results that are exact.
\begin{enumerate}
\item \emph{Zero-dimensional sum rules.} In zero dimensions every integral equals one, so that $\sum_{G}S_G$ over a
complete list is the coefficient of $g^L$ in the corresponding vertex function of the zero-dimensional $O(N)$ model,
known exactly from its moments (Appendix~\ref{app:sumrules}). Summed over all graphs, one-vertex-reducible ones
included, this tests the completeness of a list and its weights without any integral; for lists without self-loops the
bare mass is shifted so that every self-loop cancels. Every list passes it exactly, with the full $O(N)$
polynomial, through ten loops (Table~\ref{tab:sumrules})~\check{G2}.
\item \emph{Independent lists.} Converted to Nickel indices, our lists equal those of Ref.~\cite{Phi4tools2024}
(generated with \textsc{feyngen}~\cite{Borinsky2014feyngen}) as sets, for $\Gamma^{(4)}$ at $L\le7$ and $\Gamma^{(2)}$ at
$L\le8$~\check{G3}, and those of a brute-force generator in our code where that is practical ($\Gamma^{(4)}$ to five loops,
$\Gamma^{(2)}$ to six). Two independent canonicalizers (one in the generator's string format, one on Nickel indices)
agree on every graph, and all graphs of every list are distinct~\check{G1}.
\item \emph{Weights.} The weights and $O(N)$ polynomials equal those of the independent implementation of
Refs.~\cite{Ribeca2003thesis,GuidaRibeca2006} on all 260 graphs with $L\le5$ (its 251 and the nine graphs with self-loops its lists lacked) and on all 47
graphs of its four-loop tables~\check{G4}.
\item \emph{Integrands.} The expansion~\eqref{eq:integrand} agrees pointwise with an independent evaluation of the
forest formula in 100-digit arithmetic (Sec.~\ref{sec:methods-validation})~\check{I1}; vanishing graphs
vanish~\check{I2}, twins agree pointwise~\check{I3}, and the exact two-loop sums $B_2=1/2$, $H_2=1/81$ and $C_2=1/12$
are reproduced to a few $10^{-7}$ by the sampled integrals (Sec.~\ref{sec:results-validation})~\check{N3}.
\item \emph{Large order.} Each new coefficient is compared with its leading large-order estimate
(Sec.~\ref{sec:methods-validation}), a test of the complete chain that catches an error of a few per cent (Sec.~\ref{sec:methods-validation})~\check{S3}.
\end{enumerate}

\subsection{Numerical approach}
\label{sec:methods-numerics}

\subsubsection{Choosing the method}
\label{sec:methods-choice}

The earlier evaluations integrated each graph separately in momentum space, with one-loop subgraphs replaced by
analytic effective vertices, on deterministic meshes~\cite{NMB1977}, by adaptive
cubature~\cite{Ribeca2003thesis,GuidaRibeca2006}, or with VEGAS~\cite{Lepage1978,Phi4tools2024}. Because of the
cancellations of the $Z$-factor route, and of the precision lost in finding the zero of $\beta$ and in the resummation,
eight to ten digits per graph were considered necessary~\cite{Ribeca2003thesis}; high-precision
methods based on integration-by-parts reduction~\cite{ChetyrkinTkachov1981,Laporta2000} are available for massive vacuum integrals only to five
loops~\cite{LutheSchroeder2016}. The cost of that strategy is set by its hardest graphs: every graph must reach many
digits, and both the number of graphs and their dimension grow with the order, so that each new order costs more than
the last by a large and growing factor.

We made the opposite choice deliberately: a method whose precision depends as little as possible on the order. Every
integrand, whatever its size, is sampled by the same algorithm. Like any Monte Carlo estimate, an order sum then
reaches a relative error $\epsilon$ in a time proportional to $1/\epsilon^2$; what the design controls is the constant of
proportionality, and it aims at one that grows as slowly as possible with the order. Three choices serve that aim. The
samples are shared among the graphs of an order in proportion to their spread (Sec.~\ref{sec:methods-estimators}),
so that the constant is set by an average over the graphs of the relative spread and the cost of one sample, and by a
factor that measures the cancellations between graphs of opposite sign: the number of graphs does not enter it by itself. Each
graph is sampled with a density that follows its singularities (Sec.~\ref{sec:methods-tropical}), so that the weights
stay bounded at every order. And each term of the forest formula~\eqref{eq:forestterm} costs only a polynomial in $L$
to evaluate, since its polynomials are determinants of $L\times L$ matrices (Kirchhoff's matrix-tree theorem). At low
orders such a method cannot match the per-graph treatment, where effective vertices reduce each integral to a few
dimensions and deterministic integration reaches eight to ten digits. At high orders, where the dimension grows with
the number of lines and deterministic rules slow down accordingly, it should become far cheaper. How slowly its
constant actually grows could only be measured: empirically, the time to a fixed relative precision grows by a factor of about
\costgrowthrange\ per loop (Sec.~\ref{sec:results-performance}). The
construction of the integrands, done once per graph, still grows with their number, that is factorially. It is a
one-off cost on ordinary CPUs, but at high enough orders it becomes the dominant one: at the tenth order it already
exceeds the sampling (Sec.~\ref{sec:results-performance}). A precision that hardly depends on the order also keeps
the statistical errors of the highest coefficients, which carry the information on the large-order behaviour, from
limiting the resummation; at the present orders they are already negligible there (Sec.~\ref{sec:resummation-errors}).

The lower-order terms that enter a new coefficient together with its new order sums are known far more precisely
and do not cancel them, so that the relative error of a coefficient is that of its order sums. What matters is therefore the relative error of each order sum, which no
cancellation amplifies. Monte Carlo integration then becomes adequate: it requires no choice of variables per
graph, it gives reliable error estimates when the weights are bounded, and it is naturally parallel. The parametric
integrands, however, are singular on every face of $\Delta$ on which the parameters of a subgraph vanish, and the faces
overlap at the corners; under uniform sampling the variance is already infinite on the face of a one-loop bubble. Sector
decomposition~\cite{BinothHeinrich2000,pySecDec2018} removes these singularities, but in our tests it required $10^5$--$3\times10^5$ sectors for a six-loop
graph and $(1$--$5)\times10^6$ for a seven-loop four-point graph, and lattice rules on the sector
integrands~\cite{pySecDecQMC2019} remained pre-asymptotic up to $10^9$ points at 9--11 dimensions. We use instead
tropical sampling~\cite{Borinsky2023tropical,feyntrop2024}, a recent technique, increasingly used for Feynman
integrals~\cite{Balduf2023,feyntrop2024,Borinsky2025global}, which samples the Hepp sectors~\cite{Hepp1966} rather
than enumerating them. A sampler of the same
kind, with an exponent for every subset of lines and the sector drawn by a recursion over the subsets, was used
for the forest-subtracted parametric integrands of the electron anomalous magnetic moment, on CPUs and then on
GPUs~\cite{Volkov2017,Volkov2018,Volkov2024}; our exponents follow the forest terms of the zero-momentum subtraction
(Sec.~\ref{sec:methods-tropical}). Each graph is sampled with its own density. Sampling a whole order with one
density~\cite{Borinsky2025global} brings no gain for the unsubtracted integrands, which are all positive: in our tests at
four and five loops a one-sample global sampler came within 3--4\% of the optimally stratified per-graph sampling
described below. Its advantage lies elsewhere: it draws graphs and points together, without enumerating the graphs, in
polynomial time and memory, and it has been applied to finite correlators of a massive super-renormalizable theory in
three dimensions~\cite{Borinsky2025global}.

\subsubsection{Forest-aware tropical sampling}
\label{sec:methods-tropical}

In the Hepp sector $x_{\sigma(1)}<\dots<x_{\sigma(\ell)}$, with $S_j=\{\sigma(1),\dots,\sigma(j)\}$ and
$t_j=x_{\sigma(j)}/x_{\sigma(j+1)}$, the density with respect to the projective measure $\prod_j dt_j/t_j$ is
\begin{gather}
p(x)=\frac{1}{J(E)}\prod_{j=1}^{\ell-1}t_j^{\,\omega(S_j)},\nonumber\\
J(S)=\sum_{e\in S}\frac{J(S\setminus e)}{\omega(S\setminus e)},\qquad J(\{e\})=1 ,
\label{eq:tropical}
\end{gather}
where $J(E)$, a generalized Hepp bound~\cite{Panzer2022Hepp}, sums $\prod_j\omega(S_j)^{-1}$ over all sectors, and the
exponents $\omega(S)$ follow the integrand. When the parameters of a set $S$ of lines go to zero together,
$x_e\to\lambda x_e$ for $e\in S$, an atom behaves as $\lambda^{h_a(S)}$, where $h_a(S)$ is the loop number of the
lines of $S$ that belong to it, and the term $k$ of Eq.~\eqref{eq:integrand} as $\lambda^{\deg_S k}$, with
$\deg_S k=\sum_a p_{ka}\,h_a(S)/2+n_{k}(S)$ and $n_k(S)$ the number of its insertions on lines of $S$. The term
converges on this face if $|S|+\deg_S k>0$. We take
\begin{equation}
\omega(S)=s\,\Bigl[\,|S|+\min_k\deg_S k+\delta(S)\Bigr],
\label{eq:omega}
\end{equation}
with the minimum over all terms that survive the pruning, $\delta(S)=1$ where $|S|+\min_k\deg_S k\le0$ and zero
elsewhere, and $s\in(0,1]$ a softening factor: $s=1$ gives the tropical density, and $s<1$ lowers every exponent by the
same fraction, which tames the tails of the weights at the price of more samples in the bulk; we use $s=0.7$, chosen as
described in point (iii) below. Both
$\omega$ and $J$ are tabulated on the host for all $2^\ell$ subsets of lines, so that the cost grows as $2^\ell$ rather
than as the number $\ell!$ of sectors; the two tables take 1\,MB per graph at $\ell=16$. A point is drawn by removing, from
$S=E$ downwards, the largest line $e$ of $S$ with probability $J(S\setminus e)/[\omega(S\setminus e)\,J(S)]$, and setting
the ratio of the largest remaining parameter to $x_e$ to $t=v^{1/\omega(S\setminus e)}$ with $v$ uniform~\cite{Borinsky2023tropical};
a sample takes $2(\ell-1)$ uniform numbers. Its weight $w=f_G(x)\prod_e x_e/p(x)$, with $f_G$ in projective form, has expectation
$\int_\Delta d\mu\,f_G$, and the estimator is unbiased for any positive $\omega$.

Three features of Eq.~\eqref{eq:omega} adapt the tropical density to the subtracted integrands. (i) The minimum runs over
all forest terms: a counterterm brings in the faces of its cograph, and a density built from the unsubtracted graph alone
produced weights up to $8\times10^4$ times the mean. (ii) On faces where every term diverges, $|S|+\min_k\deg_S k\le0$, the
sum converges only through a $t^2$ subtraction, which raises the true exponent by two units (for the sunset, $|S|+\deg_S$ goes from 0
to 2, the measure from $\lambda^{-1}d\lambda$ to $\lambda^{+1}d\lambda$) and which the term-by-term minimum does not see; the shift $\delta=1$ keeps the density
integrable and below the true behaviour, as we checked in extended precision on every shifted face of 280 integrands at
four and five loops (no violation); a shift by two units changes the per-sample relative standard deviation of the order sum [$\rho$ of Eq.~\eqref{eq:rho}
below] by less than 0.5\%. On no face is the exponent negative before the shift, for any integrand up to ten loops: the builder refuses such
integrands, on which the double-precision evaluation of the expanded $t^2$ terms would have round-off of infinite variance,
and it built every class of every order~\check{G5}. (iii) The tropical density falls off faster than the integrand before the
asymptotic regime: the positive powers of atoms in the $t^2$ terms give large but bounded prefactors that it does not see,
and it ignores the factor $(\sum_e x_e)^{-(L/2+1)}$, which plausibly explains why softening helps the unsubtracted
integrands as well. With $s=1$ the weight distributions had tails with a Hill index~\cite{Hill1975}
as low as 1.9 and five-loop error estimates were unreliable. Softening all exponents by $s<1$ makes the weights vanish on
every face. The value $s=0.7$ is close to optimal for every kind of integrand at four and five loops, where it halves the
per-sample standard deviation of both the subtracted and the unsubtracted integrands (any positive set of exponents
gives a valid density, which can be fitted to the integrand~\cite{Salvatori2025}; the optimum found there for the
corresponding uniform factor is 0.72); the smallest tail index is then 3.87,
below 4, so that the fourth moment, and with it the error of the error below, is only indicative. We have not repeated
the scan of $s$ and of the shift beyond five loops. At six and seven loops replica ensembles test the resulting error
estimates directly (Sec.~\ref{sec:results-validation})~\check{N7}, and at eight loops the independent runs of $B_8$ and
$H_8$ do (Table~\ref{tab:newsums}); beyond that they rest on the within-run statistics alone.

\subsubsection{Estimators, error estimates and allocation}
\label{sec:methods-estimators}

Every class is sampled independently; a class is one graph, since twins (Sec.~\ref{sec:methods-construction}) were not
merged. Its contribution to the order sum
$X_L=\sum_G a_G\,I_G$, with $a_G=\Pi_L S_G$ and $I_G$ the integral in Eq.~\eqref{eq:BHC} summed over the positions of the insertions, is estimated as
$a_G\bar w_G$ with variance $a_G^2\sigma_G^2/n_G$, where $\bar w_G$ and $\sigma_G$ are the mean and standard deviation of its
$n_G$ weights. For each class we also record the kurtosis $\mathcal K_G$ of the weights, from which the relative standard error
of $\sigma_G$, $[(\mathcal K_G-1)/4n_G]^{1/2}$, is propagated to that of the error of $X_L$ (the error of the error); the largest
weight relative to the mean; and the largest round-off scale of a single sample (Sec.~\ref{sec:methods-rng}). An order is
characterized by
\begin{equation}
\rho=\frac{\sum_G|a_G|\,\sigma_G}{|X_L|},\qquad \rho_\tau=\frac{\sum_G|a_G|\,\sigma_G\sqrt{\tau_G}}{|X_L|},
\label{eq:rho}
\end{equation}
the per-sample relative standard deviation under optimal allocation, such that $n$ samples give a relative error
$\rho/\sqrt n$, and its counterpart per unit of computing time, with $\tau_G$ the measured cost per sample.

After a pilot of $2^{14}$ to $2^{16}$ samples per class in the runs of Sec.~\ref{sec:results}, the samples are allocated by Neyman's
rule~\cite{Neyman1934}, $n_G\propto|a_G|\sigma_G$, or, in a time-boxed
run, $n_G\propto|a_G|\sigma_G/\sqrt{\tau_G}$ with $\tau_G$ in core or device seconds, which minimizes the error at fixed
computing time. The main phase runs in rounds (eight by default), each of which brings every class to a fixed fraction of
its allocation; the complete state is written after every round, and a run that reaches its time limit stops at the end of
a unit of work. Since the samples of a class always form a prefix of its random-number stream, an interrupted run is
identical to a shorter one. At $s=0.7$ the pulls of single runs were Gaussian over 340 replica runs on CPUs at two to four loops,
with no single run beyond three standard deviations, and over 400 replicas at three and four loops with the final code
the mean pulls were within $\pm0.1$ of zero and their spread between 0.91 and 1.06; at $s=1$ fatter tails had been
seen. Since the sample variance of a heavy-tailed class converges slowly, we nevertheless judge agreement over
independent seeds rather than from the error estimate of a single run (Sec.~\ref{sec:methods-validation}).

\subsubsection{Pooling the runs: the ledger}
\label{sec:methods-ledger}
Each class draws its random numbers from its own stream of a counter-based generator (Sec.~\ref{sec:methods-rng}),
whose key is derived from the seed and from the class (kind, loop order, Nickel index). Two runs with different seeds therefore sample every
class independently, whatever their hardware, allocation or duration and whenever they are contributed, and their per-class statistics pool exactly:
with $n_i$ samples, weight mean $m_i$ and weight variance $s_i^2$ from run $i$, the pooled mean is
$\sum_i n_i m_i/\sum_i n_i$ and the pooled sum of squares is $\sum_i [(n_i-1)s_i^2+n_i m_i^2]$. The ledger keeps,
for every class, the registers of all its samples pooled: their number, mean and central moments up to the fourth,
updated run by run with the pairwise formulas of Welford, Chan, Golub and LeVeque, and
P\'ebay~\cite{Welford1962,ChanGolubLeVeque1979,Pebay2008}, together with the extremes of the weights, the round-off
indicators and the time per sample, so that the ledger's size does not grow with the number of runs. For every run it records
the seed, the code version and build signature, the device self-tests, the classes covered and the run's own
statistics; every run file is kept as received in an archive outside the ledger, from which the ledger can be rebuilt
bit for bit~\check{N5}. The per-order sums, the coefficients and their uncertainties, and the tables of this paper are
regenerated from the ledger~\check{S2}. A run is admitted to the ledger only if its seed is new for that
order, its build is a known one whose devices passed the start-up self-test against the host, it has no non-finite
samples, and its classes are statistically consistent with the pooled ones (a $\chi^2$ over the classes it shares with
the ledger, which detects a wrong build or a failing device)~\check{N5}. The ledger and the tools are public with the code; a
refinement is one command that chooses the order with the largest relative uncertainty, allocates the time across its
classes by their measured cost and spread, runs with a fresh seed and submits its contribution. The precision of every
coefficient can thus be improved indefinitely, by anyone, without recomputing what exists.

\subsubsection{Random numbers and arithmetic}
\label{sec:methods-rng}

The random numbers come from the counter-based generator Philox4x32-10~\cite{Salmon2011Random123}, a pure function of a
key and a counter; our implementation reproduces its published known-answer vectors~\check{N1}. Each class has its own key, a hash of
the seed and of a string made of the kind of integrand, the loop order and the Nickel index, and the counter is the sample
index. The random numbers of a class therefore depend only on the seed, the kind, $L$ and the graph, not on its position in
a list, on how an order is split into runs, or on the hardware. Runs at different loop orders are independent even with the
same seed (the largest measured correlation between index-matched classes, $5.4\times10^{-3}$, is consistent with zero), and replicas differ by the seed.

The evaluation is entirely in double precision (FP64), in logarithmic form. For each atom, the leading monomial at the
sampled point is the product of the parameters of the chords of a minimum spanning tree, found by Kruskal's algorithm with
the lines in increasing order of $x_e$; the loop matrix of the corresponding fundamental cycles, rescaled by the chord
parameters, has entries that are ratios of parameters not larger than one and a determinant between one and the number of
spanning trees, computed by an $LDL^{\mathsf T}$ factorization. Hence $\log P_a$ is obtained without overflow or underflow at
any point the sampler can produce, however close to a face. The terms of Eq.~\eqref{eq:integrand} are added with
Neumaier's compensated summation~\cite{Neumaier1974}, and the sums and moments of the weights are accumulated with
compensation as well. Pruning removes the terms that cancel identically; what remains are the cancellations of the
subtractions close to the faces they protect, where double precision loses all relative accuracy pointwise (within
about $3\times10^{-7}$ of a sunset face). The absolute round-off of a sample remains of the order of the machine
precision times $\sum_k|\text{terms}|$, which is negligible against $\sigma_G$: double and 80-bit extended precision at
the same points differ by at most $1.7\times10^{-8}$ standard deviations (rms) at four and five loops, and every run
reports the worst single-sample ratio of this scale to $\sigma_G$, with a warning above $10^{-3}$. Single precision is not
used, as it is not adequate for subtracted integrands: already for the sunset-subtracted five-loop integrals of the
intermediate scheme its error reaches $10^{-2}$ standard deviations per sample.

\subsubsection{A single kernel for CPUs and GPUs}
\label{sec:methods-kernel}

The integrator is written in C++. The sampler and the evaluation of Eq.~\eqref{eq:integrand} are written once, as
header code compiled both by the host C++ compiler and by the CUDA compiler; the expansion of Sec.~\ref{sec:methods-rbar} runs on the host and produces, for
each class, a flat image of its atoms, terms and tables. The unit of computation is a chunk of 4096 consecutive samples
of one class, evaluated by a block of 128 lanes of 32 samples each. Each lane accumulates in registers with compensated
summation, the block reduces its lanes along a fixed binary tree, chunks are merged in index order within units of 16
chunks, and units in order. Compensated merges are not associative bit for bit, so this order is part of the definition
of the result: the accumulators of a class depend only on its key and on its number of samples. On CPUs they are bitwise
identical for any number of threads; an emulator that runs the device program on the host, with the data layout of the
device, reproduces them bitwise for any number of devices and blocks, any launch size and any order of execution. The
kernel body is written against an executor that either loops over the lanes (CPU) or maps them onto threads (GPU), in
phases that end with a barrier by construction; running the lanes in reverse order gives the same bits, which excludes a
missing barrier. A GPU can therefore differ from the CPU only through the device mathematical library and the contraction
of multiplications and additions into fused operations (the host code is compiled without contraction).

This is tested on every machine by golden runs: for 13 configurations (subtracted integrands of all three kinds,
and intermediate-scheme four-point integrands, at three to six loops), reference files computed on a CPU with fixed seeds
and 65\,536 samples per class must be reproduced on the GPUs with identical sample counts and per-class means within 100
units of the round-off of the per-sample terms and within $10^{-4}$ standard deviations, on one device, on each device
separately and with the work forcibly split among all of them~\check{N2}. On the CPU, perturbing every exponential and logarithm by up
to 2 units in the last place moves the per-class means by less than one such unit. Before production each GPU also passes
a self-test: device exponential and logarithm within 2 units in the last place of the host, exact square root and fused
multiply-add, the Philox known answers, a signature of all compile-time limits and data layouts, 4096 samples of each of up to
three classes compared one by one with the host (same Hepp sector, same weight to $10^{-11}$ of the sum of the absolute values of the
terms), and one chunk through the production kernel.

On a GPU the kernel runs as persistent blocks that take chunks from an atomic counter, one class per block at a time, so
that the threads of a warp follow the same integrand and chunks of classes with different costs balance automatically.
Every GPU is driven by its own host thread, which takes units of 65\,536 samples from a common queue in class order and
keeps the images of the classes it uses in device memory, with least-recently-used eviction. Launches hold about half a
second of work but at least two full waves of the grid; their measured durations are apportioned to the classes by their
cycle counts and give the costs $\tau_G$ of the time-based allocation. The sample count of every unit is verified, uploads
are read back, and every wait on a device is guarded by a watchdog; a device that fails is removed and its units are
requeued on the others. Since a unit's result depends only on its counters, the result does not depend on how the work is
split among the devices. The kernel's limits are compile-time constants, so the larger graphs need larger builds of the
same sources: the default one handles up to 16 lines and 768 atoms per class, enough through eight loops; a variant with
up to 18 lines and 8192 atoms covers nine loops (17 lines for $\Gamma^{(2)}$, 18 for $\Gamma^{(4)}$), and one with up to
20 lines, subgraphs of up to 11 loops and 16\,384 atoms covers ten. The sector table of a class has $2^\ell$ entries,
about a million at ten loops. Each build has its own signature, and the ledger accepts runs only from known
ones (Sec.~\ref{sec:methods-ledger}).

\subsection{Resummation}
\label{sec:methods-resummation}

The critical exponents follow from the RG functions at the non-trivial zero $\ubar^*$ of $\beta$: $\eta=\eta(\ubar^*)$, $\nu=\nu(\ubar^*)$ and $\omega=\beta'(\ubar^*)$. The series are asymptotic, with
coefficients that grow like $k!$ at large order $k$~\cite{Lipatov1977,BrezinLeGuillouZinnJustin1977}, and $\ubar^*$
is of order one, so they must be resummed. The eighth-order series of Ref.~\cite{Phi4tools2024} were not resummed
there (``We postpone such study to a future work''), and to our knowledge they have not been resummed since; the
standard fixed-dimension estimates rest on the shorter series analysed in
Refs.~\cite{LeGuillouZinnJustin1977,LeGuillouZinnJustin1980,GuidaZinnJustin1998}. We resum the eighth-order series,
from Ref.~\cite{Phi4tools2024} and from our own coefficients, and our ninth- and tenth-order series with the same
method and the same settings, so that the effect of the new terms, and that of the more precise eighth-order
coefficient of $\eta$, can be separated from that of the method.

\subsubsection{Borel--Leroy transform and conformal mapping}
\label{sec:resummation-method}

We follow Refs.~\cite{LeGuillouZinnJustin1977,LeGuillouZinnJustin1980,GuidaZinnJustin1998} (see also
Ref.~\cite{ZinnJustinBook}). For a series $F(\ubar)=\sum_k F_k\,\ubar^k$ the Borel--Leroy transform
\begin{gather}
F(\ubar)=\int_0^\infty dt\;t^{b}\,e^{-t}\,\mathcal B_F^{(b)}(\ubar t),\nonumber\\
\mathcal B_F^{(b)}(x)=\sum_k\frac{F_k}{\Gamma(k+b+1)}\,x^k,
\label{eq:borel}
\end{gather}
has a convergent expansion, since at large order~\cite{Lipatov1977,BrezinLeGuillouZinnJustin1977,BrezinParisi1978,ZinnJustinBook}
\begin{equation}
F_k\simeq c_F\,k!\,(-a)^k\,k^{b_F}\bigl[1+O(1/k)\bigr].
\label{eq:largeorder}
\end{equation}
The constant $a=48\pi/[(N+8)A]$, fixed by the action $A=\rsInstanton$ of the instanton
(Sec.~\ref{sec:methods-validation}), is the same for all RG functions ($a=\rsAone$ at $N=1$), while $b_F$ and $c_F$
depend on $F$: $b_F=3+N/2$ for $\beta$, for $\eta_{\phi^2}$ and hence for $\nu$ and $1/\nu$, and $b_F=2+N/2$ for
$\eta$~\cite{ZinnJustinBook,Ribeca2003thesis}. Assuming that $\mathcal B_F^{(b)}$ is analytic in the plane cut along
$(-\infty,-1/a]$, the mapping
\begin{equation}
w(x)=\frac{\sqrt{1+a x}-1}{\sqrt{1+a x}+1}
\label{eq:conformal}
\end{equation}
sends the cut plane onto the unit disk, and the series in $\ubar$ is replaced by
\begin{equation}
F(\ubar)\approx\sum_{k=0}^{K}d_k(\alpha,b)\int_0^\infty dt\;t^{b}\,e^{-t}\,
\frac{[w(\ubar t)]^k}{[1-w(\ubar t)]^{\alpha}},
\label{eq:resummed}
\end{equation}
where the $d_k(\alpha,b)$ are fixed by matching the expansion of the right-hand side to $F_0,\dots,F_K$. The factor
$[1-w]^{-\alpha}$ makes the approximant of the Borel transform grow like $x^{\alpha/2}$ at large $x$ (it is the
parameter $\lambda=\alpha/2$ of Ref.~\cite{KompanietsPanzer2017}). Dividing $F$ by $\ubar^s$ is equivalent to replacing
$(b,\alpha)$ with $(b-s,\alpha+2s)$, so the scan over $\alpha$ below contains the shifted series of
Ref.~\cite{GuidaZinnJustin1998}, which also used a homographic transformation of the coupling and a factor
$1+r\ubar$ that we do not use.

\subsubsection{Resummed functions, parameters and errors}
\label{sec:resummation-criteria}

The fixed point $\ubar^*$ is the zero of the resummed $\beta$, and $\omega$ is the slope of the resummed $\beta$ there;
$\eta$ is resummed and evaluated at $\ubar^*$. For $\nu$ we resum the series of $1/\nu=2-\eta+\eta_{\phi^2}$ as a
single series and take its inverse: the Callan--Symanzik relations~\eqref{eq:csrel} give $1/\nu$ directly, as a
product of series, whereas the series of $\nu$ follows from it by an inversion, which turns any zero of $1/\nu$ in the
complex $\ubar$ plane into a pole. As Ref.~\cite{GuidaZinnJustin1998} summed both $S$ and $1/S$ for each exponent,
we also resum the series of $\nu$ itself as a second determination; the two differ by at most $\rsNuDirMax$, and the
difference enters the error~\check{S2}.

We scan $b=0,\frac12,\dots,\rsBmax$ and $\alpha=\rsAlphaMin,\dots,\rsAlphaMax$ in steps of $\frac12$; the range of $b$
contains the large-order values $b_F$, and beyond $\alpha\approx\rsAlphaMax$ the successive truncations oscillate with
growing amplitude (those of $\eta$ from $\alpha\approx12$; see Sec.~\ref{sec:resummation-results}). For each
$(b,\alpha)$ we compute the results $F^{[K-2]}$, $F^{[K-1]}$ and $F^{[K]}$ of the last three truncations of
Eq.~\eqref{eq:resummed} and, over the neighbourhood $|b'-b|\le1$, $|\alpha'-\alpha|\le1$,
the largest change $\delta$ between successive truncations and half the range $\sigma$ of $F^{[K]}$. As in
Ref.~\cite{GuidaZinnJustin1998}, the central parameters $(b^*,\alpha^*)$ are those for which the result depends least
on the truncation order and on the parameters, here the minimum of $\delta+\sigma$; taking the largest change over a
neighbourhood rather than at the point itself discards the isolated points where successive truncations cross by
accident, which are frequent at large $\alpha$. The region of comparable stability is the set $R$ of grid points
where $\delta+\sigma$ is at most twice its minimum. The resummation error at fixed $\ubar^*$ is the largest of
$\delta$, $\sigma$ and $\max_R|F^{[K]}-F^{[K]}(b^*,\alpha^*)|$ (for $\nu$ also the difference from the series of $\nu$); as in
Ref.~\cite{GuidaZinnJustin1998}, the change of the result when $\ubar^*$ moves by its own error is added to it. Every
series is treated with the same grid and the same criteria. The truncation order is $K=8$ for the eighth-order series,
$K=9$ for the ninth-order ones and, at tenth order, $K=10$ for $\eta$ and $1/\nu$ and $K=11$ for $\beta$.

\subsubsection{Propagating the Monte Carlo uncertainties}
\label{sec:resummation-errors}

The coefficients carry statistical errors, which we propagate through the resummation and keep separate from its
systematic error. Every per-order sum is stored class by class (Appendix~\ref{app:data}), with independent random
numbers for every class and loop order (Sec.~\ref{sec:methods-rng}), and is linear in the class integrals: if these
are drawn from Gaussians with their quoted errors, each sum at fixed $N$ is a Gaussian variable with the propagated
error of Sec.~\ref{sec:results-new}, independent of the other sums. We therefore draw the sums at each $N$ (for the
orders taken from Ref.~\cite{Phi4tools2024}, and for its own coefficients in the eighth-order row of
Table~\ref{tab:exponents}, with their quoted errors), rebuild the coefficients through Eq.~\eqref{eq:csrel}, and resum
every draw with the central parameters; the standard deviation over \rsDraws\ draws, with a fixed seed, is the
statistical error. The sums at different $N$ are correlated through the shared class integrals, but no entry of
Table~\ref{tab:exponents} combines different $N$. Drawing the class integrals themselves, for the ledger inputs of the
eighth-order series at $N=1$, gives the same spreads within \rsClassDev, as expected from the sampling error of
\rsDraws\ draws, and a linearized propagation agrees with the draws within \rsLinDev~\check{S2}.

The statistical errors of our results are at most $\rsStatMaxUstar$ for $\ubar^*$, $\rsStatMaxNu$ for
$\nu$, $\rsStatMaxEta$ for $\eta$ and $\rsStatMaxOmega$ for $\omega$, at most $\rsStatRatio$ times the resummation
errors (the largest ratios, for $\ubar^*$ at $N=2$ and~3, come mostly from the coefficient of $\ubar^{11}$ in
$\beta$). With the quoted errors of the eighth-order coefficients of Ref.~\cite{Phi4tools2024}, larger than ours, the
ratio is at most $\rsStatRatioA$~\check{S2}. The resummed exponents are therefore limited by the resummation, not
by the precision of the coefficients: more computing time spent on the present orders would not change them. Further
progress needs more orders, which have so far reduced the errors only gradually (Table~\ref{tab:exponents}), and a
better control of the non-analyticity at $\ubar^*$ (Sec.~\ref{sec:resummation-results}).

\subsection{Validation strategy}
\label{sec:methods-validation}

Individual subtracted integrals have no external reference; only the order sums do. The checks are therefore organized as
follows; their results are given in Sec.~\ref{sec:results}.

\paragraph{Integrands.}
The expansion~\eqref{eq:integrand} agrees pointwise with an independent implementation in Python (to
$1.4\times10^{-14}$ of the sum of the absolute values of the terms, on 39 graphs at two to five loops) and with an
independent multiprecision implementation with its own forests, Kirchhoff polynomials from the matrix-tree theorem and no
pruning (all graphs at three to five loops and 20 graphs at six loops); pruned and unpruned expansions agree to the
round-off of extended precision~\check{I1}. The vanishing graphs, the twins
(Sec.~\ref{sec:methods-graphs})~\checks{I2,I3} and the shifted faces
(Sec.~\ref{sec:methods-tropical}) are checked as described above.

\paragraph{Exact values and known orders.}
The exact values of $B_1$, $B_2$, $H_2$, $C_1$ and $C_2$, and that of the subtracted integral of one three-loop graph (a
sunset inserted in a ring), are reproduced at high statistics. The coefficients known from
Ref.~\cite{Phi4tools2024}, $B_L$ to $L=7$ and $H_L$, $C_L$ to $L=8$, follow from its series through
Eqs.~\eqref{eq:csrel} and~\eqref{eq:series}, with errors propagated from those quoted there~\check{S1}; a run is compared with them
through the pull
\begin{equation}
z=\frac{X-X_{\mathrm{ref}}}{\sqrt{\delta X^2+\delta X_{\mathrm{ref}}^2}} ,
\label{eq:pull}
\end{equation}
where $X\pm\delta X$ is our value of an order sum and $X_{\mathrm{ref}}\pm\delta X_{\mathrm{ref}}$ the reference, at
$N=1$ and, by recombining the same integrals with their $O(N)$ polynomials, at $N=0$, 2 and 3. The unsubtracted intermediate-scheme integrals, computed by the
same code, are compared graph by graph with the values of Ref.~\cite{Phi4tools2024} and with their order sums.

\paragraph{Error bars.}
Replica ensembles with independent seeds test the error estimates: the mean and spread of their pulls, and the scatter of
every class among replicas against its quoted error ($\chi^2$ per degree of freedom). The same per-class $\chi^2$ over
repeated runs tests classes with large kurtosis or large maximal weights individually. Since every class's integral is
stored, two runs can be compared class by class, independently of any reference.

\paragraph{Hardware.}
Golden runs and device self-tests (Sec.~\ref{sec:methods-kernel}) establish that the GPUs compute the same numbers as the
CPU, up to the round-off of the device library, before any physics run is accepted~\check{N2}.

\paragraph{New orders.}
For $B_8$--$B_{10}$, $H_9$, $H_{10}$, $C_9$ and $C_{10}$, which have no reference, and for $H_8$ and $C_8$, whose
precision we improve, we rely on the following checks.
\begin{enumerate}
\item \emph{A second route} (route B). The coefficients are also obtained from the intermediate-scheme
integrals of $\Gamma^{(4)}(0)$, $\partial\Gamma^{(2)}/\partial p^2(0)$ and $\Gamma^{(2)}(0)$, in which only the sunset
subgraphs are subtracted, through the $Z$ factors and the reversion of the coupling series~\cite{Ribeca2003thesis}, as in
Ref.~\cite{Phi4tools2024}. In this scheme $\Gamma^{(2,1)}(0;0)$ follows from $\Gamma^{(2)}(0)$ through the exact
identity~\cite{Ribeca2003thesis}
\begin{equation}
\bigl(2-u_I\,\partial_{u_I}\bigr)\hat\Gamma^{(2)}_I(0)
=\Bigl(2+\tfrac{N+2}{3}\,u_I-\tfrac{2(N+2)}{9}\,u_I^2\Bigr)\Gamma^{(2,1)}_I(0;0),
\label{eq:intermediate}
\end{equation}
with $u_I=g_I/8\pi$, where $g_I=\lambda_0/m_I$ is the bare coupling in units of the mass $m_I$ of the intermediate
scheme, $\hat\Gamma^{(2)}_I=\Gamma^{(2)}_I/m_I^2$ and integrals in Nickel's normalization; the same identity
checks intermediate-scheme data order by order. Route B shares with route A only the graph lists, the weights and the sampler, and,
because of the cancellations of Sec.~\ref{sec:methods-cs}, it reaches a lower precision at equal cost.
The formulas are derived to all orders and reproduce every published coefficient of Ref.~\cite{Phi4tools2024}
through eighth order from its per-graph values, at $N=0$--$4$~\check{S4}. We evaluated the intermediate-scheme
integrals with eight to ten vertices on GPUs, in \routeBhours\ GPU hours, and took those with fewer vertices from
Ref.~\cite{Phi4tools2024}; the ninth- and tenth-order coefficients of the two routes agree within \routeBzmax\
standard deviations (Sec.~\ref{sec:results-new}, Table~\ref{tab:routes})~\check{S4}. Route B resolves them only to
\routeBrelmin--\routeBrelmax\ relative, and it has no check of the coefficient of $\ubar^{11}$ in $\beta$, which would
need the four-point graphs with eleven vertices.
\item \emph{Sum rules}: the eight- to ten-loop lists and weights pass the zero-dimensional sum rules
(Sec.~\ref{sec:methods-graphs})~\check{G2}.
\item \emph{Large $N$}: the leading power of $N$ of $B_L(N)$, $H_L(N)$ and $C_L(N)$ at each order is compared with the
$1/N$ expansion in the same scheme~\cite{ZinnJustinBook,Ribeca2003thesis}. At $N=\infty$ with $\ubar$ fixed only the
bubble chains survive, and $\beta=-\ubar+\ubar^2$, $\eta=0$, $\eta_{\phi^2}=-\ubar$, $\nu=1/(2-\ubar)$: the terms in
$N^L$ of $B_L$, $H_L$ and $C_L$ must vanish for $L\ge2$. No class of the ledger has an $O(N)$ weight of degree $L$, so
they vanish identically, and in the normalization of Eq.~\eqref{eq:series-form} (Sec.~\ref{sec:results-new})
$p_{\beta,k}$ has degree $k-2$, $p_{\eta,k}$ and
$p_{\eta_{\phi^2},k}$ degree $k-1$, and the coefficient of $N^k$ in $p_{\nu,k}$ is $2^{-k-1}$. The terms in $N^{L-1}$
follow from the $1/N$ corrections to $Z$, $\Gamma^{(4)}(0)$ and $\Gamma^{(2,1)}(0;0)$, one-dimensional integrals over
the momentum of the bubble chain, which reproduce the leading coefficients of the polynomials of
Ref.~\cite{Phi4tools2024} through eighth order to \largeNphi~\check{S5}.
\item \emph{Large order}: every coefficient is compared with its leading large-order estimate
$c\,(-a)^k\,\Gamma(k+b+1)$, equivalent at large $k$ to Eq.~\eqref{eq:largeorder}~\cite{Lipatov1977,BrezinLeGuillouZinnJustin1977,BrezinParisi1978,ZinnJustinBook},
with the instanton constants of $d=3$~\cite{BrezinLeGuillouZinnJustin1977,ZinnJustinBook}, $A=113.3835$ and the
amplitudes $c$, so that in Nickel's coupling $\ubar$ the constant is $a=48\pi/[(N+8)A]=0.147774232\times9/(N+8)$
[Ref.~\cite{ZinnJustinBook}, Eq.~(28.73)], and $b=3+N/2$ for $\beta$ and $\eta_{\phi^2}$, $2+N/2$ for $\eta$, as
implemented for the seven-loop computation~\cite{Ribeca2003thesis}.
The leading estimate alone catches an error of a few per cent in a new coefficient~\check{S3}.
\end{enumerate}

\section{Results}
\label{sec:results}

The numbers of Secs.~\ref{sec:results-validation} and~\ref{sec:results-performance} come from the runs recorded in
the ledger (Sec.~\ref{sec:methods-ledger}), on nodes with NVIDIA H200 GPUs: device self-tests, comparisons of the GPU
and CPU code paths on identical random numbers, validation runs at two to five loops, replica ensembles,
measurements of throughput, tuning and scaling, and complete orders at four to seven loops in production mode.
Section~\ref{sec:results-new} gives the new coefficients, from eight to ten loops.

Unless stated otherwise, values refer to $N=1$, and $B_L$, $H_L$, $C_L$ are the per-order sums of
Eqs.~\eqref{eq:series} and~\eqref{eq:BHC}. Errors are one standard deviation of the Monte Carlo estimate. At two loops
we compare with the exact values of Eq.~\eqref{eq:exactsums}, $1/2$, $1/81$ and $1/12$ at $N=1$. From three loops on,
we compare with reference values derived from the $O(N)$
series of $\beta$, $\eta$ and $\nu$ of Ref.~\cite{Phi4tools2024}: we invert the Callan--Symanzik
relations~\eqref{eq:csrel} order by order~\check{S1} and propagate the quoted uncertainties of the series coefficients to first
order, adding their effects in absolute value, which is conservative. Agreement is measured by the pull $z$ of
Eq.~\eqref{eq:pull}.

\subsection{Validation on known orders}
\label{sec:results-validation}

A computation of this kind can fail in three ways that a single run does not reveal: the integrands, or the code
that evaluates them, can be wrong; the error bars can be too small; and the round-off of the cancellations inside the
subtracted integrands can bias the means. Comparisons with known values test the first, but only as far as the error
bars are right, and round-off at the level that matters lies below the precision of any reference. We therefore test
the error bars and the round-off separately, without references, after establishing that runs on GPUs and on CPUs can
be pooled.

\paragraph{GPU against CPU.}
On identical random numbers the GPU and CPU code paths evaluate the same points and can differ only through the
device's mathematical library and the contraction of multiplications and additions (Sec.~\ref{sec:methods-kernel}).
Over the 13 configurations of the golden runs, 868 classes at three to six loops, the per-class means agree to within
0.59 units of the double-precision round-off scale of the per-sample terms, $\epsilon_{\rm mach}\sum|\text{terms}|$,
and the order sums to within $6\times10^{-10}$ standard deviations, on each device alone and with the work split over
all of them~\check{N2}; fused multiply-add contraction, which only the earliest runs used, stays within the same
bound. The hardware therefore adds nothing to the error budget, and runs are pooled whatever device produced them.

\paragraph{Known values.}
The two-loop sums reproduce their exact values, $B_2=1/2$, $H_2=1/81$ and $C_2=1/12$ at $N=1$, within their errors and
to relative uncertainties of a few $10^{-7}$ (Table~\ref{tab:validation})~\check{N3}.

\begin{table*}
\caption{\label{tab:validation}%
Per-order sums at $N=1$ compared with exact values ($L=2$) and with values derived from the
series of Ref.~\cite{Phi4tools2024} ($L\ge3$; uncertainties not shown are below $10^{-9}$ relative). The reference for
$C_7$ was derived in the same way. This work: every contribution of the ledger (validation runs with a
pilot of 65\,536 samples per class, production-mode runs of complete orders with a pilot of 16\,384 or 32\,768,
replicas; all with Neyman allocation by measured device time, within a fixed time box or up to a target uncertainty)
pooled class by class.
Classes: graphs with a non-vanishing subtracted integrand (twins were not merged). $z$: pull,
Eq.~(\ref{eq:pull}). Cost: the integrand evaluations of all its runs, in Tf (1~Tf $=10^{12}$ function evaluations), a
measure of the cost that does not depend on the hardware (Table~\ref{tab:cost}). Generated from the ledger~\checks{N3,N4,S2}.}
\begin{ruledtabular}
\begin{tabular}{lrlrlllr}
Sum & $L$ & Reference & Classes & This work & $\delta X/|X|$ & Cost [Tf] & $z$ \\
\hline
$B_L$ & 2 & $1/2$ & 1 & 0.50000006(33) & $6.7\times10^{-7}$ & 1.9 & $+0.18$ \\
 & 3 & 0.598603706 & 6 & 0.59860364(29) & $4.8\times10^{-7}$ & 18 & $-0.23$ \\
 & 4 & 0.967192687 & 19 & 0.96719250(71) & $7.3\times10^{-7}$ & 23 & $-0.26$ \\
 & 5 & 1.89652536(49) & 96 & 1.8965261(12) & $6.2\times10^{-7}$ & 77 & $+0.57$ \\
 & 6 & 4.3006836(49) & 502 & 4.3006914(59) & $1.4\times10^{-6}$ & 33 & $+1.02$ \\
 & 7 & 10.917594(45) & 3150 & 10.917628(31) & $2.8\times10^{-6}$ & 17 & $+0.63$ \\
\hline
$H_L$ & 2 & $1/81$ & 1 & 0.0123456808(24) & $2.0\times10^{-7}$ & 2.7 & $+0.75$ \\
 & 3 & $-0.001542750058$ & 1 & $-0.0015427487(12)$ & $8.0\times10^{-7}$ & 34 & $+1.09$ \\
 & 4 & 0.004699104918 & 4 & 0.0046991056(34) & $7.1\times10^{-7}$ & 15 & $+0.21$ \\
 & 5 & 0.002443913(10) & 11 & 0.002443919(10) & $4.2\times10^{-6}$ & 8 & $+0.37$ \\
 & 6 & 0.008020564(89) & 50 & 0.008020530(36) & $4.5\times10^{-6}$ & 4.3 & $-0.36$ \\
 & 7 & 0.01454955(80) & 209 & 0.014550034(71) & $4.9\times10^{-6}$ & 9.7 & $+0.60$ \\
\hline
$C_L$ & 2 & $1/12$ & 1 & 0.083333395(56) & $6.8\times10^{-7}$ & 1.5 & $+1.09$ \\
 & 3 & 0.07785997192 & 2 & 0.077860025(39) & $5.0\times10^{-7}$ & 9.7 & $+1.35$ \\
 & 4 & 0.1006770275 & 5 & 0.100677024(76) & $7.6\times10^{-7}$ & 24 & $-0.04$ \\
 & 5 & 0.170551919(82) & 16 & 0.17055175(21) & $1.2\times10^{-6}$ & 24 & $-0.77$ \\
 & 6 & 0.34591073(86) & 62 & 0.34591043(98) & $2.8\times10^{-6}$ & 9.8 & $-0.23$ \\
 & 7 & 0.801121(11) & 265 & 0.8011143(39) & $4.9\times10^{-6}$ & 6.5 & $-0.53$ \\
\end{tabular}
\end{ruledtabular}
\end{table*}

All \valncmp\ comparisons at three to seven loops in Table~\ref{tab:validation} lie within \valzmax\ standard
deviations, with $\chi^2=\valchisq$ for \valncmp\ values ($p=\valpval$)~\check{N4}. The reference errors, added
linearly over the coefficients of the $N$ polynomials of Ref.~\cite{Phi4tools2024}, are conservative, and the pulls are
correspondingly small. The complete seven-loop orders
$B_7$ ($\beta$ at $\ubar^{8}$, \nclBseven\ classes), $H_7$ ($\eta$ at $\ubar^{7}$, \nclHseven) and $C_7$ ($\nu$ at
$\ubar^{7}$, \nclCseven) reach relative uncertainties of \relBseven, \relHseven\ and \relCseven, below those of the
references by factors of about \precBseven, \precHseven\ and \precCseven; at six loops, with the computing time we
allocated, the precision of the references is matched (ratios of their uncertainty to ours of \precBsix,
\precHsix\ and \precCsix). The four-loop references are known to $2\times10^{-11}$ or better, which makes the
four-loop sums the most stringent test of a bias common to all classes: with relative uncertainties of at most
\fourrelmax, $B_4$, $H_4$ and $C_4$ deviate jointly from the references by \fourjointdev, so any common bias is below
\fourbiasbound\ at 95\% confidence, and that of $H_4$ alone below \hfourbiasbound.

Every class carries its $O(N)$ weight polynomial, so the same runs give the sums at any $N$, with errors that follow
from the class errors without further assumptions. Recombined at $N=0$, 2, 3 and 4, the same \valncmp\ comparisons
give $\chi^2$ between \valNchisqmin\ and \valNchisqmax\ and no pull larger than \valNzmax\ in
magnitude~\checks{N4,S2}; they come from the same samples as those at $N=1$ and are correlated with them.

\paragraph{Error bars.}
The value of a class is the mean of its weights, and its error the standard deviation of the weights divided by the
square root of their number (Sec.~\ref{sec:methods-estimators}). The mean is unbiased whatever the distribution of the
weights, but the error is only as reliable as the sample variance, and the subtractions make that the delicate part.
Near a face of the parameter space where the lines of a renormalization part, a two- or four-point subgraph, shrink to
zero, the subtracted integrand stays finite only through cancellations between its forest terms, and the tropical
density, which keeps the variance finite (Sec.~\ref{sec:methods-tropical}), follows it there only approximately. The
weights of classes with renormalization parts therefore have heavy tails: at seven loops the kurtosis reaches
\kurtmaxseven\ (a class of \kurtmaxsevenorder) and the largest weight is up to \wmaxmaxseven\ times the class mean (a
class of \wmaxmaxsevenorder). With such tails the sample variance converges slowly, and the rare large weights that
carry part of the variance may not yet have been drawn: in a short run the quoted error of such a class is more often
too small than too large. Comparisons with references cannot detect this at the precision we reach, so we test the
error bars directly, by repeating computations with independent seeds and comparing the scatter of the results with
their quoted errors.

We made this test where it is most likely to fail, in short runs: eight independent replicas of each of $B_6$, $H_6$,
$C_6$, $B_7$, $H_7$ and $C_7$, each sampling for four to eight minutes on one GPU, with medians of $10^{7}$ to $10^{9}$
samples per class. The replicas scatter as their quoted errors say: for the order sums $\chi^2=31.6$ for 42 degrees of
freedom ($p=0.88$), and class by class, over the 29\,666 degrees of freedom of the classes sampled in every replica,
$\chi^2/{\rm dof}=1.003$ ($p=0.36$); the classes that carry half of the variance of each order, the heavy-tailed ones
among them, give $\chi^2/{\rm dof}$ between 0.93 and 1.20~\check{N7}. A second test needs no reference either, and
grows with every run: each run merged into the ledger is compared, class by class, with the pool of the runs before
it (Sec.~\ref{sec:methods-ledger}). Over the \mergeruns\ runs so compared, in \mergeorders\ orders from \mergeLmin\
to \mergeLmax\ loops, the per-class $\chi^2$ is \mergechisq\ for \mergedof\ degrees of freedom
($\chi^2/{\rm dof}=\mergeratio$); the deviations of the runs' $\chi^2$ from their expectations, in units of their
standard deviations, have mean \mergezmean\ and spread \mergezsd, as they should for correct errors; and the order
sums on the common classes give $\chi^2=\mergesumchisq$ for \mergesumn\ runs
($p=\mergesump$)~\checks{N5,S2}.

Beyond seven loops the tails are heavier: the kurtosis reaches \kurtmaxhigh\ (a class of \kurtmaxhighorder) and the
largest weight \wmaxmaxhigh\ times the class mean (a class of \wmaxmaxhighorder). At eight loops the independent runs
of $B_8$ and $H_8$ scatter as their errors say (Table~\ref{tab:newsums}); the ten runs of $B_7$ scatter less than
theirs, which suggests conservative rather than optimistic errors, and we quote the pooled errors without rescaling.
For the single runs of $C_8$, $H_9$ and $C_9$, and for the slices of $B_9$, $B_{10}$, $H_{10}$ and $C_{10}$, no replica
test exists: their errors rest on the within-run statistics, and the runs driven to a target uncertainty stopped on
their own error estimates, a data-dependent rule whose effect we have not quantified.

\paragraph{Round-off in the subtractions.}
The forest formula, Eq.~\eqref{eq:integrand}, writes a subtracted integrand as a sum of terms, the bare integrand and
one subtraction for each forest of renormalization parts, that diverge separately on the faces where the lines of a
part shrink; only their sum is integrable. Close to such a face the terms are large and nearly cancel, and the
double-precision sum carries an absolute error of the order of $\epsilon_{\rm mach}\sum|\text{terms}|$
(Sec.~\ref{sec:methods-rng}). Unlike the sampling error, this error does not decrease with more samples and is not part
of the quoted errors; with a preferred sign it would bias the class means, below the precision of any reference. It
grows with the number of renormalization parts, hence with the order, and with the number of samples, which come ever
closer to the faces. Among the four classes of $H_4$, for instance, the only one with a two-point part, a sunset, has a
single-sample round-off scale of up to $6\times10^{-4}$ of its per-sample standard deviation, against about
$10^{-13}$ in the other three.

Every run records, for each class, the largest round-off scale of a single sample and its root mean square over the
samples, in units of the per-sample standard deviation. At seven loops the single-sample scale exceeds $10^{-3}$ in
\rmaxcountBseven\ of the \nclBseven\ classes of $B_7$ (largest \rmaxmaxBseven), \rmaxcountHseven\ of the \nclHseven\ of
$H_7$ and \rmaxcountCseven\ of the \nclCseven\ of $C_7$. The rms scale, which is what could shift a mean, stays far
smaller: through seven loops it is at most \rmsmaxlow\ of the per-sample spread, and at eight to ten loops it exceeds
$10^{-3}$ in \rmscounthigh\ classes, up to \rmsmaxhigh\ in a class of \rmsmaxhighorder. If every sample of every class
were shifted coherently by its rms scale, the order sums would move by at most \rcohmax\ of their statistical errors,
at any order from four to ten loops (the largest for \rcohmaxorder).

Since the scales are estimates, we also measured the round-off itself where it is largest: the same $2\times10^{5}$
sample points of each of 55 classes, the 43 most exposed of $B_8$, $H_8$, $C_8$, $H_9$, $C_9$ and $C_{10}$ and two
controls per order at six and seven loops, were evaluated in double and in extended (64-bit-mantissa) precision on the
CPU ($B_9$, $B_{10}$ and $H_{10}$, completed later, are covered by the bound above). The measured rms round-off is
typically 25 times below the scale (median), but in two classes of $C_{10}$ it reaches $0.1$ and $1.0$ of the
per-sample spread; there it adds to the sample variance, which the quoted errors include. No class shows a significant
mean difference between the two precisions (the largest is $1.9$ standard errors among the 55), and summed over the
classes of an order the possible bias is at most $4\times10^{-6}$ of the order's statistical error~\check{N6}. Since
the GPU and CPU code paths agree on identical samples to a fraction of the round-off scale~\check{N2}, the conclusion
carries over to the GPU runs. No run produced a non-finite weight.

\subsection{Performance}
\label{sec:results-performance}

\paragraph{Hardware and kernel.}
The bulk of the numerical evaluations of this work were run on computing nodes with NVIDIA H200 GPUs and Intel Xeon
Platinum 8568Y+ host CPUs. Each H200 GPU has compute capability~9.0, 132 streaming multiprocessors at 1.98~GHz, 140~GiB of memory with ECC enabled,
and a double-precision rate of half the single-precision rate; the host CPUs also build the classes. The code was compiled with CUDA~12.1 for sm\_90. The
integration kernel uses 110 registers per thread and 1104 bytes of stack, so four blocks of 128 threads fit on a
multiprocessor, and the grid has 528 blocks per device. Launches are sized to fill whole waves of the grid, with mean
fill factors of 0.99--1.00. In the production-mode runs the devices were busy
96--99\% of the wall time. The host builds the classes. For the \nclBseven\ classes of $B_7$ this takes
\costcpus{b7}~s summed over classes.

\begin{table}
\caption{\label{tab:throughput}%
Throughput with equal allocation over the classes of an order: samples per second summed over four NVIDIA H200
GPUs, and mean and maximum over the classes of the
device time per sample on one GPU. $b$, $h$, $c$: the subtracted integrands of $B_L$, $H_L$, $C_L$.
int4: the four-point integrands of the intermediate scheme of Ref.~\cite{Phi4tools2024}. bare4: unsubtracted
four-point integrands.}
\begin{ruledtabular}
\begin{tabular}{lrrll}
Kind & $L$ & Classes & Rate [$10^9$/s] & ns/sample \\
\hline
$b$    & 4 & 19  & 2.42  & 1.65 / 2.25 \\
       & 5 & 96  & 1.22  & 3.29 / 6.05 \\
       & 6 & 502 & 0.652 & 6.13 / 15.5 \\
$h$    & 4 & 4   & 2.00  & 2.00 / 2.51 \\
       & 5 & 11  & 0.889 & 4.50 / 6.45 \\
       & 6 & 50  & 0.431 & 9.25 / 17.0 \\
$c$    & 4 & 5   & 0.997 & 4.01 / 5.16 \\
       & 5 & 16  & 0.394 & 10.2 / 17.3 \\
       & 6 & 62  & 0.162 & 24.5 / 54.3 \\
int4   & 4 & 27  & 7.00 & 0.571 / 0.804 \\
       & 5 & 129 & 4.58 & 0.874 / 1.61 \\
       & 6 & 660 & 3.06 & 1.31 / 2.36 \\
bare4  & 4 & 23  & 7.38 & 0.542 / 0.574 \\
       & 5 & 105 & 4.92 & 0.812 / 0.844 \\
\end{tabular}
\end{ruledtabular}
\end{table}

\paragraph{Throughput.}
Table~\ref{tab:throughput} gives the sampling rates. For the subtracted integrands the cost per sample about doubles
with every loop: 1.65, 3.29, 6.13~ns per sample and GPU for $b$ at four, five and six loops, and
4.0, 10.2, 24.5~ns for $c$. At the same orders, a subtracted four-point integrand costs
2.9, 3.8 and 4.7 times as much per sample as the corresponding intermediate-scheme integrand. In the
seven-loop production-mode runs, with Neyman allocation, the four GPUs drew $7.3\times10^{8}$,
$3.1\times10^{8}$ and $1.1\times10^{8}$ samples per second for $B_7$, $H_7$ and $C_7$.

\paragraph{Tuning.}
The default kernel is limited by registers to a quarter of the maximum thread occupancy. Capping the registers so that
six or eight blocks fit on a multiprocessor raises the throughput of the five-loop $b$ and $c$ kernels by 20--22\% and
30--33\%. This holds even though the six- and eight-block builds spill
96 and 232~bytes per thread to local memory. The deviations from the CPU do not change. The production-mode runs
of this section used the default build, so their costs would drop by roughly a quarter if the five-loop gains carry
over. At five loops, disabling FMA contraction (the default from version 0.4.0) costs 1\% of throughput and brings the
GPU about five times closer to the CPU, which is compiled without contraction: the worst per-class deviation from the
CPU on identical random numbers moves from 0.072 to 0.015.

\paragraph{Scaling.}
Classes are independent: each GPU draws units of work from a shared queue, and the host only merges per-class
accumulators. On one, two and four GPUs the five-loop
$B_5$ run drew $3.03$, $6.01$ and $12.1\times10^{8}$ samples per second of wall time, a parallel
efficiency of 99\% at two and at four GPUs. Beyond one node no communication is needed: the larger orders were
split into slices of their lists, each run as an independent job on one or four GPUs and pooled in the ledger
(Sec.~\ref{sec:results-new}).

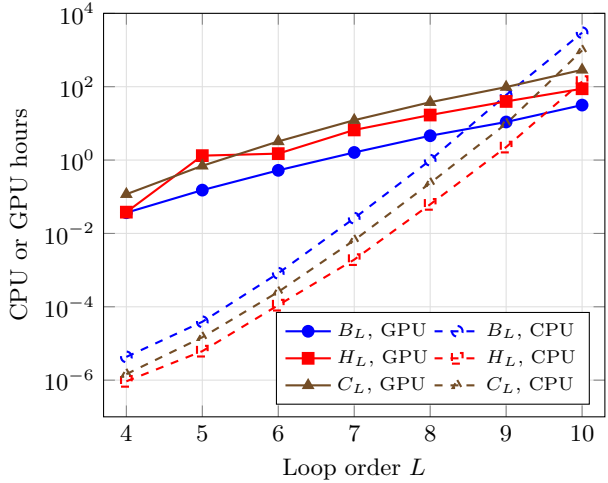
\begin{figure}
\centering
\ifdefined\pgfplotsversion
\begin{tikzpicture}
\begin{semilogyaxis}[
  width=0.95\columnwidth, height=0.8\columnwidth,
  xlabel={Loop order $L$}, ylabel={CPU or GPU hours},
  xtick={4,5,6,7,8,9,10}, xmin=3.7, xmax=10.3, ymin=1e-7, ymax=1e4,
  legend pos=south east, legend columns=2, legend cell align=left, legend style={font=\scriptsize},
  grid=major, grid style={gray!25}]
\costcurves
\end{semilogyaxis}
\end{tikzpicture}
\else
(This figure needs the \texttt{pgfplots} package.)
\fi
\caption{\label{fig:cost}%
The two costs of an order: the CPU time of the class build (dashed; the sum over the classes of the single-thread
build times recorded in the ledger, in hours of one core of an Intel Xeon Platinum 8568Y+) and the GPU time of the sampling to a relative uncertainty
of $10^{-5}$ (solid; the measured $T_{\rm dev}(10^{-5})$ of Table~\ref{tab:cost}, in hours of one NVIDIA H200 GPU,
which scales as the inverse square of the target uncertainty). From seven loops on the sampling grows by a factor
of \costgrowthrange\ per loop, the build by \cpugrowthrange, since the
sector table of a class spans the $2^\ell$ subsets of its lines and the minimum over its terms grows with them. For
$H_L$ and $C_L$ the sampling dominates through nine loops and the build at ten; for $B_L$, whose classes are many and
small, the build dominates from nine loops on. Generated from the ledger~\check{S2}.}
\end{figure}

\paragraph{Cost per decade of precision.}
For a Monte Carlo estimate the relative uncertainty after a device time $T_{\rm dev}$ (summed over GPUs) is
$\delta X/|X|=\rho_\tau\,T_{\rm dev}^{-1/2}$, with $\rho_\tau$ of Eq.~\eqref{eq:rho} under optimal allocation. Each
decade of precision therefore costs a factor of 100 in time. We take the realized value
$\rho_\tau=(\delta X/|X|)\sqrt{T_{\rm dev}}$ of each order, with $\delta X/|X|$ its uncertainty from all its runs pooled
and $T_{\rm dev}$ the device time of those runs; for a single production-mode run it is within about 1\% of the value
predicted by Eq.~\eqref{eq:rho}. Table~\ref{tab:cost} and Fig.~\ref{fig:cost} give the device time
$T_{\rm dev}(10^{-5})=(\rho_\tau/10^{-5})^2$ needed for a relative uncertainty of $10^{-5}$. From one order to the next,
$\rho_\tau^2$ grows by \costratios{b} for $B_L$ (from \costratiorange{b} loops) and by \costratios{c} for $C_L$ (from
\costratiorange{c}): the growth factors fall from four to six at low orders to about three from seven loops on. For
$H_L$ the sequence (\costratios{h}) is irregular at low orders, because $H_5$ has an unusually large per-sample
spread. At seven loops a relative uncertainty of $10^{-5}$ costs \costTm{b7}, \costTm{h7} and \costTm{c7} GPU hours
for $B_7$, $H_7$ and $C_7$. Matching the precision of the reference value of $B_6$ ($\costrefbsix\times10^{-6}$) would
take \costmatchbsix\ GPU hours, about \costmatchbsixfour~h on four GPUs.

\begin{table}
\caption{\label{tab:cost}%
Cost of the sums on NVIDIA H200 GPUs. $\delta X/|X|$: the relative uncertainty of the sum, all
its runs pooled. $\rho_\tau=(\delta X/|X|)\sqrt{T_{\rm dev}}$, with $T_{\rm dev}$ the device time of those runs summed
over their GPUs, does not depend on how long they ran; $T_{\rm dev}(10^{-5})=\rho_\tau^2/10^{-10}$ is the device time
for a relative uncertainty of $10^{-5}$, in GPU hours, the quantity to compare between orders (Fig.~\ref{fig:cost}).
$\rho_n=(\delta X/|X|)\sqrt{n}$ and $n(10^{-5})=\rho_n^2/10^{-10}$ are the same with $n$, the number of integrand
evaluations of those runs, in place of the device time, in Tf ($10^{12}$ evaluations): they do not depend on the hardware.
The numbers of classes are in Tables~\ref{tab:validation} and~\ref{tab:newsums}. Generated from the
ledger~\check{S2}.}
\footnotesize
\begin{ruledtabular}
\begin{tabular}{lrrrrrr}
Sum & $L$ & $\delta X/|X|$ & $\rho_\tau$ & $T_{\rm dev}(10^{-5})$ & $\rho_n$ & $n(10^{-5})$ \\
    &     & [$10^{-5}$] & [$10^{-4}\,$s$^{1/2}$] & [h] & & [Tf] \\
\hline
\costtablerows
\end{tabular}
\end{ruledtabular}
\end{table}

\begin{figure*}
\centering
\ifdefined\pgfplotsversion
\definecolor{colL6}{named}{blue}\definecolor{colL7}{named}{red}\colorlet{colL8}{green!55!black}
\definecolor{colL9}{named}{orange}\definecolor{colL10}{named}{violet}
\pgfplotsset{costhist/.style={scale only axis, width=0.285\textwidth, height=0.22\textwidth, xlabel={$\log_{10}\kappa_G$},
  xmin=-11.5, xmax=-3, ymin=0, ymax=36, ytick={0,10,20,30}, grid=major, grid style={gray!25},
  legend cell align=left,
  legend style={at={(0.02,1)}, anchor=north west, yshift=-0.4mm, font=\tiny, draw=none, fill=none, row sep=-1.2pt,
    inner sep=1pt, nodes={inner ysep=0.4pt}},
  legend image code/.code={\draw[#1, thick] (0cm,0cm) -- (0.3cm,0cm);},
  title style={font=\small, yshift=-1ex}}}
\begin{tikzpicture}
\begin{axis}[costhist, title={$B_L$}, ylabel={classes per bin [\%]}]
\costhistB
\end{axis}
\end{tikzpicture}\hfill
\begin{tikzpicture}
\begin{axis}[costhist, title={$H_L$}, yticklabels={}]
\costhistH
\end{axis}
\end{tikzpicture}\hfill
\begin{tikzpicture}
\begin{axis}[costhist, title={$C_L$}, yticklabels={}]
\costhistC
\end{axis}
\end{tikzpicture}
\else
(This figure needs the \texttt{pgfplots} package.)
\fi
\caption{\label{fig:costhist}%
Distribution over the classes of an order of the Neyman cost weight $\kappa_G = |a_G|\sigma_G\sqrt{\tau_G}/|X_L|$,
where $a_G\sigma_G$ is the standard deviation per sample of the weighted contribution of class $G$, $\tau_G$ its device
time per sample and $X_L$ the order sum: the device time that brings $X_L$ to the relative uncertainty $\epsilon$ is
$T(\epsilon) = (\sum_G\kappa_G)^2/\epsilon^2$, and class $G$ receives the fraction $\kappa_G/\sum_{G'}\kappa_{G'}$ of it
[Eq.~\eqref{eq:rho}]. Each histogram is normalized to the number of classes of its order (in the legends); the bins are a
quarter of a decade wide, and $\kappa_G$ is in $\sqrt{\rm s}$ of H200 time, from the pooled per-class statistics. From one order to the next the
distributions move to the left by about \kappashift\ decades while the number of classes grows by a factor of
\graphgrowthrange, so that $\sum_G\kappa_G$ grows only by a factor of \kappagrowthrange\ per loop. The resulting $T(10^{-5})$, $\costT{b8}$,
$\costT{h9}$ and $\costT{c9}$ GPU hours for $B_8$, $H_9$ and $C_9$, reproduces the measured
times of Table~\ref{tab:cost} and Fig.~\ref{fig:cost} to within \costkappadev\%. Generated from the ledger~\check{S2}.}
\end{figure*}
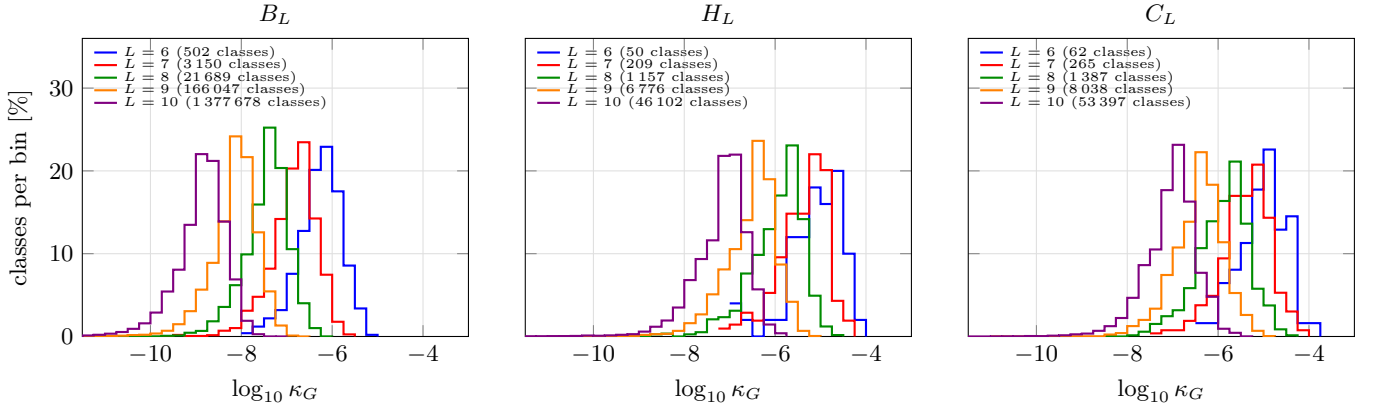

\paragraph{Scaling with the loop order.}
Figure~\ref{fig:cost} shows that the two costs of an order grow in different ways. The CPU time of the class build
curves upwards on the logarithmic scale: its growth per loop increases with the order, to a factor of \cpugrowthrange\
from seven loops on---the superexponential growth expected of a number of graphs that grows factorially times a sector
table that doubles with every line. The GPU time of the sampling curves the other way: its growth per loop decreases
with the order and settles, from seven loops on, at a factor of about \costgrowth{b} for $B_L$, \costgrowth{h} for
$H_L$ and \costgrowth{c} for $C_L$, so that the device time for a fixed relative uncertainty grows at most geometrically,
not factorially as the number of graphs does (from seven to ten loops the lists grow by factors of 6.7 to 8.1 per loop
for $\Gamma^{(4)}$ and 5.1 to 6.5 for $\Gamma^{(2)}$). Counted in
integrand evaluations, which do not depend on the hardware, the growth is slower still: the evaluations needed for a
relative uncertainty of $10^{-5}$ grow by a factor of \evalgrowthrange\ per loop (Table~\ref{tab:cost}), and the rest of
the growth of the device time is the growing cost of one evaluation. Two effects combine. The cost of a sample grows slowly,
because a class of $\ell$ lines costs a spanning-tree pass and an $L\times L$ factorization per atom and the number of
atoms per class stays small after the pruning, 19 on average at six loops and 26 at seven
(Sec.~\ref{sec:methods-construction}); and
the relative variance of an order grows slowly, because the sampling density follows the forest-subtracted
integrand sector by sector, so that the cancellations between forest terms happen inside each sample.
Figure~\ref{fig:costhist} shows the second effect graph by graph: under the Neyman allocation of Eq.~\eqref{eq:rho}
the device time of an order is $(\sum_G\kappa_G)^2/\epsilon^2$, with $\kappa_G$ as in Fig.~\ref{fig:costhist}, and
reconstructing it from the ledger reproduces the measured times to within \costkappadev\%. The distribution of
$\kappa_G$ is broad, about one decade wide, and the effort is not concentrated in a few classes: at every order
\costninetyrange\% of the classes are needed to account for 90\% of the cost. The median
$\kappa_G$ falls by about \kappashift\ decades per loop, because the typical graph shrinks faster than the
order sum grows, while the number of classes grows by a factor of \graphgrowthrange, so that $\sum_G\kappa_G$ grows by
\kappagrowthrange\ per loop and the cost by its square. At this rate
the sums of the tenth order cost up to a few hundred GPU hours at $10^{-5}$ (from $\costTm{b9}$ for $B_9$ to
$\costTm{c10}$ for $C_{10}$), and the limiting factor becomes the
generation and the symbolic preparation of the graphs rather than the sampling. Figure~\ref{fig:cost} shows the
crossing: the class build, which is dominated by the sector table over the $2^\ell$ subsets of lines, took
$\costcpu{c9}$ CPU hours for $C_9$ against $\costTm{c9}$ GPU hours of sampling, but grows by a
factor of \cpugrowthrange\ per loop and increasingly so with the order, so that at ten loops it exceeds the sampling
and is run
on CPU nodes, once per class, into the class cache of Sec.~\ref{sec:methods-construction}. For the three sums of the
tenth order, $B_9$, $H_{10}$ and $C_{10}$, the build took \costtenthcpu\ core-hours, against \costtenthgpu\ GPU hours
of sampling to the precisions of Table~\ref{tab:newsums}; in node-hours, with 32 cores per CPU node and the eight H200 GPUs of a GPU node, that is \costtenthnodecpu\ against \costtenthnodegpu.

\paragraph{Route A against route B.}
With the integrator of Sec.~\ref{sec:methods-numerics} the per-sample relative standard deviation of the order sum,
$\rho$ of Eq.~\eqref{eq:rho}, is 3.38 and 5.22 for $B_4$ and $B_5$, and 2.79 and 11.6 for $H_4$ and $H_5$, against 19.2,
40.5, 13.8 and 101 for the intermediate-scheme input of route B alone at the same order: the subtracted integrands
need 32, 60, 24 and 76 times fewer samples or, as each sample costs 2.4 to 3.9 times more on the CPU, 14, 22, 8 and 19
times less work at equal instruction count. These gains are lower bounds, since the route-B figures include only the
highest-order input, and they grow with the loop order. On the GPUs, where a subtracted four-point sample costs 2.9 and
3.8 times more at four and five loops (Table~\ref{tab:throughput}), the GPU time for $B_4$ and $B_5$ drops by factors of
about 11 and 16. At the ninth and tenth orders the route-B runs of Table~\ref{tab:routes}, \routeBhours\ GPU hours
in all, give coefficients whose uncertainties are \routeBratiomin\ to \routeBratiomax\ times those of route A.

\begin{table*}
\caption{\label{tab:effort}%
Computing effort of complete fixed-dimension computations. The entries are not like for like: they differ in the
quantities computed (individual graphs in Refs.~\cite{NMB1977,Phi4tools2024}, complete per-order sums here), in the
nature of the uncertainty (deterministic convergence estimates, VEGAS errors per graph, statistical errors per order),
and in the hardware. This work: per loop order, the sums computed and their classes, the device time for a relative
uncertainty of $10^{-5}$ of each sum (GPU h, Table~\ref{tab:cost}) and, from eight loops on, the CPU time of the class
construction (CPU h). The row of this work is generated from the ledger~\check{S2}.}
\begin{ruledtabular}
\raggedcolumns
\begin{tabular}{p{2.2cm}p{4.7cm}p{2.35cm}p{4.7cm}p{2.35cm}}
Computation & Quantities & Hardware & Time & Uncertainty \\
\hline
Nickel et al.~\cite{NMB1977,BNM1978} & $\Gamma^{(2)}$, $\Gamma^{(4)}$ to six loops, graph by graph & IBM 370/155, CDC 6600 & $\approx20$~h on the IBM and $\approx3$~h on the CDC, for the whole evaluation & $\sim10^{-8}$ absolute per integral \\
Phi4tools~\cite{Phi4tools2024} & $\Gamma^{(4)}$ to seven loops, $\Gamma^{(2)}$ to eight, graph by graph & two CPU clusters & 1--2 days for each integral of dimension $\ge6$ & a few $10^{-5}$ ($\Gamma^{(2)}$, $\Gamma^{(4)}$), a few $10^{-4}$ ($\partial_{p^2}\Gamma^{(2)}$) per graph \\
\effortrows
\end{tabular}
\end{ruledtabular}
\end{table*}

\begin{table*}
\caption{\label{tab:newsums}%
Per-order sums at $N=1$ from all runs pooled in the ledger. Enters: the function and the power of $\ubar$ whose
coefficient the sum gives. Graphs: the number of graphs of the list (without self-loops, one-vertex-reducible ones
included; Table~\ref{tab:graphcounts}). Classes: the number of non-vanishing integrands, every one of them sampled. $\delta X_{\rm s}/|X|$ and
$\chi^2/{\rm dof}$: where several seeds each cover the whole order, the $\chi^2$ of their order sums about the
inverse-variance mean, and the relative error of the pool rescaled by the scatter, $(\delta X/|X|)\sqrt{\chi^2/{\rm dof}}$;
the per-seed values are in the ledger. Previous: derived from the series of
Ref.~\cite{Phi4tools2024} as in Table~\ref{tab:validation}. Cost: the integrand evaluations of all its runs, in Tf, as in
Table~\ref{tab:validation}. The tenth- and eleventh-order sums were split
into disjoint slices of their lists, run separately. Generated from the ledger~\check{S2}.}
\begin{ruledtabular}
\begin{tabular}{llrrllllll}
Sum & Enters & Graphs & Classes & Previous & This work & $\delta X/|X|$ & Cost [Tf] & $\delta X_{\rm s}/|X|$ & $\chi^2/{\rm dof}$ \\
\hline
$B_{7}$ & $\beta$, $\ubar^{8}$ & 3986 & 3150 & 10.917594(45) & 10.917628(31) & $2.8\times10^{-6}$ & 17 & $1.4\times10^{-6}$ & 2.3/9 \\
$B_{8}$ & $\beta$, $\ubar^{9}$ & 26\,540 & 21\,689 & --- & 30.469859(76) & $2.5\times10^{-6}$ & 39 & $3.6\times10^{-6}$ & 4.3/2 \\
$B_{9}$ & $\beta$, $\ubar^{10}$ & 197\,022 & 166\,047 & --- & 92.27709(91) & $9.8\times10^{-6}$ & 4.5 & --- & --- \\
$B_{10}$ & $\beta$, $\ubar^{11}$ & 1\,593\,254 & 1\,377\,678 & --- & 300.5548(35) & $1.2\times10^{-5}$ & 6.2 & --- & --- \\
$H_{7}$ & $\eta$, $\ubar^{7}$ & 317 & 209 & 0.01454955(80) & 0.014550034(71) & $4.9\times10^{-6}$ & 9.7 & $4.6\times10^{-6}$ & 7.8/9 \\
$H_{8}$ & $\eta$, $\ubar^{8}$ & 1622 & 1157 & 0.037491(20) & 0.03750774(17) & $4.4\times10^{-6}$ & 20 & $4.5\times10^{-6}$ & 2.0/2 \\
$H_{9}$ & $\eta$, $\ubar^{9}$ & 9184 & 6776 & --- & 0.09845268(98) & $9.9\times10^{-6}$ & 7.7 & --- & --- \\
$H_{10}$ & $\eta$, $\ubar^{10}$ & 59\,614 & 46\,102 & --- & 0.2879452(41) & $1.4\times10^{-5}$ & 5.8 & --- & --- \\
$C_{7}$ & $\nu$, $\ubar^{7}$ & 317 & 265 & 0.801121(11) & 0.8011143(39) & $4.9\times10^{-6}$ & 6.5 & $6.6\times10^{-6}$ & 16.2/9 \\
$C_{8}$ & $\nu$, $\ubar^{8}$ & 1622 & 1387 & 2.07009(11) & 2.070192(21) & $9.9\times10^{-6}$ & 3.1 & --- & --- \\
$C_{9}$ & $\nu$, $\ubar^{9}$ & 9184 & 8038 & --- & 5.859068(58) & $1.0\times10^{-5}$ & 5.3 & --- & --- \\
$C_{10}$ & $\nu$, $\ubar^{10}$ & 59\,614 & 53\,397 & --- & 17.95856(52) & $2.9\times10^{-5}$ & 1.2 & --- & --- \\
\end{tabular}
\end{ruledtabular}
\end{table*}

\paragraph{Comparison with earlier computations.}
Table~\ref{tab:effort} sets these times against those of the two complete computations of the fixed-dimension
series~\cite{NMB1977,Phi4tools2024}. Three differences must be kept in mind. First, both computations produce individual graph values, from which the RG
functions follow after the cancellations of the $Z$-factor route (route B). In that route the four-point input at four to six loops
exceeds the net coefficient by factors of 29 to 74 (Sec.~\ref{sec:methods-cs}), so per-graph relative errors are not
relative errors of the RG coefficients. Our uncertainties are those of the per-order sums themselves, from which the
coefficients follow algebraically. Second, the errors are of different kinds: deterministic convergence estimates,
VEGAS errors per
graph, and statistical errors of an order sum. Third, the hardware spans five decades. What the comparison does show is
that a complete order, with all its graphs and subtractions and with an uncertainty of $10^{-5}$, now takes minutes
to a day on one node of four GPUs through nine loops, and up to a few hundred GPU hours, split into independent jobs,
at ten, without any graph-specific work. Whether our values are more precise is a
separate question: at six loops, with the computing time we allocated, ours are about as precise as the existing ones or
more, and at seven loops more precise, by factors of about \precBseven, \precHseven\ and \precCseven\ for $B_7$, $H_7$ and $C_7$ (Table~\ref{tab:validation}).

\subsection{New coefficients}
\label{sec:results-new}

Once the integration scheme had been validated on the known orders (Sec.~\ref{sec:results-validation}), several
production runs computed the four-point sums $B_8$--$B_{10}$ and the two-point sums $H_8$--$H_{10}$ and
$C_8$--$C_{10}$, and repeated $B_7$, $H_7$ and $C_7$ at higher precision (Table~\ref{tab:newsums}). $B_8$ and $H_8$ were
computed as independent replicas with fresh seeds, pooled in the ledger, with a pilot of $16\,384$ samples per class,
allocation by measured device time and density parameter $s=0.7$; $H_9$, $C_9$, $C_8$ and the repetitions of
$B_7$, $H_7$ and $C_7$ were run in one queue driven to target relative uncertainties, $10^{-5}$ for $H_9$, $C_9$ and
$C_8$, $3\times10^{-6}$ for $B_7$ and $5\times10^{-6}$ for $H_7$ and $C_7$, with checkpoints every 15 minutes from
which a run resumes across sessions. The graph lists (Appendix~\ref{app:graphs}) contain 26\,540 four-point graphs at
eight loops, and 1622 and 9184 two-point graphs at eight and nine loops. Without the one-vertex-reducible graphs and,
for $H_L$, the generalized tadpoles, 21\,689 graphs remain for $B_8$, 1157 and 1387 for $H_8$ and $C_8$, and 6776 and
8038 for $H_9$ and $C_9$ (Table~\ref{tab:graphcounts}); every one of them is a class, since twins were not merged.
Their builds take \costcpus{b8}, \costcpus{h8}, \costcpus{c8}, \costcpus{h9} and \costcpus{c9} core-seconds. $H_8$
reaches a relative uncertainty of \relHeight, so that the coefficient of $\ubar^8$ in $\eta$ is \etaEightGain\ times
more precise than the value derived from Ref.~\cite{Phi4tools2024}, whose relative uncertainty, $5.4\times10^{-4}$, is
the largest among the eighth-order coefficients (Table~\ref{tab:sota}). The improvement also matters for $\beta$:
through the term $2\eta u$ of the Callan--Symanzik relation, the coefficient of $\ubar^{9}$ in $\beta$ depends on
$H_8$ as well as on $B_8$. The sums of the tenth order, $B_9$, $H_{10}$ and $C_{10}$, with \nclBnine, \nclHten\ and
\nclCten\ classes, and that of the eleventh, $B_{10}$, with \nclBten, were too large for one node: each list was split into disjoint slices, each slice was built into a class
cache on CPU nodes and then sampled by its own GPU job, and the slices were pooled in the ledger like any other runs.
Since every class has its own random-number stream, the result does not depend on how the list was split or where
the slices ran.

\begin{table*}
\caption{\label{tab:newpoly}%
Coefficients $a_j$ of the polynomials $p_{X,k}(N)=\sum_j a_j N^j$ of Eq.~(\ref{eq:series-form}) for the new and
improved terms of $\beta$ and $\eta$. Powers beyond the degree of a polynomial are left empty. The coefficients are
correlated; their covariance is in the Supplemental Material (\texttt{series/generated/polynomials.json}). Generated
from the ledger~\check{S2}.}
\footnotesize
\newpolybetaeta
\end{table*}
\begin{table*}
\caption{\label{tab:newpolynu}%
As Table~\ref{tab:newpoly}, for $\eta_{\phi^2}$ and $\nu$. The coefficient of $N^k$ in $p_{\nu,k}$ is exactly
$2^{-k-1}$, and $p_{\eta_{\phi^2},k}$ has degree $k-1$, as the limit $N\to\infty$ requires~\check{S5}. Generated from
the ledger~\check{S2}.}
\footnotesize
\newpolyphinu
\end{table*}

\paragraph{Series coefficients.}
Below the orders we compute anew, each order sum entering the polynomials $p_{X,k}(N)$ of Eq.~\eqref{eq:series-form}
is the more precise of ours and the one derived from Ref.~\cite{Phi4tools2024} (at present \refinputs\ from
Ref.~\cite{Phi4tools2024}, the exact values at one and two loops, and all the others from the ledger); with our own
sums at every order the coefficients change by at most \ledgeronlyzmax\ of their standard deviations.
Tables~\ref{tab:newpoly} and~\ref{tab:newpolynu} give the new polynomials. Their coefficients are correlated, so we also give the coefficients evaluated at $N=0$--$4$
directly from the classes (Tables~\ref{tab:newcoeff}, \ref{tab:newcoeffeta} and~\ref{tab:newcoeffnu}). The full
covariance of the coefficients of all the polynomials, propagated linearly from the per-class integrals and from the
errors quoted in Ref.~\cite{Phi4tools2024}, is in the Supplemental Material, as the file
\texttt{series/generated/polynomials.json} and, one row per coefficient, \texttt{polynomials.tsv}
(Appendix~\ref{app:data}). The uncertainties of the
previous values in Tables~\ref{tab:newcoeffeta} and~\ref{tab:newcoeffnu} are the errors quoted in Ref.~\cite{Phi4tools2024} for the coefficients of the
polynomials, added linearly.

At $N=1$, with the quoted errors propagated through the Callan--Symanzik relations~\check{S2}, the coefficient of
$\ubar^{8}$ in $\eta$ is \etaEight, against $0.0029093(16)$ from Ref.~\cite{Phi4tools2024}, a difference of
\etaEightPull\ of its standard deviation (\routeBzphietaeight\ with the per-graph errors of Ref.~\cite{Phi4tools2024}
added in quadrature, see below), and that of $\ubar^{8}$ in $\nu$ is \nuEight, against $-0.0317852(17)$, a
difference of \nuEightPull; ours are \etaEightGain\ and \nuEightGain\ times more precise. The new coefficients are in
Tables~\ref{tab:newcoeff}--\ref{tab:newcoeffnu}: for instance, the coefficients of $\ubar^{9}$, $\ubar^{10}$ and
$\ubar^{11}$ in $\beta$ are \betaNine, \betaTen\ and \betaEleven, and those of $\ubar^{10}$ in $\eta$ and $\nu$ are
\etaTen\ and \nuTen. The ratios $|\beta_{k+1}/\beta_k|$ of successive coefficients of $\beta$ at $N=1$ are \betaratios\
for $k=\betaratiokmin$--$\betaratiokmax$. They grow as the large-order behaviour of Sec.~\ref{sec:methods-resummation}
requires: Eq.~\eqref{eq:largeorder} has them approach $a(k+b+1)$, which runs from 1.40 at $k=5$ to 2.14 at $k=10$
(Table~\ref{tab:largeorder} gives the quotients of the coefficients by their leading estimates).

\paragraph{Large-order and large-$N$ tests.}
Against the leading large-order estimate of Sec.~\ref{sec:methods-validation} (Table~\ref{tab:largeorder}, $N=1$) the signs agree from $k=4$ on; the ratio coefficient/estimate is
$1.025$, $1.022$, $1.030$, $1.043$ and $1.058$ for $\beta$ at $k=7$ to $11$, so the new orders lie within six per cent
of the leading asymptote, with a shallow minimum at $k=8$; for $\eta_{\phi^2}$ the ratio lies between $0.81$ and $0.86$ from
$k=6$ on, and for $\eta$, whose asymptotic regime sets in late, between $0.54$ and $0.65$~\check{S3}. The ratios are
not yet asymptotic, since all of them should tend to 1. The subleading corrections to Eq.~\eqref{eq:largeorder} are
powers of $1/k$, so we fitted the ratios of $\beta$ and $\eta_{\phi^2}$, at every $N$ from 0 to 4, with
$c_0(1+d_1/k+d_2/k^2)$: the form reproduces them from $k=\lofitkmin$ on, to $\lofitresbeta$ for $\beta$ and
$\lofitresphi$ for $\eta_{\phi^2}$, but with large coefficients, $d_1$ between $\lofitdonemin$ and $\lofitdonemax$
and $d_2$ between $\lofitdtwomin$ and $\lofitdtwomax$, nearly equal for $\beta$ and $\eta_{\phi^2}$ at each $N$, and
with a limit $c_0$ between $\lofitcomin$ and $\lofitcomax$ instead of 1: with corrections this large, orders this low
cannot fix the limit, and the fit describes the known terms rather than extrapolating them. As a test of the new
coefficients, though, the fit is much sharper than the leading estimate: predicted from the lower orders alone, the coefficients
of $\beta$ from $k=9$ on agree with the computed ones to $\lofitpredbeta$ at every $N$ ($\lofitpredbetaone$ at
$N=1$) and those of $\eta_{\phi^2}$ to $\lofitpredphi$, against Monte Carlo errors of about $10^{-5}$, so that an error
of several times $\lofitpredbeta$ in a new coefficient of $\beta$ would stand out~\checks{S2,S3}.
The leading coefficients of the \largeNorders\ order sums of the ledger,
$L=\largeNLmin$--$\largeNLmax$, agree with the $1/N$ expansion (Sec.~\ref{sec:methods-validation}) within their errors: $\chi^2=\largeNchisq$ for \largeNorders\ degrees
of freedom ($p=\largeNpvalue$), with no pull above \largeNzmax~\check{S5}. At the highest orders the leading
coefficients carry relative errors of tens of per cent (Tables~\ref{tab:newpoly} and~\ref{tab:newpolynu}), and the
test resolves them only at that level.

\begin{table}
\caption{\label{tab:largeorder}%
Large-order test at $N=1$: the coefficient of $\ubar^k$ in $\beta$, $\eta$ and $\eta_{\phi^2}$ divided by its leading
large-order estimate (Sec.~\ref{sec:methods-validation}). Generated from the ledger~\checks{S2,S3}.}
\begin{ruledtabular}
\begin{tabular}{lrrr}
$k$ & $\beta$ & $\eta$ & $\eta_{\phi^2}$ \\
\hline
3 & 1.751 & -0.548 & 1.165 \\
4 & 1.313 & 1.120 & 0.937 \\
5 & 1.121 & 0.368 & 0.838 \\
6 & 1.051 & 0.622 & 0.812 \\
7 & 1.025 & 0.542 & 0.808 \\
8 & 1.022 & 0.598 & 0.821 \\
9 & 1.030 & 0.617 & 0.839 \\
10 & 1.043 & 0.652 & 0.860 \\
11 & 1.058 & --- & --- \\
\end{tabular}
\end{ruledtabular}
\end{table}

\begin{table}
\caption{\label{tab:newcoeff}%
Coefficients $\beta_k$ of $\ubar^{k}$ in $\beta$ [Eq.~(\ref{eq:series-form})] for $N=0$--$4$; $\beta_{10}$ follows from $B_9$ and $H_9$, $\beta_{11}$ from $B_{10}$ and $H_{10}$. Generated from the
ledger~\check{S2}.}
\footnotesize
\begin{ruledtabular}
\begin{tabular}{rlll}
$N$ & $\beta_9$ & $\beta_{10}$ & $\beta_{11}$ \\
\hline
0 & \betaNineNzero & \betaTenNzero & \betaElevenNzero \\
1 & \betaNineNone & \betaTenNone & \betaElevenNone \\
2 & \betaNineNtwo & \betaTenNtwo & \betaElevenNtwo \\
3 & \betaNineNthree & \betaTenNthree & \betaElevenNthree \\
4 & \betaNineNfour & \betaTenNfour & \betaElevenNfour \\
\end{tabular}
\end{ruledtabular}
\end{table}

\begin{table}
\caption{\label{tab:newcoeffeta}%
Coefficients $\eta_k$ of $\ubar^{k}$ in $\eta$ for $N=0$--$4$, in units of $10^{-3}$: $\eta_8$ from the polynomials of
Ref.~\cite{Phi4tools2024} (their quoted errors added linearly) and from this work, and the new $\eta_9$ and $\eta_{10}$
(from $H_{10}$). Generated from the ledger~\check{S2}.}
\footnotesize
\begin{ruledtabular}
\begin{tabular}{rllll}
$N$ & $10^3\eta_8$~\cite{Phi4tools2024} & $10^3\eta_8$ & $10^3\eta_9$ & $10^3\eta_{10}$ \\
\hline
0 & 3.4748(18) & \etaEightNzeroK & \etaNineNzeroK & \etaTenNzeroK \\
1 & 2.9093(16) & \etaEightNoneK & \etaNineNoneK & \etaTenNoneK \\
2 & 2.2635(13) & \etaEightNtwoK & \etaNineNtwoK & \etaTenNtwoK \\
3 & 1.7200(11) & \etaEightNthreeK & \etaNineNthreeK & \etaTenNthreeK \\
4 & 1.3013(10) & \etaEightNfourK & \etaNineNfourK & \etaTenNfourK \\
\end{tabular}
\end{ruledtabular}
\end{table}

\begin{table}
\caption{\label{tab:newcoeffnu}%
Coefficients $\nu_k$ of $\ubar^{k}$ in $\nu$ for $N=0$--$4$, in units of $10^{-2}$: $\nu_8$ from the polynomials of
Ref.~\cite{Phi4tools2024} (their quoted errors added linearly) and from this work, and the new $\nu_9$ and $\nu_{10}$
(from $H_{10}$ and $C_{10}$). Generated from the ledger~\check{S2}.}
\footnotesize
\begin{ruledtabular}
\begin{tabular}{rllll}
$N$ & $10^2\nu_8$~\cite{Phi4tools2024} & $10^2\nu_8$ & $10^2\nu_9$ & $10^2\nu_{10}$ \\
\hline
0 & $-3.89135(17)$ & \nuEightNzeroC & \nuNineNzeroC & \nuTenNzeroC \\
1 & $-3.17852(17)$ & \nuEightNoneC & \nuNineNoneC & \nuTenNoneC \\
2 & $-2.41413(15)$ & \nuEightNtwoC & \nuNineNtwoC & \nuTenNtwoC \\
3 & $-1.79044(14)$ & \nuEightNthreeC & \nuNineNthreeC & \nuTenNthreeC \\
4 & $-1.32044(13)$ & \nuEightNfourC & \nuNineNfourC & \nuTenNfourC \\
\end{tabular}
\end{ruledtabular}
\end{table}

\begin{table}
\caption{\label{tab:routes}%
Route B at $N=1$. Top: the sums over the intermediate-scheme graphs at $L$ loops of $\Gamma^{(4)}(0)$, which
have $L+1$ vertices, and of $\Gamma^{(2)}(0)$ and $\partial_{p^2}\Gamma^{(2)}(0)$, which have $L$, in the normalization of
Ref.~\cite{Phi4tools2024}, from our runs and from the per-graph values of Ref.~\cite{Phi4tools2024} (their errors
added in quadrature). Bottom: the ninth- and tenth-order coefficients from route A and from route B, which takes the sums of
Ref.~\cite{Phi4tools2024} below eight vertices and ours from eight to ten. $z$: the difference in units of the
combined uncertainty. Generated from the stored runs and the ledger~\check{S4}.}
\footnotesize
\begin{ruledtabular}
\begin{tabular}{lrlll}
Sum & $L$ & This work & Ref.~\cite{Phi4tools2024} & $z$ \\
\hline
\routesumrows
\hline
Coefficient & & Route A & Route B & $z$ \\
\hline
\routecoefrows
\end{tabular}
\end{ruledtabular}
\end{table}

\paragraph{Two independent routes.}
As a check on the subtraction and on the Callan--Symanzik relations, we also obtained the ninth- and tenth-order
coefficients along route B (Sec.~\ref{sec:methods-validation}), from the intermediate-scheme vertex functions through
the $Z$ factors and the reversion of the coupling series, as in Refs.~\cite{NMB1977,Phi4tools2024}. The graph
lists, the weights and the sampler are shared, but the integrands are the intermediate-scheme ones, and the route
shares no subtraction with route A. We sampled the four-point graphs, the two-point graphs and their $p^2$
derivatives with eight to ten vertices, and took the sums with fewer vertices from the per-graph values of
Ref.~\cite{Phi4tools2024}. Table~\ref{tab:routes} compares the two
routes. Our eight-vertex sums agree with those of Ref.~\cite{Phi4tools2024}, and the ninth- and tenth-order coefficients
of the two routes agree within \routeBzmax\ standard deviations~\check{S4}. Because of the cancellations of route B
(Sec.~\ref{sec:methods-cs}), its uncertainties are \routeBratiomin\ to \routeBratiomax\ times larger, \routeBrelmin\
to \routeBrelmax\ relative, and the comparison is limited by them. With the eight-vertex sums of Ref.~\cite{Phi4tools2024} in place of ours, the
differences become \routeBzphibeta, \routeBzphieta, \routeBzphietaphi\ and \routeBzphinu\ standard deviations for
$\beta_9$, $\eta_9$, $\eta_{\phi^2,9}$ and $\nu_9$. Route B then carries the per-graph errors of
Ref.~\cite{Phi4tools2024} added in quadrature, which for $\eta_8$ give an uncertainty \routeBquadratio\ times smaller
than the one quoted there; measured with it, the $\eta_8$ of Ref.~\cite{Phi4tools2024} differs from ours by
\routeBzphietaeight\ standard deviations, and the difference carries over to $\eta_9$.

\subsection{Critical exponents}
\label{sec:resummation-results}

\begin{table*}
\caption{\label{tab:exponents}%
Fixed point and critical exponents from the resummed series (Sec.~\ref{sec:resummation-criteria}). For each $N$, in
the chronological order of the series: the fixed-dimension estimates of Ref.~\cite{GuidaZinnJustin1998} (Table~1,
$d=3$; series through $\ubar^7$; for $N=3$ the source prints $\omega=0.782\pm0.0013$, an uncertainty ten times
smaller than those of the other $N$, which we read as $0.782(13)$, as does Ref.~\cite{PelissettoVicari2002});
the eighth-order series ($\beta$, $\eta$ and $\nu$ through $\ubar^8$) with the coefficients of
Ref.~\cite{Phi4tools2024} and with ours, in one row where the two give the same digits
(Sec.~\ref{sec:resummation-results}); our ninth-order series (through $\ubar^9$) and our tenth-order series ($\eta$ and
$\nu$ through $\ubar^{10}$, $\beta$ through $\ubar^{11}$), all resummed with the same method; and non-perturbative
values: self-avoiding walks for $N=0$ ($\eta=2-\gamma/\nu$, $\omega=\Delta_1/\nu$), the conformal bootstrap for $N=1$
($\omega=\Delta_{\epsilon'}-3$), and Monte Carlo simulations of improved lattice models and the conformal bootstrap for $N=2$
and~3 (for the latter $\eta=2\Delta_\phi-1$, $\nu=1/(3-\Delta_s)$ and, for $N=2$, $\omega=\Delta_{s'}-3$). The errors
of our results are those of the resummation, including that of $\ubar^*$; the statistical errors are at most
$\rsStatRatio$ times as large (Sec.~\ref{sec:resummation-errors}). Generated from the ledger~\check{S2}.}
\begin{ruledtabular}
\begin{tabular}{rlllll}
$N$ & Series & $\ubar^*$ & $\nu$ & $\eta$ & $\omega$ \\
\hline
\exprows
\end{tabular}
\end{ruledtabular}
\end{table*}

Resummed as described in Sec.~\ref{sec:methods-resummation}, the series give the fixed point and the exponents of
Table~\ref{tab:exponents}. The eighth-order coefficients of Ref.~\cite{Phi4tools2024} and ours give
exponents that differ by at most $\rsABdiff$, $\rsABmax$ of the resummation errors, with the same central parameters:
the more precise eighth-order coefficient of $\eta$ changes nothing at this level, and the table gives the two in one
row where their digits coincide. From the eighth to the tenth order the resummation errors decrease, for $N=1$ from
$\rsUstarBNone$ to $\rsUstarDNone$ for $\ubar^*$, from $\rsNuBNone$ to $\rsNuDNone$ for $\nu$, from $\rsEtaBNone$ to
$\rsEtaDNone$ for $\eta$ and from $\rsOmegaBNone$ to $\rsOmegaDNone$ for $\omega$. Over $N=0$--3 the errors decrease by
factors of \rsGainNu\ for $\nu$ and \rsGainEta\ for $\eta$, and by factors that grow with $N$, \rsGainUstar\ for
$\ubar^*$ and \rsGainOmega\ for $\omega$ (the error of $\omega$ does not decrease for $N=0$). The coefficient of
$\ubar^{11}$ in $\beta$ moves the tenth-order $\ubar^*$ for $N=1$ from $\rsUstarDtenNone$ (with $\beta$ through
$\ubar^{10}$) to $\rsUstarDNone$. Our estimates agree with those of Ref.~\cite{GuidaZinnJustin1998} within the errors
of the latter for every entry, and they share their pattern of deviations from the non-perturbative values: the ninth
and tenth orders do not remove those deviations, but make them more visible.

In units of the combined standard deviation, the tenth-order $\nu$ differs from the non-perturbative values by
$\rsPullNuNzero$, $\rsPullNuNone$, $\rsPullNuNtwo$ and $\rsPullNuNthree$ for $N=0$--3: it agrees for $N=0$ and~1 (for
$N=1$, $\nu=\rsNuDNone$ against $\rsBenchCBNuNone$ from the conformal bootstrap~\cite{ChangEtAl2025}) and is
low for $N=2$ and~3. The tenth-order $\eta$ is low for every $N$ (differences $\rsPullEtaNzero$, $\rsPullEtaNone$,
$\rsPullEtaNtwo$ and $\rsPullEtaNthree$), and it increases with the order; a low $\eta$ is a known feature of the
fixed-dimension series~\cite{LeGuillouZinnJustin1980,GuidaZinnJustin1998}. For $\eta$ the optimum lies at the upper
end of the range of $\alpha$, but extending the range to $\alpha=12$ moves $\eta$ by at most $\rsEtaWindow$. For
$\omega$ the differences, $\rsPullOmegaNzero$, $\rsPullOmegaNone$, $\rsPullOmegaNtwo$ and $\rsPullOmegaNthree$, change
sign with $N$: the resummed $\omega$ depends on $N$ less than the non-perturbative one. For $N=3$ the successive
truncations agree closely ($\omega=\rsOmegaDNthree$) on a value well above the Monte Carlo result
$\rsBenchMCOmegaNthree$~\cite{Hasenbusch2020}, where the seven-loop resummation had already
settled~\cite{GuidaZinnJustin1998}, and the same holds, to a lesser degree, for $\nu$. The Monte Carlo values are not
subject to the asymptotic nature of the series, and the conformal bootstrap, where it determines the same exponents,
agrees with them to their quoted errors (Table~\ref{tab:exponents}). In these cases, then, our error estimates, which
measure only the apparent convergence of the resummation, do not cover its systematic error.

\paragraph{Limitations and cross-checks.}
The deviations of Table~\ref{tab:exponents} point to a limitation specific to fixed dimension: the RG functions are
expected to be non-analytic at $\ubar^*$, because of confluent corrections to
scaling~\cite{Nickel1982,Nickel1991,PelissettoVicari1998,CalabreseEtAl2000,PelissettoVicari2002}. A non-analytic term of the form $(\ubar^*-\ubar)^{\Delta}$ with
non-integer $\Delta$ slows the convergence of any resummation that assumes analyticity there, and the successive
truncations can agree with each other well before they agree with the exact value: apparent convergence, which is
what the criteria of Sec.~\ref{sec:resummation-criteria} measure, does not guarantee accuracy, and the derivative
that gives $\omega$ is the most exposed. A longer series helps to quantify this systematic error
but cannot remove it: from the eighth to the tenth order the uncertainties of $\nu$, $\eta$ and $\omega$ at $N=1$
shrink only slowly (Table~\ref{tab:exponents}), and that of $\eta$, the exponent farthest from its non-perturbative
value, no faster than that of $\nu$, unlike in the $\varepsilon$ expansion, where the uncertainty of $\eta$ was found
to fall rapidly with the loop order while those of $\nu$ and $\omega$ did not~\cite{Shapoval2026}. Other
resummation methods can serve as cross-checks: Pad\'e approximants of the Borel transform,
as in the original analyses of the fixed-dimension series~\cite{BNGM1976,BNM1978,ZinnJustinBook}; conformal mapping
combined with a homographic transformation of the expansion parameter, with the parameters and the errors chosen by the
criteria of recent resummations of the $\varepsilon$ expansion~\cite{KompanietsPanzer2017,Shapoval2026}; and variational
perturbation theory~\cite{KleinertBook}, which in its strong-coupling form works with the bare coupling and determines the
approach exponent $\omega$ self-consistently, and has been applied to the seven-loop series~\cite{Kleinert1999}. We have
not applied them here: the first two share the assumption of analyticity at $\ubar^*$, and the comparison with the
non-perturbative values already shows the size of the systematic error.

\section{Summary and outlook}
\label{sec:summary}

We have extended the fixed-dimension RG functions $\beta$, $\eta$ and $\nu$ of the three-dimensional $O(N)$ model to
tenth order, and $\beta$ to eleventh, for general $N$, and re-evaluated all lower orders, by sampling on GPUs the
$\Rbar$-subtracted parametric integral of every graph. At $N=1$ the new coefficients have relative uncertainties between
\newrelmin\ and \newrelmax; they agree with the large-$N$ limit and with the large-order behaviour, and at the ninth and
tenth orders with the independent route B, within its coarser precision. Through seven loops the per-order sums
reproduce the published ones, and the eighth-order coefficients of $\eta$ and $\nu$ are now \etaEightGain\ and
\nuEightGain\ times more precise than before.

The design goal, a cost that grows slowly with the order, is met: from seven loops on, the time needed to integrate an
order to a fixed relative precision grows by a factor of \costgrowthrange\ per loop on average, whereas the number of graphs grows
factorially. The bottleneck is now the construction of the integrands, which grows faster than the number of graphs and at the
tenth order already costs more than the sampling. A sampler that draws graphs and points together, without enumerating
the graphs, as the global tropical sampler of Ref.~\cite{Borinsky2025global} does for unsubtracted integrands, would
remove it once extended to the forest-subtracted integrands used here.

The resummed exponents are limited by the resummation, not by the coefficients: the statistical errors are at most
\rsStatRatio\ times the resummation errors, and the remaining deviations from the non-perturbative values, largest
for $\omega$ at $N=3$, point to the non-analyticity of the RG functions at the fixed point. More orders, or resummations
that account for this non-analyticity, are needed to go further.

Every per-class result is pooled in a public ledger, from which the generated numbers of this paper are rebuilt and
checked (Appendix~\ref{app:reproducibility}). Further runs can be contributed and folded into it exactly, and a new
order needs only its graph list and computing time.

\begin{acknowledgments}
I wish to thank Jean Zinn-Justin and Riccardo Guida for proposing to me, in 2000, the problem of computing critical
exponents in scalar quantum field theories, and for their patience since. I am greatly indebted to Riccardo Guida for
several years of mentoring and discussions on this topic, from 2000 to 2005, during which many of the materials and
methods used here to generate and check results were devised; it is much to my regret that he declined to be an author
of this paper. I sincerely thank my institutions for the generous permission to use their computational infrastructure
for this project. The work described here was carried out with the invaluable assistance of Claude (Anthropic).
\end{acknowledgments}

\appendix

\section{Conventions and normalizations}
\label{app:conventions}

These conventions are also summarized, with the names used by the code, in the file \texttt{doc/CONVENTIONS.md} of
the code repository (Appendix~\ref{app:data}).

\paragraph{Couplings.}
The renormalized coupling $g$ is defined by $\Gamma^{(4)}_R(0)=m\,g$ [Eq.~\eqref{eq:normcond}], with the $O(N)$ tensor
$(\delta_{ij}\delta_{kl}+\delta_{ik}\delta_{jl}+\delta_{il}\delta_{jk})/3$ factored out. The integrals are expanded in
$u=g/8\pi$ and the RG functions are quoted in
\begin{gather}
\ubar=\frac{N+8}{6}\,u=\frac{N+8}{48\pi}\,g ,\nonumber\\
\beta(\ubar)\equiv m\frac{\partial\ubar}{\partial m}=\frac{N+8}{6}\,\beta_u ,
\label{eq:ubar}
\end{gather}
the normalized coupling of Refs.~\cite{BNGM1976,BNM1978}, which is the $\tilde u$ of Ref.~\cite{Phi4tools2024}. With
it $\beta(\ubar)=-\ubar+\ubar^2+O(\ubar^3)$ for every $N$. We write $\eta_{\phi^2}$ for the anomalous dimension of $\phi^2$ ($\eta_2$ in
Ref.~\cite{Ribeca2003thesis}), so that $1/\nu=2-\eta+\eta_{\phi^2}$, as in Ref.~\cite{Phi4tools2024}.

\paragraph{Integrals.}
Our integrals have one $d^3k/\pi^2$ per loop~\cite{NMB1977}, that is $8\pi$ times the measure $d^3k/(2\pi)^3$; with
this choice each loop comes with one power of $u$, the one-loop bubble at zero momentum equals one, and the low-order
coefficients are rational. (In Ref.~\cite{Ribeca2003thesis} ``Nickel's normalization'' refers to the coupling
$\ubar$ rather than to the measure; integrals normalized with $d^3k/(2\pi)^3$ per loop are $(8\pi)^{-L}$ times ours.)
The propagators of $B_L$, $H_L$ and $C_L$ have the renormalized mass, $m=1$, whereas those of
Refs.~\cite{NMB1977,Ribeca2003thesis,Phi4tools2024}, and of the intermediate-scheme integrals we use for comparison,
have the mass of the intermediate scheme, $m_I=1$. Individual integrals in the two conventions differ by powers of
$m^2/m_I^2$; the RG functions do not. Reference~\cite{Phi4tools2024} quotes graph values $V_G$ that include its weight
factor and a factor $(16\pi)^{L}$ per graph with $L$ loops, with the measure $d^3k/(2\pi)^3$. At $N=1$ the sum of
its intermediate-scheme four-point values at $L$ loops equals $(2^{L}/3)\sum_G S_G I_G$, where $I_G$ is the integral of
$G$ in our normalization, in the same scheme and with $m_I=1$; this is the relation behind the comparisons of
intermediate-scheme order sums in Sec.~\ref{sec:results-validation}.

\paragraph{Weights.}
The weight $S_G=T_G(N)\,E!/|\mathrm{Aut}\,G|$ of Sec.~\ref{sec:methods-graphs} includes the $O(N)$ factor $T_G(N)$,
normalized to $T_G(1)=1$ as are the symmetry factors of Ref.~\cite{Phi4tools2024}. The insertion of $\tfrac12\phi^2$ is
normalized by $\Gamma^{(2,1)}_R(0;0)=1$, and $\partial\Gamma^{(a,b)}/\partial m_0^2=\Gamma^{(a,b+1)}$ at zero insertion
momentum, so that, e.g., $C_1>0$.

\paragraph{From the per-order sums to the series.}
Writing $[\ubar^k]F$ for the coefficient of $\ubar^k$ in $F$, Eqs.~\eqref{eq:csrel}, \eqref{eq:series}
and~\eqref{eq:ubar} give, at each order,
\begin{align}
[\ubar^{L+1}]\beta&=2(-1)^{L+1}\Bigl(\frac{6}{N+8}\Bigr)^{L}\bigl(B_L-2H_L\bigr)+\dots,\nonumber\\
[\ubar^{L}]\eta&=2(-1)^{L}\Bigl(\frac{6}{N+8}\Bigr)^{L}H_L+\dots,\nonumber\\
[\ubar^{L}]\eta_{\phi^2}&=2(-1)^{L}\Bigl(\frac{6}{N+8}\Bigr)^{L}C_L+\dots,\label{eq:linear}\\
[\ubar^{L}]\nu&=\frac{(-1)^{L+1}}{2}\Bigl(\frac{6}{N+8}\Bigr)^{L}\bigl(C_L-H_L\bigr)+\dots,\nonumber
\end{align}
where the dots stand for products of sums of lower order ($H_1=0$). The coefficient of $\ubar^{L+1}$ in $\beta$ thus
depends on $B_L$ and $H_L$, those of $\ubar^{L}$ in $\eta$ on $H_L$, and in $\nu$ on $C_L$ and $H_L$; the errors of the
new sums enter linearly, with the prefactors of Eq.~\eqref{eq:linear}. The reference values of $B_L$, $H_L$, $C_L$ used
in Sec.~\ref{sec:results} are obtained by solving these relations order by order for the sums, starting from the
series of Ref.~\cite{Phi4tools2024}~\check{S1}.

At the lowest orders the sums are known exactly,
\begin{gather}
B_1=\frac{N+8}{12},\qquad B_2=\frac{5N+22}{54},\qquad H_2=\frac{N+2}{243},\nonumber\\
C_1=\frac{N+2}{12},\qquad C_2=\frac{N+2}{36},
\label{eq:exactsums}
\end{gather}
that is $3/4$, $1/2$, $1/81$, $1/4$ and $1/12$ at $N=1$, and give
\begin{align}
\beta(\ubar)&=-\ubar+\ubar^2-\frac{4(41N+190)}{27(N+8)^2}\,\ubar^3+O(\ubar^4),\nonumber\\
\eta(\ubar)&=\frac{8(N+2)}{27(N+8)^2}\,\ubar^2+O(\ubar^3),\nonumber\\
\eta_{\phi^2}(\ubar)&=-\frac{N+2}{N+8}\,\ubar+\frac{2(N+2)}{(N+8)^2}\,\ubar^2+O(\ubar^3),\label{eq:lowseries}\\
\nu(\ubar)&=\frac12+\frac{N+2}{4(N+8)}\,\ubar\nonumber\\
&\quad+\frac{(N+2)(27N-38)}{216(N+8)^2}\,\ubar^2+O(\ubar^3),\nonumber
\end{align}
in agreement with Ref.~\cite{Phi4tools2024}. At $N=1$ the coefficients are $-308/729$ ($\ubar^3$ in $\beta$),
$8/729$ ($\ubar^2$ in $\eta$), $-1/3$ and $2/27$ ($\eta_{\phi^2}$), and $1/12$ and $-11/5832$ ($\nu$). In the notation of
Eq.~\eqref{eq:series-form}, $p_{\beta,3}=-4(41N+190)/27$, $p_{\eta,2}=8(N+2)/27$, $p_{\nu,1}=(N+2)/4$ and
$p_{\nu,2}=(N+2)(27N-38)/216$.

\section{Graphs}
\label{app:graphs}

\paragraph{Counts.}
Table~\ref{tab:graphcounts} gives, for every loop order, the numbers of 1PI graphs and of the integrands of
Eq.~\eqref{eq:BHC} that do not vanish identically. One-vertex-reducible (1VR)
graphs vanish for all three kinds, and so do, for $H_L$, the graphs with both external legs on one vertex
(Sec.~\ref{sec:methods-rbar})~\check{I2}; through seven loops no other graph vanishes, so that the numbers of integrands are the
numbers of the remaining graphs; at eight and nine loops the expansion confirmed that every remaining graph has a
non-vanishing integrand (no class was dropped at the build). At ten loops the table gives the numbers of remaining
graphs~\check{G5}. In the runs reported in Sec.~\ref{sec:results} each class is a single graph.

\begin{table*}
\caption{\label{tab:graphcounts}%
Numbers of 1PI graphs per loop order $L$ and of non-vanishing integrands. $|\mathcal G_4(L)|$, $|\mathcal G_2(L)|$:
graphs without self-loops, one-vertex-reducible ones included; 1VR: one-vertex-reducible graphs among them; one vertex:
two-point graphs with both legs on one vertex (generalized tadpoles), 1VR ones included; $B$, $H$, $C$: integrands of
Eq.~\eqref{eq:BHC} that survive the algebraic pruning, i.e.\ graphs that are not 1VR and, for $H$, do not have both
legs on one vertex; with self-loops: complete two-point lists including graphs with self-loops, used only for
checks. \check{G1}}
\begin{ruledtabular}
\begin{tabular}{rrrrrrrrrr}
$L$ & $|\mathcal G_4(L)|$ & 1VR & $B$ & $|\mathcal G_2(L)|$ & 1VR & one vertex & $H$ & $C$ & with self-loops \\
\hline
1 & 1        & 0    & 1                      & 0    & --   & --   & --                    & 1\footnotemark[1]     & 1    \\
2 & 2        & 1    & 1                      & 1    & 0    & 0    & 1                     & 1                     & 2    \\
3 & 8        & 2    & 6                      & 2    & 0    & 1    & 1                     & 2                     & 5    \\
4 & 27       & 8    & 19                     & 6    & 1    & 2    & 4                     & 5                     & 17   \\
5 & 129      & 33   & 96                     & 19   & 3    & 7    & 11                    & 16                    & 68   \\
6 & 660      & 158  & 502                    & 75   & 13   & 21   & 50                    & 62                    & 313  \\
7 & 3986     & 836  & 3150                   & 317  & 52   & 85   & 209                   & 265                   & 1632 \\
8 & 26\,540  & 4851 & 21\,689 & 1622 & 235  & 356  & 1157  & 1387  & 9685 \\
9 & 197\,022 & 30\,975 & 166\,047\footnotemark[2] & 9184 & 1146 & 1804 & 6776  & 8038  & 63\,774 \\
10 & 1\,593\,254 & 215\,576 & 1\,377\,678\footnotemark[2] & 59\,614 & 6217 & 10\,064 & 46\,102\footnotemark[2] & 53\,397\footnotemark[2] & 463\,933 \\
\end{tabular}
\end{ruledtabular}
\footnotetext[1]{The one-loop tadpole with two insertions.}
\footnotetext[2]{Graphs that are not 1VR (and, for $H$, do not have both legs on one vertex); all their integrands
survive the pruning, and every one of them was sampled~\check{G5}.}
\end{table*}

\paragraph{Generator.}
The generation is described in Sec.~\ref{sec:methods-generation}. At eight loops the
26\,540 four-point graphs have their legs distributed as $(2,2)$ (2254 graphs), $(2,1,1)$ (14\,614) and $(1,1,1,1)$
(9672); at nine loops the 9184 two-point graphs as $(2)$ (1804) and $(1,1)$ (7380)~\check{G1}.

\paragraph{Twins.}
Besides the seven-loop numbers of Sec.~\ref{sec:methods-construction}, the four-point lists at five and six loops contain
104 and 518 distinct unsubtracted integrals for 129 and 660 graphs, and the eight-loop two-point list 1512 for 1622.
Merging twins, after a pointwise check of their subtracted integrands, would save the corresponding samples.

\paragraph{Notation.}
A Nickel index~\cite{NMB1977} lists, for each vertex in turn, an \texttt{e} for every external leg and the labels of the
vertices with equal or higher label to which it is joined, once per line, followed by a bar; the canonical index is the
lexicographic minimum over the labellings of the vertices, with $|<\texttt{e}<0<1<\dots$. For example, the one-loop
four-point bubble is \texttt{ee11|ee|} and the two-loop sunset \texttt{e111|e|}. Our generator, like the one of
Ref.~\cite{Ribeca2003thesis} and the programs built on it~\cite{GuidaRibeca2006}, writes a graph instead as its number
of vertices followed by the run-length-coded upper triangle of its symmetric matrix of line multiplicities, with
optional sections for vertex colours (external legs) and self-loops. Those programs canonicalized it as a
lexicographically minimal adjacency matrix, and their labels such as D40L5N78 (a four-point graph with five loops,
line 78 of their list) are list positions, not graph invariants. For instance, the string \texttt{6(.2 1\^{}2.2 1\^{}6.1 1\^{}2)} of D40L5N78
is the graph \texttt{ee12|345|345|e4|5|e|}. A table that maps every historical label to its Nickel index accompanies
the graph lists (Appendix~\ref{app:data}).

\section{Zero-dimensional sum rules}
\label{app:sumrules}

In zero dimensions every Feynman integral with unit propagators equals one. The sum $\sum_G S_G$ of the weights over a
complete list is therefore the corresponding coefficient of the vertex functions of the zero-dimensional $O(N)$
model~\cite{CvitanovicLautrupPearson1978,Ribeca2003thesis,Borinsky2017}, with partition function
\begin{equation}
Z=\int d^N\phi\;\exp\Bigl[-\frac{m_0^2}{2}\,\phi^2-\frac{g}{24}\,(\phi^2)^2\Bigr].
\label{eq:zerod}
\end{equation}
With $Q=\langle\phi^2\rangle/N$ and $A=\langle(\phi^2)^2\rangle/[N(N+2)]$ one has $\Gamma^{(2)}_0=1/Q$ and
$\Gamma^{(4)}_0=-3(A-Q^2)/Q^4$, and, as formal power series in $g$,
\begin{equation}
\langle(\phi^2)^k\rangle=\frac{I_k}{I_0},\quad
I_k=\sum_{n\ge0}\frac{(-g/24)^n}{n!}\,\frac{P_{k+2n}(N)}{m_0^{2(k+2n)}},
\label{eq:zerodmoments}
\end{equation}
with $P_j(N)=N(N+2)\cdots(N+2j-2)$. For the lists without self-loops (one-vertex-reducible graphs included), the bare
mass is shifted, $m_0^2=1-(N+2)g/6$, so that every graph with a self-loop cancels against a graph with a mass
counterterm; then
\begin{gather}
\Gamma^{(2)}_0=1+\sum_{L\ge2}(-1)^{L+1}\Sigma_{2,L}\,g^L,\nonumber\\
\Gamma^{(4)}_0=\sum_{L\ge0}(-1)^{L}\Sigma_{4,L}\,g^{L+1},
\label{eq:sumrule}
\end{gather}
with $\Sigma_{E,L}=\sum_{G\in\mathcal G_E(L)}S_G$. With $m_0^2=1$ the same expansion gives the sums over all 1PI graphs,
self-loops included. Table~\ref{tab:sumrules} gives the polynomials $\Sigma_{E,L}(N)$. Through eight loops they are those
of the zero-dimensional computation of Ref.~\cite{Ribeca2003thesis}, whose table stops at $\Gamma^{(4)}$ with seven
loops and $\Gamma^{(2)}$ with seven, and whose original output extends to eight loops in both. The printed table has two
sign misprints, absent from the original output: the last four terms of the six-loop $\Gamma^{(2)}$ numerator and the
term in $N^3$ of the three-loop $\Gamma^{(2,1)}$ numerator have the same sign as the other terms. The nine- and
ten-loop rows were obtained here from Eqs.~\eqref{eq:zerod}--\eqref{eq:sumrule}, which reproduce every coefficient
of the original computation at lower orders.

The lists used here reproduce every row of Table~\ref{tab:sumrules} exactly, as polynomials in $N$, through ten
loops for both $\Gamma^{(4)}$ and $\Gamma^{(2)}$, up to the 1\,593\,254 ten-loop four-point graphs; the $O(N)$ factor of
every graph is counted exactly by carrying the partial pairings of its vertices from vertex to vertex, which agrees
graph by graph with the enumeration of all $3^V$ pairings wherever both run~\check{G2}. The lists with self-loops reproduce the bare series,
whose coefficients at $N=1$ are $1/2$, $5/12$, $5/6$, $115/48$, $625/72$, $10\,775/288$, $80\,275/432$,
$21\,615\,925/20\,736$, $67\,531\,025/10\,368$ and $620\,085\,875/13\,824$ at $L=1$--10. They also reproduce the sum rules of
$\Gamma^{(2,1)}$ of Ref.~\cite{Ribeca2003thesis}, obtained by inserting $\phi^2$ on any line of a graph without
self-loops and on the self-loop of a graph with exactly one, at five and six loops for all $N$, the seven-loop
polynomial of the original output, and its eight-loop value at $N=1$.

\begin{table*}
\caption{\label{tab:sumrules}%
Zero-dimensional sum rules: $\Sigma_{E,L}(N)=\sum_{G\in\mathcal G_E(L)}S_G$ over the $L$-loop 1PI graphs without self-loops,
one-vertex-reducible graphs included [Eq.~\eqref{eq:sumrule}], written as a numerator polynomial over a denominator
$D$, and its value at $N=1$. Rows to eight loops: Ref.~\cite{Ribeca2003thesis} (printed table and original output);
the nine- and ten-loop rows: this work, from the lists and from the exact zero-dimensional series, which agree.
\check{G2}}
\footnotesize
\begin{ruledtabular}
\begin{tabular}{llp{0.62\textwidth}rr}
$E$ & $L$ & Numerator & $D$ & $N=1$ \\
\hline
4 & 1 & $8+N$ & 6 & $3/2$ \\
  & 2 & $108+26N+N^2$ & 36 & $15/4$ \\
  & 3 & $2048+724N+62N^2+N^3$ & 216 & $105/8$ \\
  & 4 & $47648+21632N+2928N^2+124N^3+N^4$ & 1296 & $893/16$ \\
  & 5 & $1297664+702864N+127392N^2+9044N^3+220N^4+N^5$ & 7776 & $8795/32$ \\
  & 6 & $40362304+24950944N+5549104N^2+546408N^3+23404N^4+358N^5+N^6$ & 46656 & $97987/64$ \\
  & 7 & $1410726400+966932160N+250483648N^2+31004768N^3+1898240N^4+53312N^5+546N^6+N^7$ & 279936 & $10951025/1152$ \\
  & 8 & $54747705344+40784744192N+11897634304N^2+1746112576N^3+137234272N^4+5654416N^5+110208N^6+792N^7+N^8$ & 1679616 & $49985915/768$ \\
  & 9 & $2336870225920+1865178040064N+598669644160N^2+100305560704N^3+9494392864N^4+511500848N^5+14961264N^6+211032N^7+1104N^8+N^9$ & 10077696 & $2245562203/4608$ \\
  & 10 & $108856221154304+92103242480128N+31990963159296N^2+5962312144640N^3+652186499392N^4+42906530592N^5+1667820720N^6+36012864N^7+379788N^8+1490N^9+N^{10}$ & 60466176 & $4057808535/1024$ \\
\hline
2 & 2 & $2+N$ & 18 & $1/6$ \\
  & 3 & $10+7N+N^2$ & 54 & $1/3$ \\
  & 4 & $280+232N+52N^2+3N^3$ & 648 & $7/8$ \\
  & 5 & $2552+2384N+688N^2+71N^3+2N^4$ & 1944 & $211/72$ \\
  & 6 & $113888+116608N+39952N^2+5680N^3+320N^4+5N^5$ & 23328 & $3413/288$ \\
  & 7 & $1498912+1650832N+641936N^2+112904N^3+9260N^4+316N^5+3N^6$ & 69984 & $48323/864$ \\
  & 8 & $90828928+106190272N+45557472N^2+9353504N^3+982000N^4+51060N^5+1134N^6+7N^7$ & 839808 & $3123017/10368$ \\
  & 9 & $778012480+956252384N+444020448N^2+102678120N^3+12884816N^4+882730N^5+31028N^6+473N^7+2N^8$ & 1259712 & $9443467/5184$ \\
  & 10 & $118963074560+152597681664N+75647848448N^2+19237783296N^3+2767156512N^4+231239968N^5+10977808N^6+273456N^7+2982N^8+9N^9$ & 30233088 & $506798407/41472$ \\
\end{tabular}
\end{ruledtabular}
\end{table*}

\section{Reproducibility}
\label{app:reproducibility}

Every claim of this paper that a script can verify is marked in the text with the identifier of its check, as in
\check{G2}. The checks are part of the code repository (Appendix~\ref{app:data}), in its directory \texttt{checks/}: one
script per identifier, which states the claim as it is made here, its inputs in the repository and the expected result,
prints the numbers it obtains, and ends with \texttt{PASS} or \texttt{FAIL}. Table~\ref{tab:checks} lists them. The
command
\begin{verbatim}
make -C checks
\end{verbatim}
runs them all and writes a report, \texttt{checks/REPORT.md}; \texttt{make -C checks quick} runs those that take less
than about a minute, and \texttt{python3 checks/run.py G2} runs one. They need Python~3.6 or later with its standard
library, GNU make and a C++17 compiler with OpenMP, which build the CPU integrator and the compiled helpers on first
use. Where a check needs what the machine lacks, a
GPU for the live part of N2 or the \textsc{mpmath} library for I1, it reports \texttt{SKIP} for that part and verifies
instead the stored outputs of the runs that did it, which the repository keeps (\texttt{results/}).

The numbers that describe the pooled Monte Carlo data are not typed in the sources of the paper. The bodies of
Tables~\ref{tab:validation}, \ref{tab:cost}, \ref{tab:newsums}--\ref{tab:newcoeffnu}, \ref{tab:routes},
\ref{tab:exponents} and~\ref{tab:largeorder}, the rows of this work in Table~\ref{tab:effort}, Figs.~\ref{fig:cost}
and~\ref{fig:costhist}, and the values quoted from them in the text are written by \texttt{make -C series} from the
ledger of pooled per-class statistics (Sec.~\ref{sec:methods-ledger}) and from the stored runs of route B. Check S2
regenerates all of them except Table~\ref{tab:routes}, which check S4 regenerates, and both verify that they are
identical to those printed here. The exceptions are the numbers of individual early runs and tests in
Sec.~\ref{sec:results-validation} (the round-off example of $H_4$ and the test of fused multiply-add contraction),
which are historical; where a check exists, it verifies them from their stored outputs.

\begin{table*}
\caption{\label{tab:checks}%
The reproducibility checks of the repository (directory \texttt{checks/}). Where: the sections and appendices whose
claims cite the check. Needs: CPU, a computation on a CPU with the graph tools or the CPU build of the integrator; GPU,
a computation on a GPU; stored, only the stored outputs of the repository (the ledger and the reference data of the
checks).}
\begin{ruledtabular}
\begin{tabular}{lp{0.2\textwidth}p{0.55\textwidth}l}
Id & Where & Claim & Needs \\
\hline
\checkid{G1} & \ref{sec:methods-generation}, \ref{sec:methods-checks}, App.~\ref{app:graphs} &
  Every graph list holds distinct, connected 1PI graphs in canonical Nickel form, in the numbers of
  Tables~\ref{tab:generation} and~\ref{tab:graphcounts} & CPU \\
\checkid{G2} & \ref{sec:methods-checks}, \ref{sec:methods-validation}, App.~\ref{app:sumrules} &
  The lists satisfy the zero-dimensional sum rules exactly, as polynomials in $N$, through ten loops, and reproduce
  every row of Table~\ref{tab:sumrules} & CPU \\
\checkid{G3} & \ref{sec:methods-checks} &
  The lists equal those of Ref.~\cite{Phi4tools2024} as sets: $\Gamma^{(4)}$ to seven loops, $\Gamma^{(2)}$ to eight
  & CPU \\
\checkid{G4} & \ref{sec:methods-checks} &
  The weights and $O(N)$ polynomials equal those of Refs.~\cite{Ribeca2003thesis,GuidaRibeca2006} on all graphs with
  $L\le5$ and on the four-loop tables & CPU \\
\checkid{G5} & \ref{sec:methods-tropical}, App.~\ref{app:graphs} &
  The counts of Table~\ref{tab:graphcounts} follow from the lists (1VR by a cut-vertex test, checked against the
  lists' flags), and the ledger holds that many classes for every order with data & CPU \\
\checkid{I1} & \ref{sec:methods-checks}, \ref{sec:methods-validation} &
  The expanded forest formula, Eq.~\eqref{eq:integrand}, agrees pointwise with an independent evaluation in 100-digit
  arithmetic & CPU\footnotemark[1] \\
\checkid{I2} & \ref{sec:methods-construction}, \ref{sec:methods-checks}, \ref{sec:methods-validation},
  App.~\ref{app:graphs} &
  One-vertex-reducible graphs, graphs with self-loops and generalized tadpoles vanish after the pruning & CPU \\
\checkid{I3} & \ref{sec:methods-construction}, \ref{sec:methods-checks}, \ref{sec:methods-validation} &
  Twins (2-isomorphic graphs) have pointwise equal subtracted integrands & CPU \\
\checkid{I4} & \ref{sec:methods-construction} &
  A run that loads the built classes from the cache is bitwise identical to one that builds them & CPU \\
\checkid{N1} & \ref{sec:methods-rng} &
  Our Philox4x32-10 reproduces the known-answer vectors of Ref.~\cite{Salmon2011Random123} & CPU \\
\checkid{N2} & \ref{sec:methods-kernel}, \ref{sec:methods-validation}, \ref{sec:results-validation} &
  On identical random numbers the GPUs reproduce the per-class means of the CPU to round-off (golden runs)
  & CPU, GPU\footnotemark[2] \\
\checkid{N3} & \ref{sec:methods-checks}, \ref{sec:results-validation} &
  The exact two-loop sums $B_2=1/2$, $H_2=1/81$ and $C_2=1/12$ are reproduced within the errors & stored \\
\checkid{N4} & \ref{sec:results-validation} &
  The sums at three to seven loops agree within the errors with those derived from Ref.~\cite{Phi4tools2024}; the
  statistics of the pulls & stored \\
\checkid{N5} & \ref{sec:methods-ledger}, \ref{sec:results-validation} &
  The ledger is rebuilt bit for bit from the archive of its runs, and every run is consistent class by class with
  the pool of the runs before it (per-class $\chi^2$) & stored\footnotemark[3] \\
\checkid{N6} & \ref{sec:results-validation} &
  FP64 round-off does not bias the order sums: paired double and extended-precision evaluation of 55 classes (the 43 most exposed of $B_8$, $H_8$,
  $C_8$, $H_9$, $C_9$ and $C_{10}$, and 12 controls) finds no significant difference, and at most $4\times10^{-6}$ of an order's error summed over its classes & CPU \\
\checkid{N7} & \ref{sec:methods-tropical}, \ref{sec:results-validation} &
  Eight replicas of each of $B_6$, $H_6$, $C_6$, $B_7$, $H_7$ and $C_7$ scatter as their quoted errors say, order
  sums and class by class & stored \\
\checkid{N8} & \ref{sec:results-validation} &
  The round-off scales of the four classes of $H_4$ quoted as an example in Sec.~\ref{sec:results-validation}
  & stored \\
\checkid{S1} & \ref{sec:methods-validation}, \ref{sec:results}, App.~\ref{app:conventions} &
  The Callan--Symanzik relations map the sums derived from Ref.~\cite{Phi4tools2024} back to its coefficients, to
  $10^{-15}$ & stored \\
\checkid{S2} & \ref{sec:methods-ledger}, \ref{sec:resummation-criteria}, \ref{sec:resummation-errors},
  \ref{sec:methods-validation}, \ref{sec:results-validation}, \ref{sec:results-performance}, \ref{sec:results-new},
  \ref{sec:resummation-results} &
  Every number of the generated tables and values is regenerated from the ledger by \texttt{make -C series} & stored \\
\checkid{S3} & \ref{sec:methods-checks}, \ref{sec:methods-validation} &
  The large-order ratios of Table~\ref{tab:largeorder} follow from the coefficients and the constants of
  Sec.~\ref{sec:methods-validation} & stored \\
\checkid{S4} & \ref{sec:methods-validation}, \ref{sec:results-new} &
  Route B reproduces the coefficients of Ref.~\cite{Phi4tools2024} from its per-graph values, its eight-vertex
  sums agree with those of Ref.~\cite{Phi4tools2024} and its ninth- and tenth-order coefficients with route A
  (Table~\ref{tab:routes}) & stored \\
\checkid{S5} & \ref{sec:methods-validation}, \ref{sec:results-new} &
  The terms in $N^L$ of every order sum vanish class by class, as the limit $N\to\infty$ requires, and those in
  $N^{L-1}$ of every order of the ledger agree with the $1/N$ expansion, which reproduces the leading coefficients of
  Ref.~\cite{Phi4tools2024}; the polynomials $p_{X,k}(N)$ have the degrees and leading coefficients that follow
  & stored \\
\end{tabular}
\end{ruledtabular}
\footnotetext[1]{With the \textsc{mpmath} library; without it, the check verifies the stored values
(\texttt{results/checks/}).}
\footnotetext[2]{Without a GPU, the check verifies the stored GPU outputs (\texttt{results/checks/}).}
\footnotetext[3]{The replay needs the archive of the runs (\texttt{ledger-archive/}, outside the repository);
without it, the check verifies the invariants of the ledger and reports the replay as skipped.}
\end{table*}

\section{Data and code availability}
\label{app:data}

The code, the data and the checks are in the repository ReBar, \url{https://github.com/PaoloRibeca/ReBar},
whose release 0.4.3 is archived at Zenodo, \url{https://doi.org/10.5281/zenodo.23750242}: the
code under the GNU General Public License, version~3 or later, the data and the documentation under the Creative
Commons Attribution 4.0 International licence, and the text of this paper under the Creative Commons
Attribution-NonCommercial-NoDerivatives 4.0 International licence. The repository is the Supplemental Material of this paper; its file
\texttt{SUPPLEMENTAL.md} says where each of the following items is.
\begin{enumerate}
\item \emph{Graph lists}: $\Gamma^{(4)}$ and $\Gamma^{(2)}$ to ten loops, in the notation of
Refs.~\cite{Ribeca2003thesis,GuidaRibeca2006} (\texttt{graphs/lists/}) and, in the form the integrator reads
(\texttt{integrator/data/lists/}), with the canonical Nickel index, the weight $E!/|\mathrm{Aut}\,G|$ at $N=1$ and
the one-vertex-reducible and self-loop flags of every graph; the $O(N)$ weight polynomials are in the ledger, class by
class, with the SHA-256 of every list the integrator ran on. The lists to seven ($\Gamma^{(4)}$) and eight
($\Gamma^{(2)}$) loops coincide with those of Ref.~\cite{Phi4tools2024}, version~1.0.2 (check G3), and the four
six- and seven-loop lists of the integrator are its files, with its columns, reproduced under its MIT licence; the
$\Gamma^{(4)}$ lists from eight loops and the $\Gamma^{(2)}$ lists from nine loops on are new.
\item \emph{Per-class results} (\texttt{ledger/}): the statistics of every class pooled over the runs (number of
samples, mean and central moments to the fourth, extremes, round-off indicators, time per sample), the exact weight
polynomials, the per-order sums with the reference values and the pulls, and a record of every run with its own
statistics (Sec.~\ref{sec:methods-ledger}); and, in the archive deposited with the release (too large for the
repository), the output of every run as received, with the integral, error, samples and diagnostics of every class. With them the coefficients can be recombined for any $N$, the error propagation of
Sec.~\ref{sec:resummation-errors} can be repeated, and two runs can be compared class by class.
\item \emph{Run records}: dated snapshots of the series and tables derived from the ledger
(\texttt{results/coefficients/}), and the reference data and inputs of the checks (\texttt{results/checks/}). The
complete records of the campaigns of Sec.~\ref{sec:results} (reports, logs and environment records) are kept outside
the repository; the run files of the ledger's contributions are deposited with its archive.
\item \emph{Series} (\texttt{series/}): the tools that validate and pool the runs and derive from the ledger the
polynomials $B_L(N)$, $H_L(N)$, $C_L(N)$ and $p_{X,k}(N)$, the coefficients and the generated tables of this paper; the
polynomials $p_{X,k}(N)$ with the full covariance of their coefficients (Sec.~\ref{sec:results-new}), in
\texttt{series/generated/polynomials.json} and, one row per coefficient, \texttt{polynomials.tsv}, with a dated copy
in \texttt{results/coefficients/}.
\item \emph{Checks} (\texttt{checks/}): the scripts of Appendix~\ref{app:reproducibility}.
\item \emph{Code} (\texttt{integrator/}, \texttt{graphs/}): the program that builds and integrates the subtracted integrands (C++17 and CUDA from a single
source; versions 0.3.0 to 0.4.3 for the runs of Sec.~\ref{sec:results}, each recorded with its build signature in
the ledger), with
its self-tests and validation driver, and, in \texttt{results/checks/}, the golden reference files of
Sec.~\ref{sec:methods-kernel}; the graph generator of
Sec.~\ref{sec:methods-generation}; and the scripts that convert the graphs to Nickel indices and evaluate the sum
rules of Appendix~\ref{app:sumrules}.
\end{enumerate}
Every results archive records the hardware, driver and compiler versions and the checksums of all sources. Since the
random numbers of a class depend only on the seed, the kind of integrand, the loop order and the Nickel index
(Sec.~\ref{sec:methods-rng}), any class of any run can be recomputed independently: bit for bit with the same build
on CPUs, for any number of threads, and within round-off with other compilers and libraries and on GPUs
(Sec.~\ref{sec:methods-kernel}).

\bibliographystyle{apsrev4-2}
\bibliography{references}

\end{document}